\documentclass[preprint,aps,superscriptaddress,nofootinbib,tightenlines,floatfix]{revtex4}
\usepackage{graphicx}
\usepackage{bm}
\usepackage{comment}
\usepackage{color}

\def\si{^1 \hskip -0.03in S _0}
\def\siii{^3 \hskip -0.025in S _1}
\def\diii{^3 \hskip -0.03in D _1}

\def\htpnn{$^{3} {\rm H}$}
\def\hetppn{$^{3} {\rm He}$}
\def\htL{$_{\Lambda} ^{3} {\rm H}$}
\def\htLtilde{$_{\Lambda} ^{3} {\rm \tilde H}$}
\def\htLL{$_{\Lambda\Lambda} ^{~\, 3} {\rm H}$}

\def\hetL{$_{\Lambda} ^{3} {\rm He}$}
\def\nnL{$nn\Lambda$}
\def\htL{$_{\Lambda} ^{3} {\rm H}$}
\def\hetS{$_{\Sigma} ^{3} {\rm He}$}

\def\htLJhalf{_{\Lambda}^3{\rm H}(1/2^+)}
\def\htLJthreehalf{_{\Lambda}^3{\rm H}(3/2^+)}

\def\hfL{$_{\Lambda} ^{4} {\rm H}$}

\def\hef{$^{4} {\rm He}$}
\def\hefL{$_{\Lambda} ^{4} {\rm He}$}
\def\hefLL{$_{\Lambda\Lambda} ^{~\, 4} {\rm He}$}
\def\nnLL{$nn\Lambda\Lambda$}
\def\hfLL{$_{\Lambda\Lambda} ^{~\, 4} {\rm H}$}

\def\hymag{$\Lambda\Xi^0pnn$}

\def\revcolor{red}
\def\revcolor{black}

\def\b{0.145}

\def\La{3.4}
\def\Lb{4.5}
\def\Lc{6.7}
\def\Lafm{\La~{\rm fm}}
\def\Lbfm{\Lb~{\rm fm}}
\def\Lcfm{\Lc~{\rm fm}}
\def\Ta{6.7}
\def\Tb{6.7}
\def\Tc{9.0}
\def\NcfgA{3822} 
\def\NscrA{\textcolor{\revcolor}{72}}
\def\NcfgB{3050} 
\def\NscrB{\textcolor{\revcolor}{48}}
\def\NcfgC{\textcolor{\revcolor}{1905}} 
\def\NscrC{\textcolor{\revcolor}{54}}

\def\blattapprox{0.145~{\rm fm}}
\def\blatt{0.1453(16)~{\rm fm}}
\def\cfga{$24^3\times 48$\ }
\def\cfgb{$32^3\times 48$\ }
\def\cfgc{$48^3\times 64$\ }

\def\mpi{ 800 }
\def\mpiMeV{ \mpi~{\rm MeV} }
\def\mpifull{ \textcolor{\revcolor}{805.9(0.6)(0.4)(8.9)} }
\def\mpifullMeV{ \mpifull~{\rm MeV} }
\def\xipi{ \textcolor{\revcolor}{1.0055 (57) (26)} }

\def\mBfull{  \textcolor{\revcolor}{ 1.634(0)(0)(18)~{\rm GeV} } } 
\def\xiB{ \textcolor{\revcolor}{1.019 (10) (03)}}

\def\Bd{  \textcolor{\revcolor}{19.5(3.6)(3.1)(0.2)}  }
\def\Bnn{  \textcolor{\revcolor}{ 15.9(2.7)(2.7)(0.2)}  }
\def\BnXJone{ \textcolor{\revcolor}{37.7(3.0)(2.7)(0.4)} }
\def\BnSJone{  \textcolor{\revcolor}{5.5(3.4)(3.7)(0.0)} }
\def\BHgs{  \textcolor{\revcolor}{74.6(3.3)(3.3)(0.8)}  }
\def\BdMeV{ \Bd~{\rm MeV}  }

\def\BnXJ1MeV{\BnXJone~{\rm MeV}  }
\def\BnSJ1MeV{\BnSJone~{\rm MeV}  }
\def\BHgsMeV{\BHgs~{\rm MeV}  }

\def\Bhet{\textcolor{\revcolor}{53.9(7.1)(8.0)(0.6)} }       % helium-3
\def\BhetGSperAapprox{24~{\rm MeV}} 
\def\BhetLJhalf{ \textcolor{\revcolor}{ 69(5)(12)(0) } } 
 
\def\BhtLJhalf{ \Bhet }                                       % hypertriton j=1/2 
\def\BhtLJthreehalf{  \textcolor{\revcolor}{82(8)(12)(1)}  } % hypertriton j=3/2 
\def\BhetS{ \textcolor{\revcolor}{55(6)(10)(1)} }
 
\def \BhtLJthreehalfminusonehalf{ \textcolor{\revcolor}{ 26.2(2.3)(5.5)(0.3)~{\rm
      MeV}}  } % hypertriton spin splitting  3/2 - 1/2

\def\BhetMeV{\Bhet~{\rm MeV}} 
\def\BhetLJhalfMeV{\BhetLJhalf~{\rm MeV}} 
 
\def\BhtLJhalfMeV{ \BhtLJhalf~{\rm MeV } } 
\def\BhtLJthreehalfMeV{\BhtLJthreehalf~{\rm MeV } }

\def\Bhef{ \textcolor{\revcolor}{107(12)(21)(1) } }
\def\BhefL{  \Bhef  } 
\def\BhefLL{\textcolor{\revcolor}{ 156(16)(21)(2)} } 
\def\BhefLLtwentyseven{ \BhefLL  } 
 
\def\BhefMeV{ \Bhef~{\rm MeV}}

\def\Bhymag{ \textcolor{\revcolor}{245(28)(81) } }

\newcount\hour \newcount\hourminute \newcount\minute 
\hour=\time \divide \hour by 60
\hourminute=\hour \multiply \hourminute by 60
\minute=\time \advance \minute by -\hourminute
\newcommand{\mydate}{\ \today \ - \number\hour :\number\minute}

\begin{document}

\title{ Light Nuclei and Hypernuclei from Quantum Chromodynamics in
  the Limit of SU(3) Flavor Symmetry }

\author{S.R.~Beane} \affiliation{Department of Physics, University of
  New Hampshire, Durham, NH 03824-3568, USA}

\author{E.~Chang} \affiliation{Dept. d'Estructura i Constituents de la
  Mat\`eria.  Institut de Ci\`encies del Cosmos (ICC), Universitat de
  Barcelona, Mart\'{\i} i Franqu\`es 1, E08028-Spain}

\author{S.D.~Cohen} \affiliation{Department of Physics, University of
  Washington, Box 351560, Seattle, WA 98195, USA}

\author{W.~Detmold} \affiliation{Department of Physics, College of
  William and Mary, Williamsburg, VA 23187-8795, USA}
\affiliation{Jefferson Laboratory, 12000 Jefferson Avenue, Newport
  News, VA 23606, USA}

\author{H.W.~Lin} \affiliation{Department of Physics, University of
  Washington, Box 351560, Seattle, WA 98195, USA}

\author{T.C.~Luu} \affiliation{N Section, Lawrence Livermore National
  Laboratory, Livermore, CA 94551, USA}

\author{K.~Orginos} \affiliation{Department of Physics, College of
  William and Mary, Williamsburg, VA 23187-8795, USA}
\affiliation{Jefferson Laboratory, 12000 Jefferson Avenue, Newport
  News, VA 23606, USA}

\author{A.~Parre\~no} \affiliation{Dept. d'Estructura i Constituents
  de la Mat\`eria.  Institut de Ci\`encies del Cosmos (ICC),
  Universitat de Barcelona, Mart\'{\i} i Franqu\`es 1, E08028-Spain}

\author{M.J.~Savage} \affiliation{Department of Physics, University of
  Washington, Box 351560, Seattle, WA 98195, USA}

\author{A.~Walker-Loud} \affiliation{Lawrence Berkeley National
  Laboratory, Berkeley, CA 94720, USA}\affiliation{Department of Physics, 
University of California, Berkeley, CA 94720, USA}

\collaboration{ NPLQCD Collaboration }

\date{\mydate}

\begin{abstract}
  \noindent
  The binding energies of a range of nuclei and
  hypernuclei with atomic number $A\le 4$ and strangeness $|s|\le 2$, 
including the deuteron, di-neutron, H-dibaryon, \hetppn, \hetL,
\hef, 
  \hefL, and \hefLL, 
are calculated in the limit of flavor-SU(3) symmetry at the
  physical strange-quark mass
with quantum  chromodynamics
(without electromagnetic interactions).  
The nuclear states are extracted from
  Lattice QCD calculations performed with $n_f=3$ dynamical light
  quarks using an isotropic clover discretization of the quark action
  in three lattice volumes of spatial extent $L\sim \Lafm, \Lbfm$
  and $\Lcfm$, and with a single lattice spacing $b\sim
  \blattapprox$.
\end{abstract}
\maketitle
%\tableofcontents
\vfill\eject
%

%%%%%%%%%%%%%%%%%%%%%%%%%%%%%%%%%%%%%%%%%%%%%%%%%%%%
\section{Introduction}
\label{sec:intro}
\noindent
The structure and interactions of the light nuclei have been the focus
of experimental and theoretical explorations since the infancy of
nuclear physics.  Yet more than one hundred years later, and despite
having made remarkable progress in describing these systems in
terms of nuclear forces that are well-constrained by experiment, we
remain unable to predict the binding and interactions of any given
nucleus with reliable estimates of the associated uncertainties.  It has long been accepted that
quantum chromodynamics (QCD) and the electroweak interactions produce
the nuclear forces, and consequently are responsible for the structure
and interactions of all nuclei.  Unfortunately, the complexity of the
QCD vacuum  has
so far prevented the calculation of low-energy and medium-energy
nuclear systems directly from QCD.  Beyond recovering the results
of decades of experimental investigation, it is crucial to
establish and verify tools with which to perform such calculations,
with quantifiable uncertainties, in order to determine
the properties and structure of exotic nuclei, and of matter in
extreme environments or in kinematic regimes where experiments are not
possible or practical.

The only known way with which to calculate the
low-energy properties of
hadronic and nuclear
systems rigorously is Lattice QCD
(LQCD).  In LQCD calculations, the quark and gluon fields are defined
on a discretized space-time of finite volume, and the path integral
over the fields is evaluated numerically.  While LQCD calculations
deviate from those of QCD due to the finite distance between points of
the grid (lattice spacing) and the finite volume of the grid (lattice
volume), such deviations can be systematically removed by reducing the
lattice spacing, increasing the lattice volume and extrapolating to
the continuum and infinite-volume limits using the known dependences
determined with effective field theory (EFT).
Calculation of important quantities in nuclear physics using LQCD 
is only now becoming practical, with  
first calculations of
simple multibaryon interactions being recently performed,
although not at the physical values of the light-quark masses.
Early exploratory quenched calculations of the nucleon-nucleon (NN) scattering
lengths~\cite{Fukugita:1994na,Fukugita:1994ve} performed more than a
decade ago have been superseded by $n_f=2+1$ calculations within the
last few years~\cite{Beane:2006mx,Beane:2009py} (and added to by
further quenched calculations~\cite{Ishii:2006ec,Aoki:2008hh,Aoki:2009ji}).
Further, the first quenched calculations of the
deuteron~\cite{Yamazaki:2011nd}, $^3$He and
$^4$He~\cite{Yamazaki:2009ua} have been performed, along with
$n_f=2+1$ calculations of $^3$He and multibaryon
systems containing strange quarks~\cite{Beane:2009gs}.  
In addition, efforts to
explore nuclei and nuclear matter using the strong coupling limit of
QCD have led to some interesting
observations~\cite{deForcrand:2009dh}.  
Recently, $n_f=2+1$ calculations~\cite{Beane:2010hg,Beane:2011iw} and
$n_f=3$ calculations~\cite{Inoue:2010es,Inoue:2011pg,Inoue:2011ai} have provided evidence
that the H-dibaryon (with the quantum numbers of $\Lambda\Lambda$) is
bound at a pion mass of $ m_\pi\sim 390~{\rm MeV}$ at the physical
value of the strange-quark mass~\cite{Beane:2010hg,Beane:2011iw} and over a
range of 
SU(3)-degenerate light-quark masses with 
$m_\pi\sim 469~{\rm MeV}$ to $1171~{\rm MeV}$~\cite{Inoue:2010es,Inoue:2011pg}. 
Extrapolations to the physical light-quark
masses suggest that a weakly bound H-dibaryon or a near-threshold resonance
exists in this
channel~\cite{Beane:2011xf,Shanahan:2011su,Beane:2011iw,Haidenbauer:2011ah}.
We have searched for bound states in other channels at $ m_\pi\sim
390~{\rm MeV}$~\cite{Beane:2010hg,Beane:2011iw}, and evidence
has been found for a bound state in the strangeness-4 $\Xi^0\Xi^0$
system.  This is consistent with model-dependent  and EFT 
predictions of a bound state at the physical pion
mass~\cite{Stoks:1999bz,Miller:2006jf,Haidenbauer:2009qn}.  
In
addition to the identification of bound states, 
calculations of hyperon-nucleon scattering
extrapolated to the physical pion mass (using leading-order (LO) EFT)
have been performed
and directly compared with the results of phase-shift
analyses of experimental data~\cite{Beane:2012ey}.

In this work we focus on the lightest nuclei and hypernuclei and present results of the 
first LQCD calculations of a number of s-shell nuclei and hypernuclei
with $A\leq5$, including \hetppn, \hetL, \hefL, \hefLL, and a
five-body state \hymag\  in the limit of exact SU(3)-flavor symmetry
(and consequently, exact isospin symmetry).  
Hypernuclear spectroscopy is
enjoying an experimental renaissance with ongoing and planned programs
at  DA$\Phi$NE,  FAIR, Jefferson Lab, J-PARC and Mainz 
providing motivation for  enhanced theoretical efforts (for a recent review,
see Ref.~\cite{Pochodzalla:2011rz}).
Our LQCD calculations are
performed using an isotropic clover quark  action at the SU(3)-flavor 
symmetric point corresponding to the physical strange-quark mass, with
$m_\pi=m_K=m_\eta \sim \mpiMeV$.  
Three lattice volumes have been
employed with spatial extent of $L\sim \Lafm, \Lbfm$ and $\Lcfm$, 
and calculations of systems with non-zero total momentum (boosted systems) have
been performed  
to investigate the volume dependence of binding energies~\cite{Bour:2011ef,Davoudi:2011md}.
\textcolor{\revcolor}{
As this is the first calculation of hypernuclei with baryon number $A>2$, 
it is prudent to establish benchmarks for future works.  
The spectra of nuclei will have the simplest structure at the SU(3)
symmetry point,  where the up, down and strange quarks have the same mass,
allowing for a relatively uncomplicated analysis.
While any common light-quark mass could have been used, the physical value of
the strange quark mass was chosen 
so that only the (common) up and down quark masses deviated from their physical
values, 
and also so that the four- and five-baryon systems
would be well contained within the three selected lattice volumes.
Further, such a large value of the pion mass, combined with the temporal extent of the
gauge-field configurations, strongly suppresses thermal effects which are
present in all calculations and can provide a systematic
uncertainty in extracting  the small energy differences present in nuclei.
}
Only one relatively coarse lattice spacing, $b\sim \blattapprox$
has been
used in the calculations, dictated by the available computational resources, 
and therefore, an extrapolation to the
continuum has not been performed.  
Further, extrapolations to the
physical light-quarks masses have not been attempted because the quark-mass
dependences of the energy levels in the light nuclei are not known.
Future calculations at smaller lattice spacings and at
lighter quark masses will facilitate such extrapolations and lead to
first predictions for the spectrum of light nuclei with completely
quantified uncertainties that can be compared with experiment.

%%%%%%%%%%%%%%%%%%%%%%%%%%%%%%%%%%%%%%%%%%%%%%%%%%%%
\section{Lattice QCD Calculations }
\label{sec:Method}
\noindent

%%%%%%%%%%%%%%%%%%%%%%%%%%%%%%%%%%%%%%%%%%%%%%%%%%%%
\subsection{Computational Overview}
\label{sec:calcs}
\noindent
Three ensembles of isotropic gauge-field 
configurations, generated with a tadpole-improved L\"uscher-Weisz gauge
action and a clover fermion action~\cite{Sheikholeslami:1985ij},
are used in this work. 
This particular lattice-action setup follows closely the anisotropic clover
action of the ensembles generated by the JLab group that we have used in our
previous calculations~\cite{Beane:2009py,Beane:2009gs,Beane:2010hg,Beane:2011iw,Beane:2011xf,Beane:2012ey}. 
The parameter tuning and scaling properties of this action will be
discussed elsewhere~\cite{ISO}.
One level of stout smearing~\cite{Morningstar:2003gk} with
$\rho=0.125$ and tadpole-improved tree-level clover
coefficient $c_{\rm SW}=1.2493$ 
are  used in the gauge-field  generation.
Studies~\cite{Hoffmann:2007nm,Edwards:2008ja,ISO} of the PCAC relation in the Schr\"odinger functional indicate
that this choice is consistent with 
vanishing ${\cal O}(b)$ violations,
leading to discretization effects that are essentially ${\cal O}(b^2)$. 
The parameters of the ensembles are listed in Table~\ref{tab:gauageparams},
and further details will be presented elsewhere~\cite{ISO}. 
As multibaryon systems are the focus of this work, relatively large lattice volumes
are employed for the calculations, with 
correspondingly large values of $m_\pi L$ and $m_\pi T$.
\textcolor{\revcolor}{
In order to convert the calculated  (binding) energies from lattice units (l.u.)
into physical units (MeV),
a lattice spacing of $b=\blatt$ has been determined for these ensembles of
gauge-field configurations
from the $\Upsilon$ spectrum~\cite{meinelPRIV}.
}
\begin{table}
\begin{center}
\begin{minipage}[!ht]{16.5 cm}
  \caption{Parameters of the ensembles of gauge-field configurations and of the measurements used in this work.
    The lattices have dimension  $L^3\times T$, a lattice spacing $b$, and
    a bare quark mass $b\ m_q$ (in lattice units) generating a pion of
    mass $m_\pi$. 
$N_{\rm src}$ light-quark sources are used (as described in the text)
to perform measurements on $N_{\rm cfg}$ configurations in each ensemble.
 \textcolor{\revcolor}{
The three uncertainties associated with the pion mass are
statistical, fitting systematic and that associated with the 
lattice spacing, respectively.
}
  }  
\label{tab:gauageparams}
\end{minipage}
\setlength{\tabcolsep}{0.3em}
\begin{tabular}{c|cccccccccccc}
\hline
      Label & $L/b$ & $T/b$ & $\beta$ & $b\ m_q$ & $b$ [fm]  & $L$ [fm] & $T$
      [fm] & $m_\pi$ [MeV] & $m_\pi L$ & $m_\pi T$ & $N_{\rm cfg}$ & $N_{\rm src}$\\
\hline
      A& 24 & 48 & 6.1 &-0.2450 & \b& \La & \Ta & 806.5(0.3)(0)(8.9) & 14.3 & 28.5 & \NcfgA & \NscrA\\
      B&       32 & 48 & 6.1 &-0.2450 & \b  &  \Lb & \Tb & 806.9(0.3)(0.5)(8.9) & 19.0 & 28.5  & \NcfgB & \NscrB\\	
      C&48 & 64 & 6.1 &-0.2450 & \b  &  \Lc & \Tc & 806.7(0.3)(0)(8.9) & 28.5 & 38.0 & \NcfgC & \NscrC\\
\hline
\end{tabular}
%noalign{\smallskip\hrule}\cr}
\begin{minipage}[t]{16.5 cm}
\vskip 0.0cm
\noindent
\end{minipage}
\end{center}
\end{table}     

The $N_{\rm cfg}$ gauge configurations in each of the ensembles are
separated by at least 10 HMC evolution trajectories to reduce
autocorrelations, 
and 
\textcolor{\revcolor}{an average of} 
$N_{\rm src}$ measurements are
performed on each configuration. 
The quark propagators were constructed with gauge-invariant Gaussian-smeared sources with  
stout-smeared gauge links. 
These sources are distributed over a grid, the center of which is
randomly distributed within the lattice volume on each configuration,
and  the quark propagators
are computed  using the BiCGstab algorithm  with a
tolerance  of $10^{-12}$ in double precision.  
The quark propagators, either  unsmeared or 
smeared at
the sink using the same parameters as used at the source, 
give rise to two sets of correlation functions for each combination of source and sink interpolating  fields,
labeled as SP and SS, respectively.  
The propagators are 
contracted to form baryon blocks projected to fixed momentum at the sink
for use
in the calculation of the correlation functions to be described
below. 
The blocks are defined as
\begin{equation}
  \label{eq:blockdef}
  {\cal B}^{ijk}_{H}({\bf p},t;x_0)= \sum_{\bf x}e^{i{\bf p}\cdot{\bf
      x}} S_{i}^{(f_1),i^\prime} ({\bf x},t;x_0)S_{j}^{(f_2),j^\prime}({\bf x},t;x_0)
  S_{k}^{(f_3),k^\prime}({\bf x},t;x_0) b^{(H)}_{i^\prime j^\prime k^\prime}
\ \ \ ,
\end{equation}
where $S^{(f)}$ is a quark propagator of flavor $f$, and the indices
are combined spin-color indices running over
$i=1,\ldots,N_cN_s$.\footnote{To be specific, for a quark spin
  component $i_s=1,\ldots,N_s$ and color component
  $i_c=1,\ldots,N_c$, the combined index $i= N_c(i_s-1)+i_c$.}  The
choice of the $f_i$ and the tensor $b^{(H)}$ depend on the spin and
flavor of the baryon, $H$, under consideration. For our calculations we used
the local interpolating fields constructed in Ref.~\cite{Basak:2005ir}, restricted to those
that contain only upper spin components (in the Dirac spinor basis). This choice
results in the simplest interpolating fields that also have the best overlap with the
octet-baryon ground states.
Blocks are constructed for all lattice momenta $|{\bf p}|^2<4$ allowing for the 
study of multibaryon systems with zero or nonzero total momentum and  
with nontrivial spatial wave functions.

%%%%%%%%%%%%%%%%%%%%%%%%%%%%%%%%%%%%%%%%%%%%%%%%%%%%
\subsection{Multibaryon Interpolating Operators and Contractions}
\label{sec:interpolatorss}
\noindent
In order to define correlation functions for the multihadron
systems, interpolating operators with  well defined quantum
numbers at the source and sink are constructed. 
As we intend to perform calculations away from the SU(3)-flavor symmetry limit
at lighter quark masses, 
the quantum numbers of parity $\pi$, angular momentum  ${\bf J}^2$ and
$J_z$, strangeness $s$, baryon number (atomic number) $A$, and isospin ${\bf I}^2$
and $I_z$ 
are used to define the interpolating operators.\footnote{
For calculations restricted to the SU(3)-flavor symmetric limit,
it would also be advantageous to work directly with
  SU(3) irreducible representations.} 
These interpolating operators
are first constructed recursively at the hadronic level from the 
octet-baryon field operators using the appropriate group products (Clebsch-Gordan
coefficients for isospin and angular momentum) 
to build an outer product wavefunction 
$|{\rm  space}\rangle\otimes|{\rm  ang. mom.}\rangle\otimes |{\rm isospin}\rangle\otimes |{\rm
  parity}\rangle$ of given strangeness and baryon
number. This approach is similar to that used in Ref.~\cite{Basak:2005ir} in the context of excited baryons.
The baryons within this wavefunction are then replaced by
appropriate quark-level wavefunctions, of which there are, in principle,  multiple choices,
and then a quark-level anti-symmetrization is performed  (as color is
included in the quark level wavefunctions).
A similar approach has been used to investigate the $\Omega^-\Omega^-$ system~\cite{Buchoff:2012ja}.

The quantum numbers defining the systems that we discuss in this paper are shown in
Table~\ref{tab:ops}. 
States are given a representative hadronic label
(first column in Table~\ref{tab:ops}) indicating one component of their hadronic level
wavefunction. 
In order to determine which SU(3) irreducible representations (irreps) 
are present in the correlation
functions, the states are acted on by the 
quadratic and cubic SU(3) Casimir operators, and by 
V-spin, U-spin and isospin raising and lowering operators, the results of
which are presented in Table \ref{tab:ops} 
(eigenvalues of the Casimir operator for relevant SU(3)
irreps are tabulated in Appendix \ref{app:su3operators}). 
Because of
the overall antisymmetric nature of allowed quark-level
wavefunctions, a number of the constructed interpolating operators 
give rise to correlation functions that contain only one SU(3) irrep,
while others contain more than one.

\begin{table}[!t]
  \centering
  \begin{ruledtabular}
    \begin{tabular}{@{} ccccccc @{}}
      Label & $A$ & $s$ & $I$ & $J^\pi$ &  Local SU(3) irreps& This work\\
      \hline
      % One body
      $N$ & 1 & 0 & $1/2$ & $1/2^+$ &  {\bf 8}  &  {\bf 8} \\
      $\Lambda$ & 1 & -1 & $0$ & $1/2^+$ &   {\bf 8}  &  {\bf 8}\\
      $\Sigma$ & 1 & -1 & $1$ & $1/2^+$ &  {\bf 8}  &  {\bf 8}\\
      $\Xi$ & 1 & -2 & $1/2$ & $1/2^+$ &   {\bf 8}  &  {\bf 8}\\
      \hline
      % Two body
      $d$ & 2 & 0 & $0$ & $1^+$ & $\overline{\bf 10}$ & $\overline{\bf 10}$ \\
      $nn$ & 2 & 0 & $1$ & $0^+$ &   {\bf 27}&  {\bf 27} \\
      $n\Lambda$ & 2 & -1 & $1/2$ & $0^+$ &   {\bf 27}&  {\bf 27}\\
      $n\Lambda$ & 2 & -1 & $1/2$ & $1^+$ &   ${\bf 8}_A,\, \overline{\bf 10}$ &--- \\
      $n\Sigma$ & 2 & -1 & $3/2$ & $0^+$ &     {\bf 27}   &   {\bf 27}  \\
      $n\Sigma$ & 2 & -1 & $3/2$ & $1^+$ &     {\bf 10} &   {\bf 10}  \\
      $n\Xi$ & 2 & -2 & $0$ & $1^+$ &        ${\bf 8}_A$ &    ${\bf 8}_A$  \\
      $n\Xi$ & 2 & -2 & $1$ & $1^+$ &      ${\bf 8}_A,\, {\bf
        10},\,\overline{\bf 10}$ & ---\\
      $H$ & 2 & -2 & $0$ & $0^+$ &   ${\bf 1},\, {\bf 27}$ & ${\bf 1},\, {\bf 27}$  \\
      \hline
      % Three body
      \htpnn, \hetppn & 3 & 0 & $1/2$ & $1/2^+$
      & $\overline{\bf 35}$ & $\overline{\bf 35}$\\
      \htL($1/2^+$) & 3 & -1 & 0 & $1/2^+$ 
      &$\overline{\bf 35}$ & ---\\
      \htL($3/2^+$) & 3 & -1 & 0 & $3/2^+$ &  $\overline{\bf 10}$& $\overline{\bf 10}$ \\
      \hetL,\htLtilde, $nn\Lambda$  & 3 & -1 & 1 & $1/2^+$ &   ${\bf
          27},\,\overline{\bf 35}$ &  ${\bf
          27},\,\overline{\bf 35}$  \\
      \hetS & 3 & -1 & 1 & $3/2^+$ &   ${\bf 27}$ &  ${\bf 27}$  \\
            \hline
      % Four body
      \hef & 4 & 0 & $0$ & $0^+$ &    $\overline{\bf 28}$&  $\overline{\bf 28}$  \\
      \hefL,\ \hfL & 4 & -1 & $1/2$ & $0^+$ &       $\overline{\bf 28}$ & ---  \\
      \hefLL & 4 & -2 & 1 & $0^+$ &    ${\bf 27}$, $\overline{\bf 28}$ &
      ${\bf 27},\ \overline{\bf 28}$\\
      \hline
      % Five body
      \hymag & 5 & -3 & $0$ & $3/2^+$ &   $\overline{\bf 10}$ + ...  &  $\overline{\bf 10}$ \\
    \end{tabular}
  \end{ruledtabular}
  \caption{
The baryon number $A$, strangeness $s$, total isospin
    $I$, total spin and parity $J^\pi$ quantum numbers of the
    states and interpolating operators studied in the current
    work. For each set of quantum numbers, the SU(3) irreps that are
    possible to construct with local interpolating operators are listed. 
The  last column lists the SU(3) irrep(s) of the interpolating
    operators used in this work, 
and the dashes indicate that the state is inferred from other states using  SU(3) symmetry.
    }
  \label{tab:ops}
\end{table}

Given the blocks discussed in the previous section and the quark- and
hadron-level wavefunctions introduced previously, the contractions are
performed using an algorithm that  is 
described in more detail in Ref.~\cite{Detmold:2012eu}.  For a given set of
quantum numbers, denoted by ${\cal Q}$, we have a basis of $N_{\rm
  wf}$ hadron-level and quark-level wavefunctions, $\Psi_i^{(h)}$ and
$\Psi_i^{(q)}$ respectively for $i=1,\ldots,N_{\rm wf}$. Note that
$N_{\rm wf}$ depends on ${\cal Q}$. 
In this work the spatial wavefunction at the source
is restricted to a single point. 
In addition, the single-baryon interpolating fields are restricted to the 
upper spin components (in the Dirac basis) only. 
These two restrictions drastically
reduce both the 
size of the space of allowed quark-level 
wavefunctions, and the number of terms each wavefunction can have. 
In all cases, an orthonormal basis
of wavefunctions consistent with the above constraints is obtained. 
The construction, as
well as the simplification of 
the wavefunctions,
is done automatically  with symbolic manipulation. 
Finally, after the 
construction of the wavefunctions,
independent checks of transformation properties of these wavefunctions were
performed, confirming that 
these wavefunctions transform as expected.  
As discussed previously,
hadron-level wavefunctions  and hadronic
blocks with a given total momentum are used at the sink. 
These basic building blocks allow for the 
construction of  more interpolating 
fields at the sink with nontrivial spatial hadronic wavefunctions. In
addition, 
hadron systems with non-vanishing total
momentum can be constructed, since  the
point sources couple to all momenta. 

The contraction algorithm is then straightforward and amounts to
selecting the appropriate indices 
in all possible ways from
the hadron blocks 
building the hadronic-level sink wavefunction, dictated by the quark-level 
wavefunction.\footnote{
\textcolor{\revcolor}{
We note that the algorithm proposed in Ref.~\cite{Doi:2012xd} is quite similar to the one we
have been using in the production of the results presented here.
}
}
For all the systems studied here, the total contraction time was an order of
magnitude less than 
the rest of the calculation.
In addition, the biggest contraction  burden was due to the large number of
terms contributing to 
the  wavefunctions with a nontrivial spatial part at the sink (moving hadrons
at the sink). 
As an example of the speed of our contraction code,
a \hef\  correlation function can be computed in $\sim 0.8~{\rm s}$ per time-slice on
a single core of 
a dual core AMD Opteron 285 
processor.

%%%%%%%%%%%%%%%%%%%%%%%%%%%%%%%%%%%%%%%%%%%%%%%%%%%%
\section{The Pion and Baryon Dispersion Relations}
\label{sec:B}
\begin{figure}[!ht]
  \centering
  \includegraphics[width=0.30\textwidth]{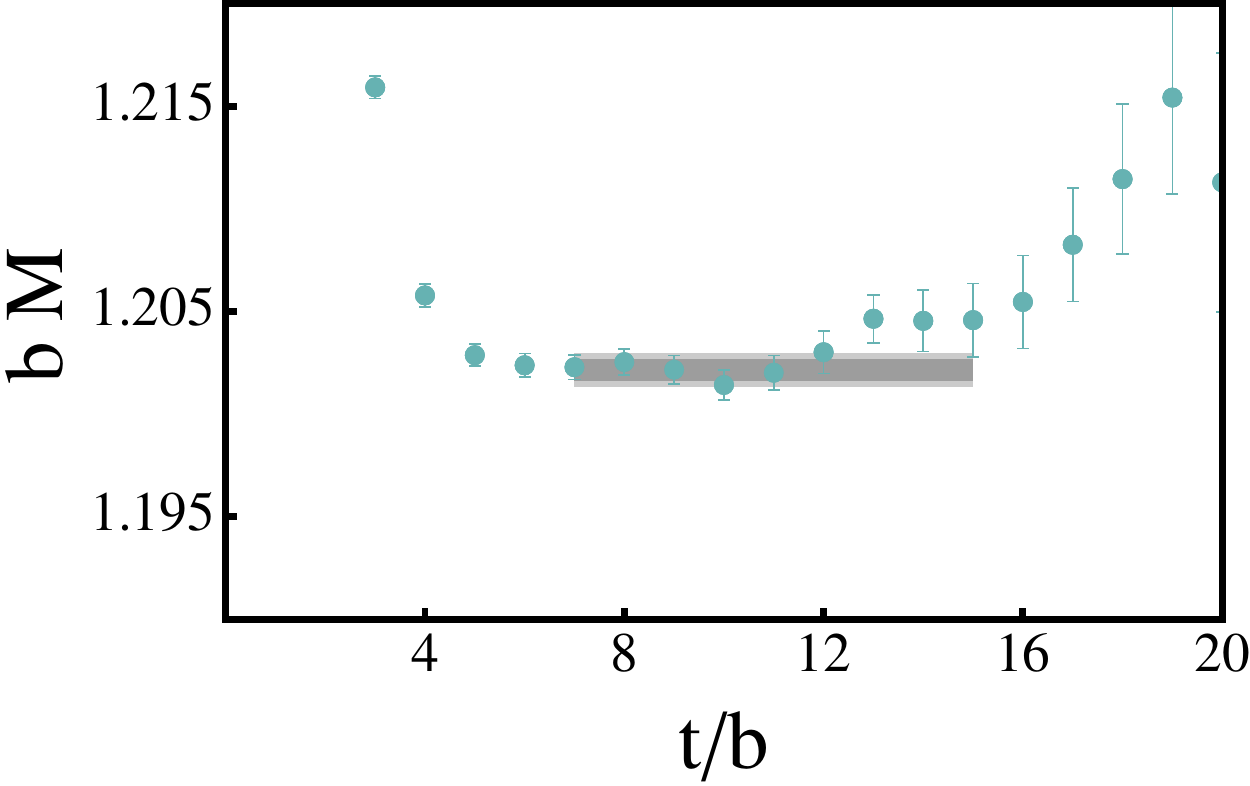}\ \ \ \ 
  \includegraphics[width=0.30\textwidth]{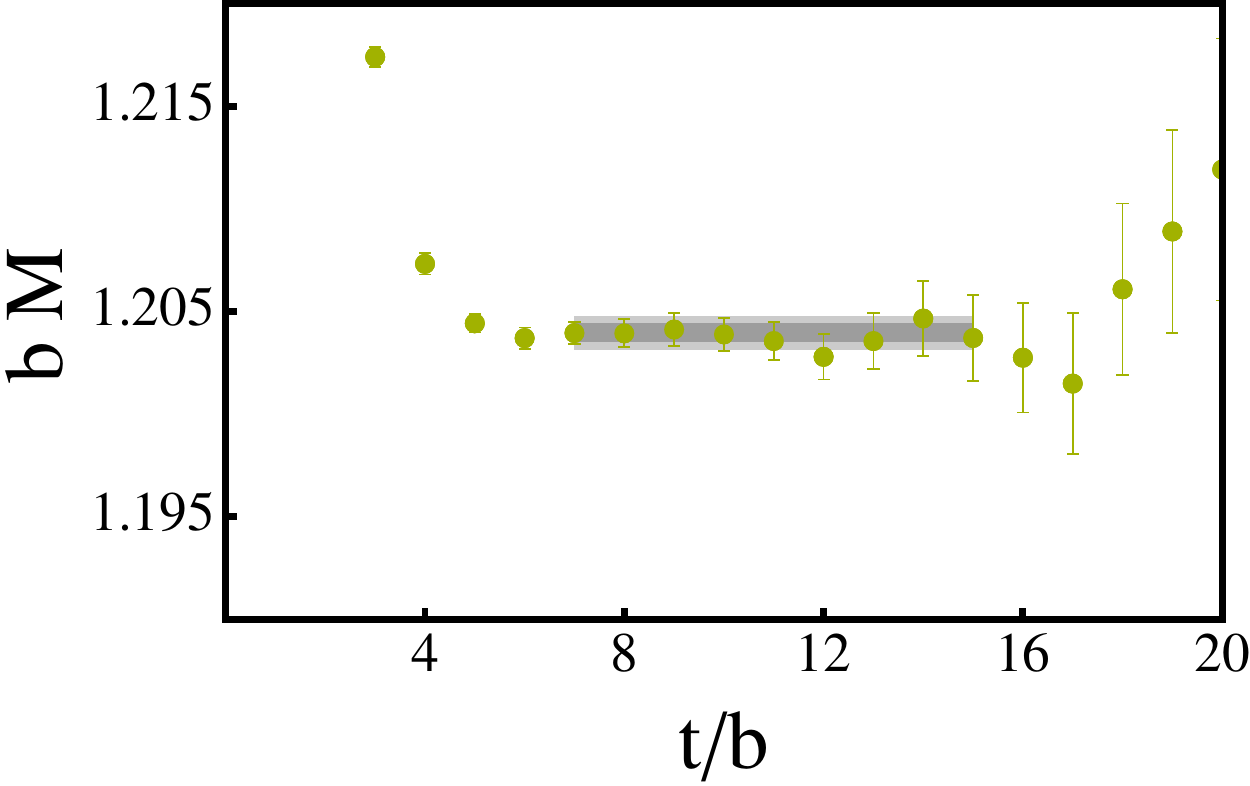}\ \ \ \ 
  \includegraphics[width=0.30\textwidth]{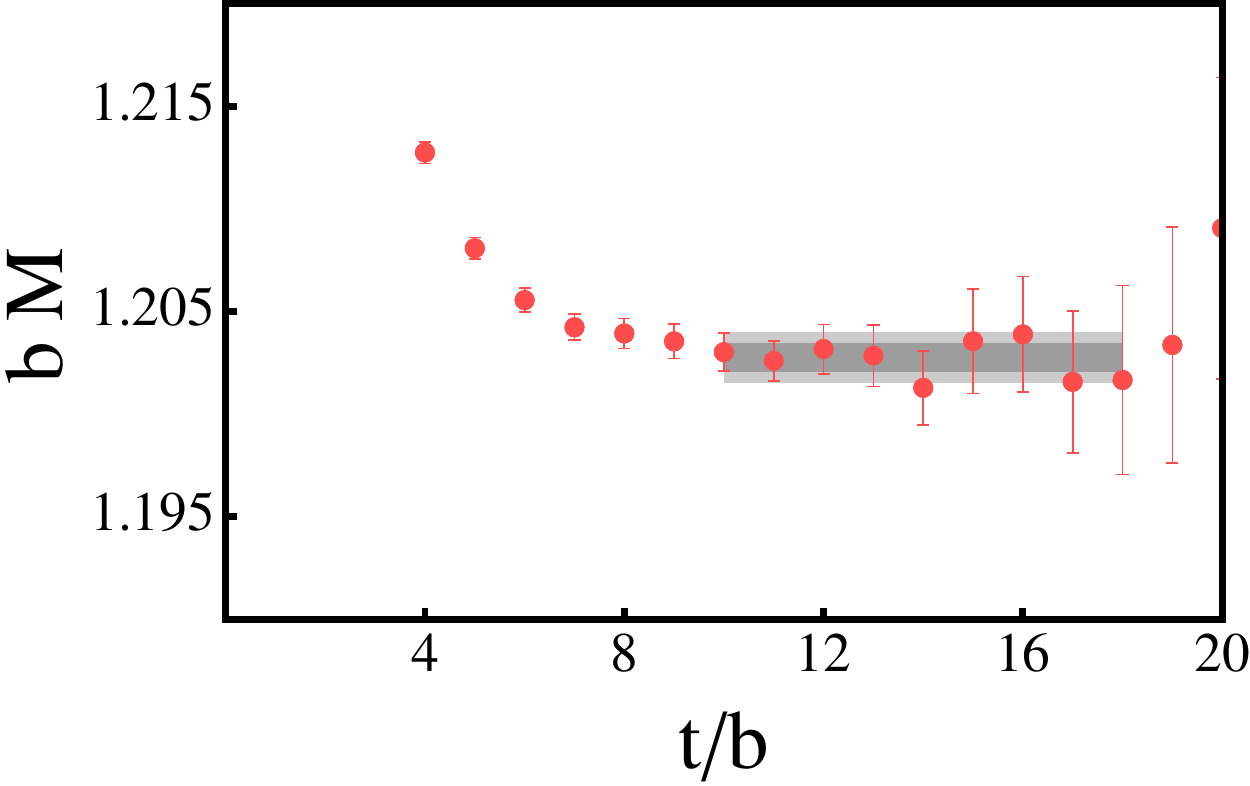}
  \caption{The EMPs associated with linear combinations of
    baryon correlation functions computed with the \cfga\ (left),
    \cfgb\ (center) and \cfgc\ (right) ensembles, with momentum $|{\bf
      P}|=0$.  
 \textcolor{\revcolor}{
The inner (darker) shaded region corresponds to the statistical uncertainty
of the extracted energy, while  the outer (lighter) shaded region
corresponds to the
statistical and fitting systematic uncertainties combined in quadrature.
}
The time-extent of each band corresponds to the choice  of the fitting interval
for each correlation function.
}
  \label{fig:A1EMPs}
\end{figure}

\noindent
In the limit of SU(3)-flavor symmetry, all members of the lightest
baryon octet have the same mass, and as such, we compute correlation
functions associated with only one of the octet baryons. Similarly all
octet pseudoscalar mesons are degenerate, and we refer to them as the
pion.  Linear combinations of single-hadron correlation functions
generated from smeared quark sources and either smeared or point sinks
are formed for hadrons with a given lattice momentum.
The lowest energy eigenvalue can be determined from these correlation functions, 
the results of which are presented in
Table~\ref{tab:A0mass} (pion) and Table~\ref{tab:A1mass} (baryon),
and the baryon  effective mass plots (EMPs) are shown in 
fig.~\ref{fig:A1EMPs}.  
\begin{table}
\begin{center}
\begin{minipage}[!ht]{16.5 cm}
  \caption{The pion energy (l.u.) as a function of momentum (l.u.),
    $|{\bf P}| = \left({2\pi\over L}\right) |{\bf n}|$,
    calculated on each ensemble of gauge-field configurations.
The infinite-volume pion mass, determined by fitting the expression in
eq.~(\ref{eq:FVfit}), is provided in the last row.
The first uncertainty is statistical and the second is the fitting systematic.
  }
\label{tab:A0mass}
\end{minipage}
\setlength{\tabcolsep}{0.2em}
\resizebox{\linewidth}{!}{%
\begin{tabular}{c|cccccc}
\hline
ensemble  
& $|{\bf n}|=0$ 
& $|{\bf n}|^2=1$ 
& $|{\bf n}|^2=2$
& $|{\bf n}|^2=3$ 
& $|{\bf n}|^2=4$ 
& $|{\bf n}|^2=5$  \\
\hline
\cfga 
& \textcolor{\revcolor}{0.59389(18)(18)} 
& \textcolor{\revcolor}{0.64652(16)(19)} 
& \textcolor{\revcolor}{0.69482(17)(29)} 
& \textcolor{\revcolor}{0.73971(20)(36)}
& \textcolor{\revcolor}{0.77800(30)(72)} 
& \textcolor{\revcolor}{0.81946(36)(78)} \\
\cfgb 
&  \textcolor{\revcolor}{0.59445(15)(17)} 
&  \textcolor{\revcolor}{0.62474(15)(18)} 
&  \textcolor{\revcolor}{0.65326(16)(20)} 
&  \textcolor{\revcolor}{0.68099(18)(25)} 
&  \textcolor{\revcolor}{0.70672(19)(28)} 
&  \textcolor{\revcolor}{0.73194(22)(31)} \\
\cfgc 
&  \textcolor{\revcolor}{0.59403(16)(14)} 
&  \textcolor{\revcolor}{0.60768(16)(15)} 
&  \textcolor{\revcolor}{0.62101(18)(17)} 
& \textcolor{\revcolor}{0.63403(19)(20)} 
& \textcolor{\revcolor}{0.64667(21)(24)} 
& \textcolor{\revcolor}{0.65915(24)(28)} \\
\hline
      $L=\infty$ &  \textcolor{\revcolor}{0.59426(12)(11)} & & & & &  \\
\hline
\end{tabular}
}
%noalign{\smallskip\hrule}\cr}
\begin{minipage}[t]{16.5 cm}
\vskip 0.0cm
\noindent
\end{minipage}
\end{center}
\end{table}     
\begin{table}
\begin{center}
\begin{minipage}[!ht]{16.5 cm}
  \caption{
    The ground-state octet-baryon energy (l.u.) as a function of momentum (l.u.),
    $|{\bf P}| = \left({2\pi\over L}\right) |{\bf n}|$,
    calculated on each ensemble of gauge-field configurations.
The infinite-volume baryon mass, determined by fitting the expression in
eq.~(\ref{eq:FVfit}), is provided in the last row.
The first uncertainty is statistical and the second is the fitting systematic.
  }
\label{tab:A1mass}
\end{minipage}
\setlength{\tabcolsep}{0.2em}
\resizebox{\linewidth}{!}{%
\begin{tabular}{c|cccccc}
\hline
ensemble  & $|{\bf n}|=0$ & $|{\bf n}|^2=1$ & $|{\bf n}|^2=2$& $|{\bf n}|^2=3$
& $|{\bf n}|^2=4$ 
& $|{\bf n}|^2=5$  \\
\hline
      \cfga  
      & \textcolor{\revcolor}{1.20317(58)(84)} 
      & \textcolor{\revcolor}{1.2282(9)(16) }
      & \textcolor{\revcolor}{1.2537(9)(23) }
      & \textcolor{\revcolor}{1.2785(11)(31) }
      & \textcolor{\revcolor}{1.3023(11)(25) }
      & \textcolor{\revcolor}{1.3254(12)(29) }   \\
      \cfgb  
      &  \textcolor{\revcolor}{ 1.20396(47)(69) }&
      \textcolor{\revcolor}{1.21821(61)(64)} &
      \textcolor{\revcolor}{1.23263(65)(70)} & 
      \textcolor{\revcolor}{1.24685(69)(79)} & 
      \textcolor{\revcolor}{1.26077(74)(94)} & 
      \textcolor{\revcolor}{1.2746(08)(11)}  \\
      \cfgc  
      &  \textcolor{\revcolor}{ 1.2032(07)(11)} & 
       \textcolor{\revcolor}{1.2096(11)(22)} & 
       \textcolor{\revcolor}{1.2162(11)(21)} &
       \textcolor{\revcolor}{1.2227(12)(22)} & 
       \textcolor{\revcolor}{1.2290(12)(21)} & 
        \textcolor{\revcolor}{1.2354(13)(21) } \\
\hline
      $L=\infty$ &  \textcolor{\revcolor}{ 1.20293(35)(20) } & & & & &  \\
\hline
\end{tabular}
}
%noalign{\smallskip\hrule}\cr}
\begin{minipage}[t]{16.5 cm}
\vskip 0.0cm
\noindent
\end{minipage}
\end{center}
\end{table}     

\begin{figure}[!ht]
  \centering
  \includegraphics[width=0.48\textwidth]{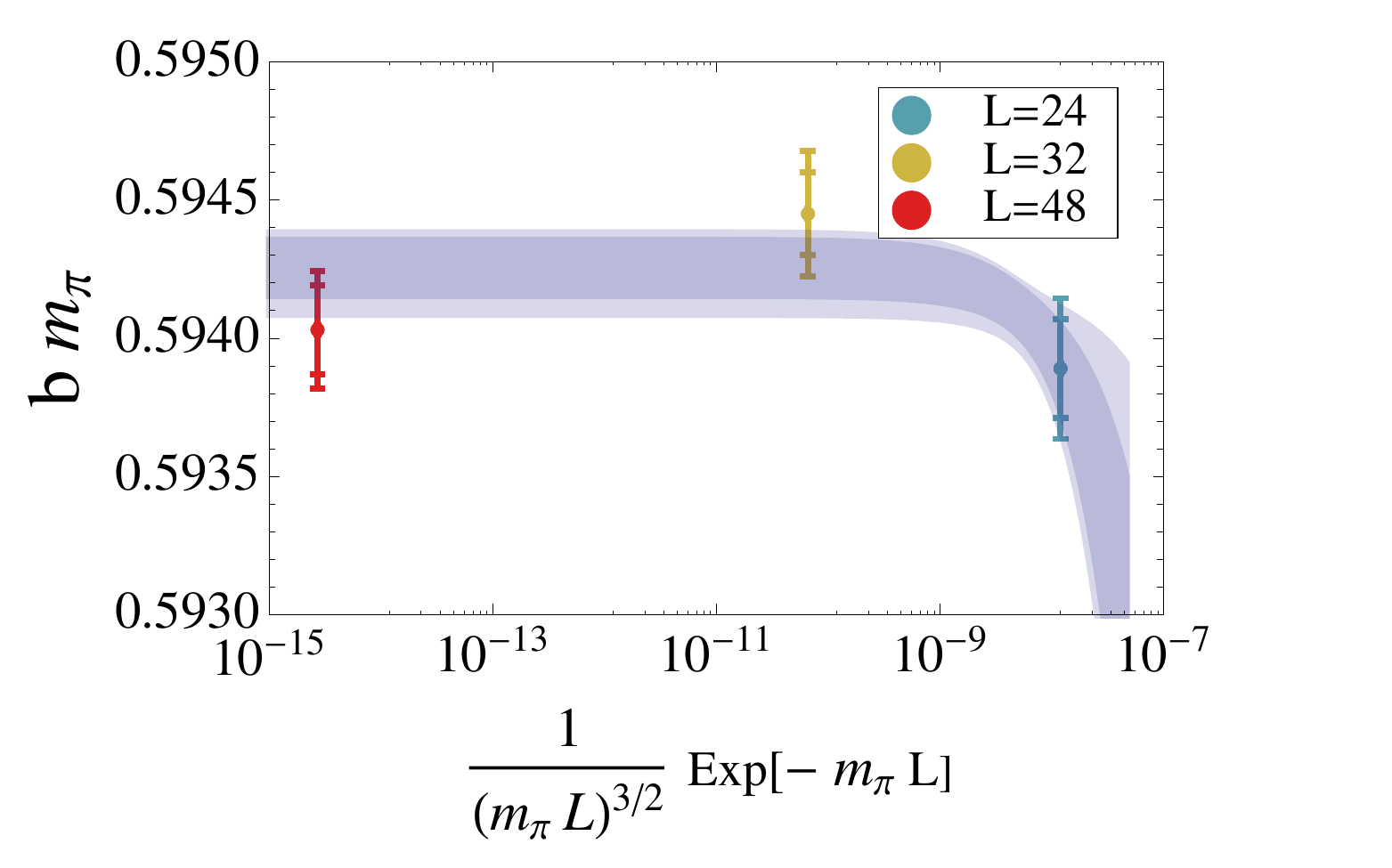}\ \ \
  \includegraphics[width=0.48\textwidth]{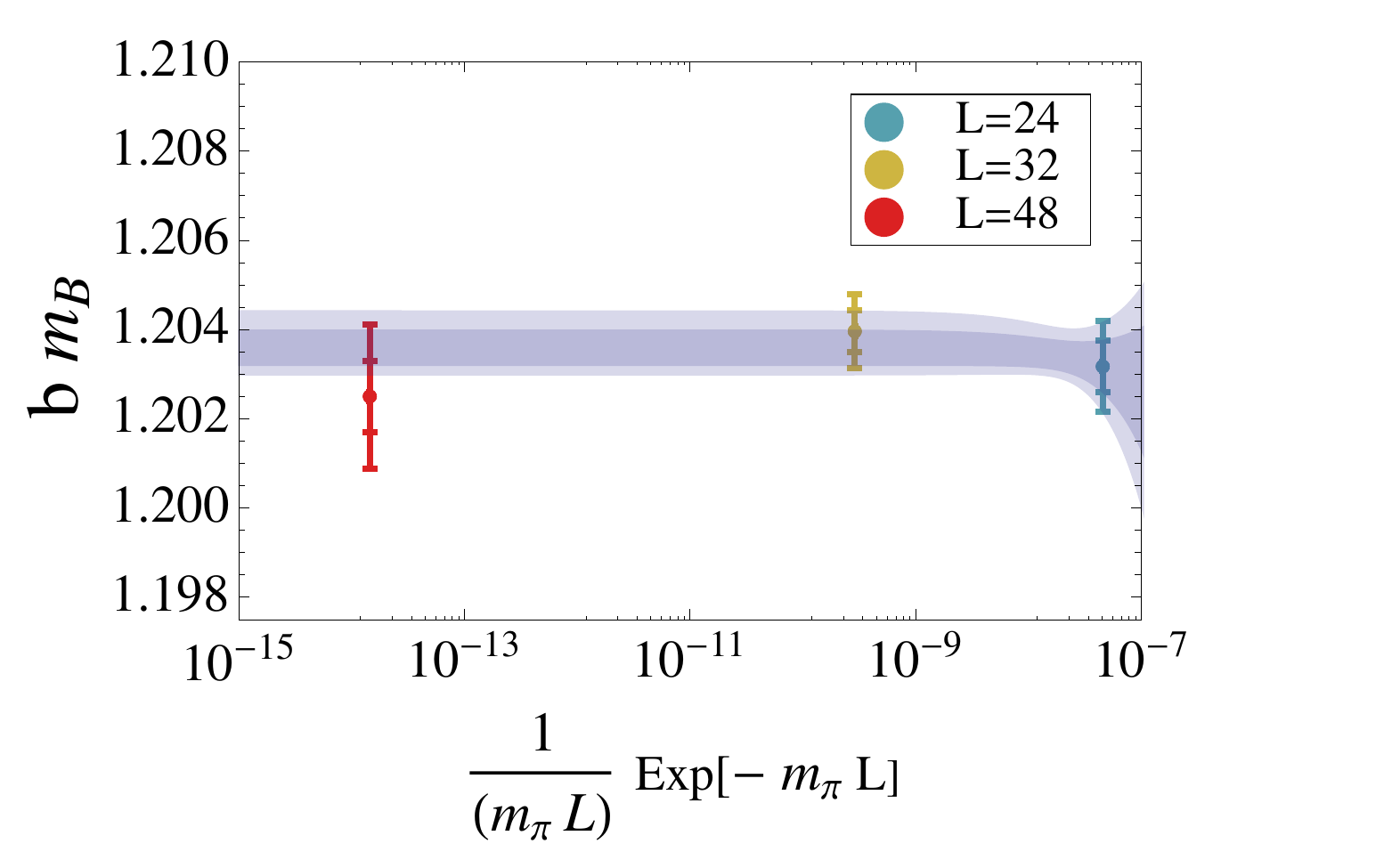}
  \caption{The volume dependence of the pion mass
    (left panel) and the baryon mass (right panel) extracted
    from the zero-momentum correlation functions.
The shaded regions are extrapolations of the form given in eq.~(\ref{eq:FVfit}).
 }
  \label{fig:A1volume}
\end{figure}

For hadrons at rest, 
the masses of the pion and baryon  in finite volume,  $m_H^{(V)} (m_\pi L)$, 
are extrapolated to infinite volume using
\begin{eqnarray}
  m_\pi^{(V)} (m_\pi L) & = & m_\pi^{(\infty)}\ +\ 
  c_\pi^{(V)}\ {e^{- m_\pi\  L}\over (m_\pi L)^{3/2}}
\ +\ ...
  \nonumber\\
  M_B^{(V)} (m_\pi L) & = & M_B^{(\infty)}\ +\ 
  c_B^{(V)}\ {e^{- m_\pi\  L}\over m_\pi L}
\ +\ ...
  \ \ \ ,
  \label{eq:FVfit}
\end{eqnarray}
where only the first terms in the finite-volume (FV) expansion are
required due to the large pion mass~\cite{Beane:2011pc}.  The
extrapolations to infinite volume are shown as the solid regions in
fig.~\ref{fig:A1volume}, and the extrapolated
values of the pion and octet-baryon mass are presented in
Table~\ref{tab:A0mass} and Table~\ref{tab:A1mass}, respectively. As
expected for calculations with large values of $m_\pi L$, the 
single-hadron FV  effects are very small.
The extrapolated pion and octet-baryon masses, using the measured
lattice spacing, are $m_\pi=\mpifullMeV$ and $m_B=\mBfull$, where the
first uncertainty is statistical, the second is the fitting
systematic, and the third is due to the uncertainty in the
lattice spacing.

\begin{figure}[!ht]
  \centering
  \includegraphics[width=0.49\textwidth]{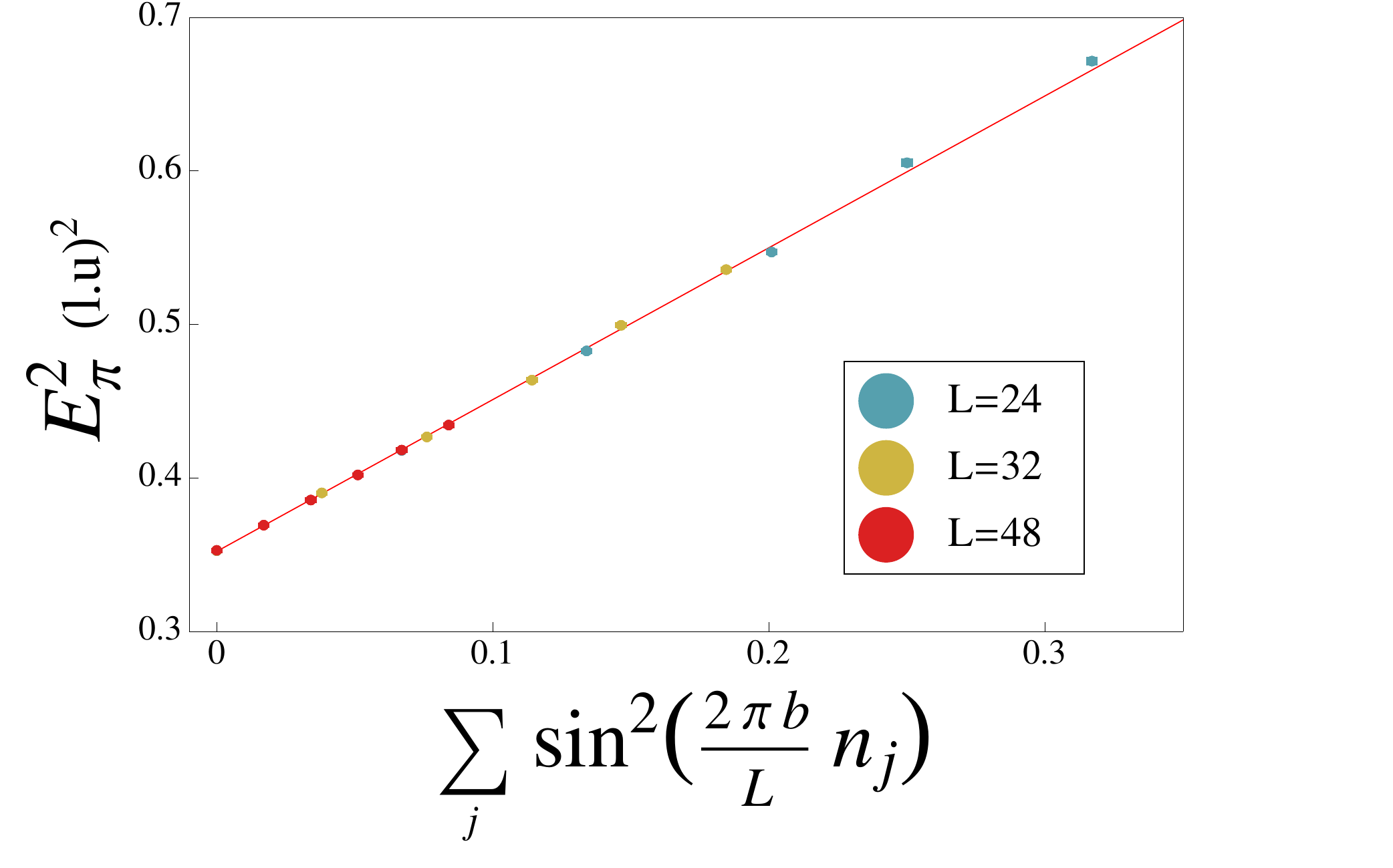}\
  \
  \includegraphics[width=0.49\textwidth]{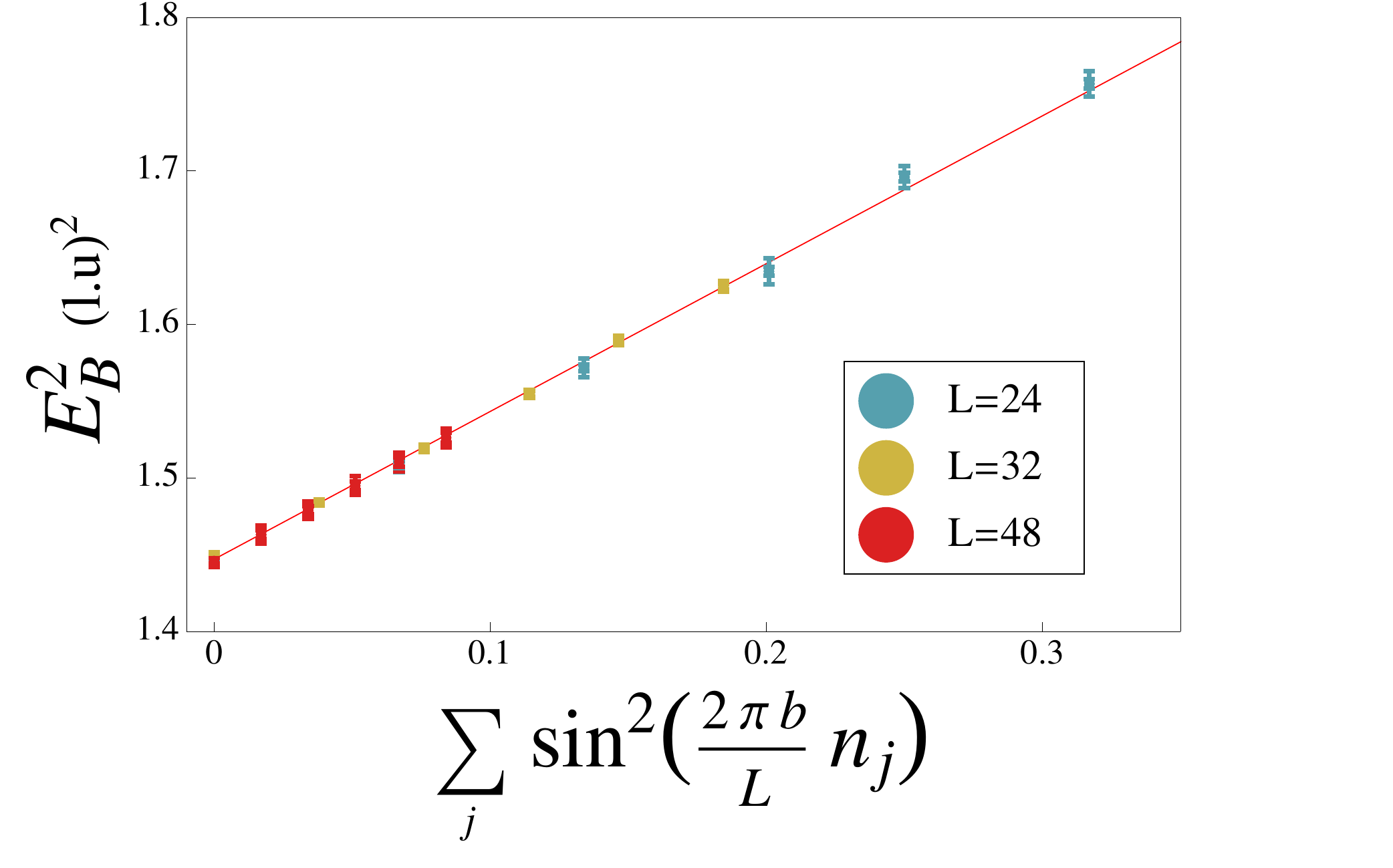}\
  \
  \caption{ The squared energy (in $({\rm l.u.})^2$) of the single
    pion and baryon as a function of $\sum\limits_j\
    \sin^2\left({2\pi b\over L} n_j\right)$.  The points are the
    results of the LQCD calculations with the inner (outer)
    uncertainties being the statistical uncertainties (statistical and
    systematic uncertainties combined in quadrature).  The red curves
    correspond to the best linear fits.  }
  \label{fig:Dispersion}
\end{figure}

In order to have confidence in the extraction of multibaryon binding
energies and to be able to quantify one of the systematic
uncertainties in these determinations, it is important to determine
the single-hadron dispersion relation.  
The energies of the pion and baryon are shown in fig.~\ref{fig:Dispersion} as a
function of $\sum\limits_j\ \sin^2\left({2\pi b\over L} n_j\right)$, 
where the triplet of integers ${\bf n}=(n_1, n_2,n_3)$ is related to the
lattice momentum via 
$|{\bf P}|^2 = \left({2\pi\over L}\right)^2 |{\bf n}|^2$.
In these LQCD calculations, the energy of the hadron can related to its
lattice momentum through a dispersion relation of the form
\begin{eqnarray}
  \left(\,b\,E_H\right)^2 
  & = &
  (b\,M_H)^2 
  \ +\
  \frac{1}{\xi_H^2}
  \ \sum_j\ \sin^2
  \left(\frac{2\,\pi\,b}{L} n_j \right) 
  \ ,
  \label{eq:latticeDisp}
\end{eqnarray}
where the anisotropy parameter, $\xi_H$ (or equivalently the speed of light
$c=1/\xi_H$), is expected to be unity in
calculations performed with isotropic lattices.\footnote{
 \textcolor{\revcolor}{
As the lattice hadronic dispersion relations are {\it a priori} unknown, 
they must be calculated. The form given in eq.~(\ref{eq:latticeDisp}) is
expected to capture the leading momentum dependence.
}
}  
Fitting $\xi_H$ to the
energy of the pion and baryon, given in Table~\ref{tab:A0mass} and
Table~\ref{tab:A1mass}, respectively, yields $\xi_\pi = \xipi$ and
$\xi_B = \xiB$. 
Therefore, the dispersion relations provide only a small uncertainty in the extraction of
multihadron energies.

%%%%%%%%%%%%%%%%%%%%%%%%%%%%%%%%%%%%%%%%%%%%%%%%%%%%
\section{Two-Body Systems}
\label{sec:BB}
\begin{figure}[!t]
  \centering
  \includegraphics[width=0.32\textwidth]{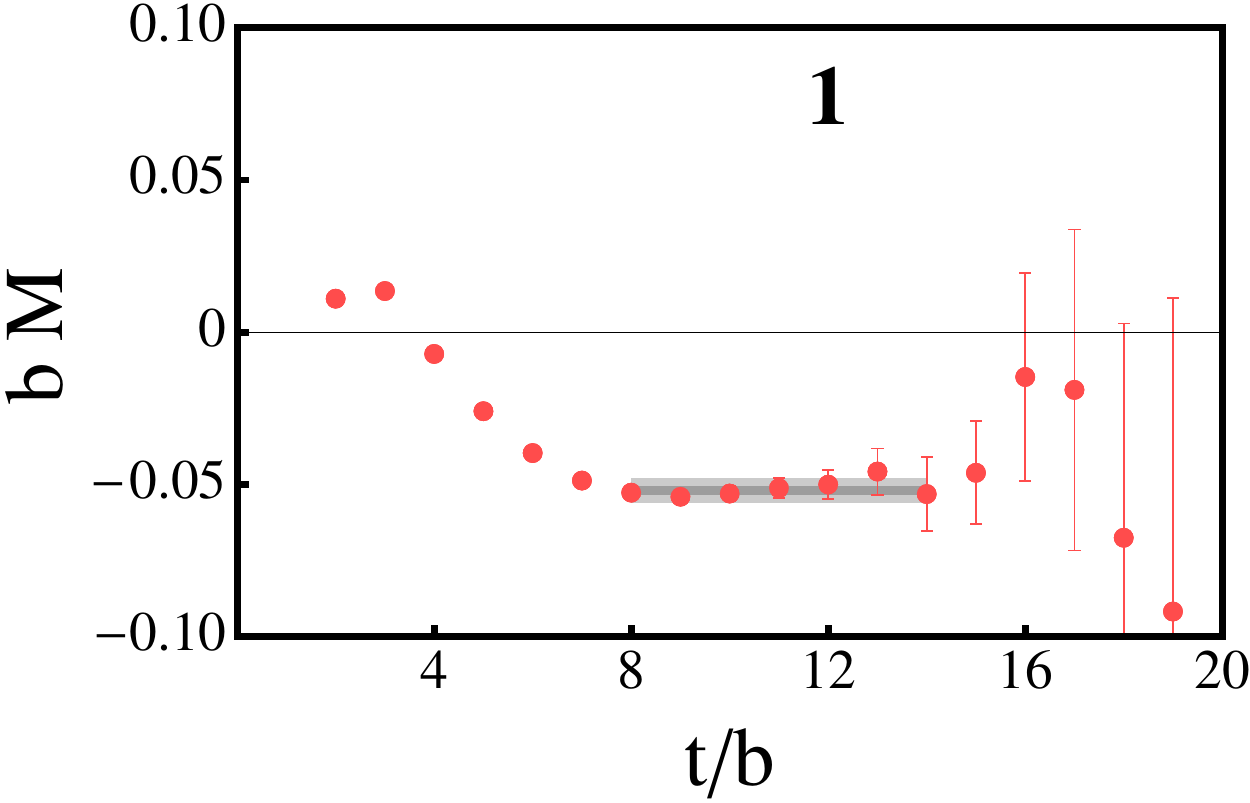}\ \  
  \includegraphics[width=0.32\textwidth]{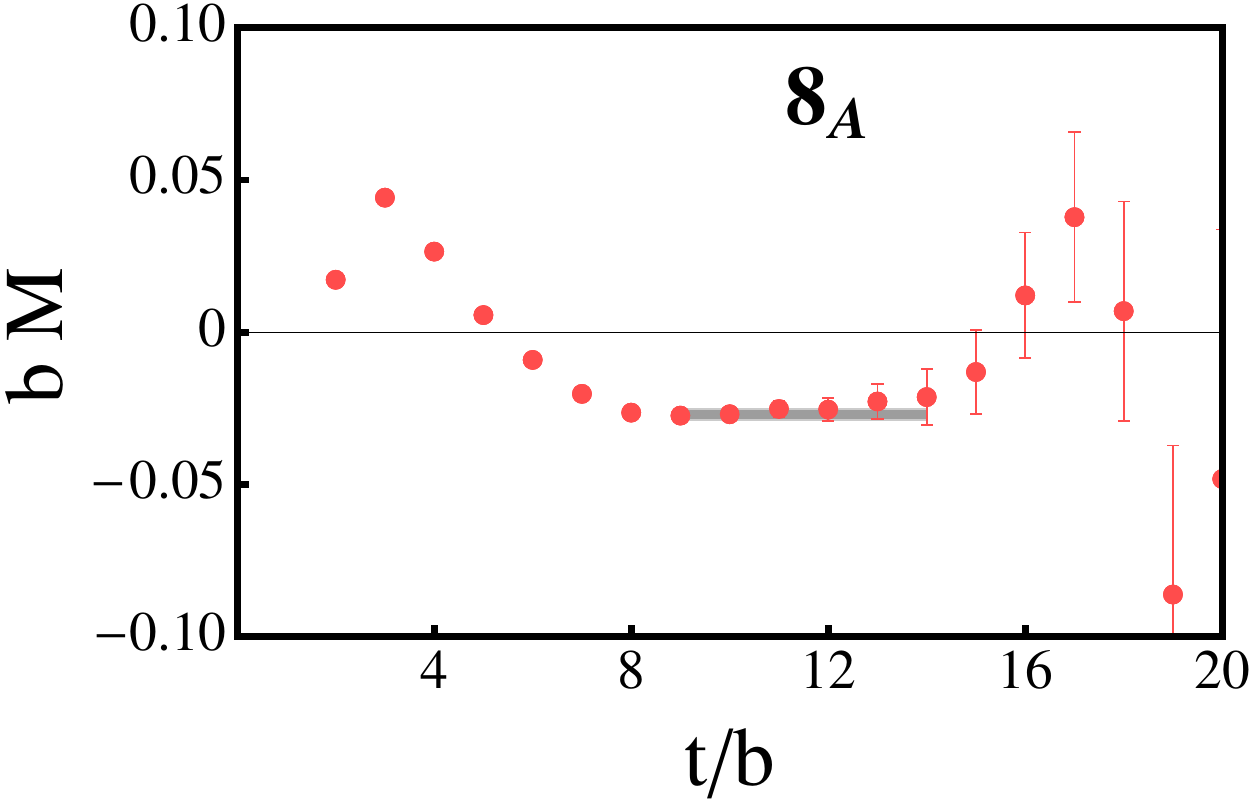}\ \  
  \includegraphics[width=0.32\textwidth]{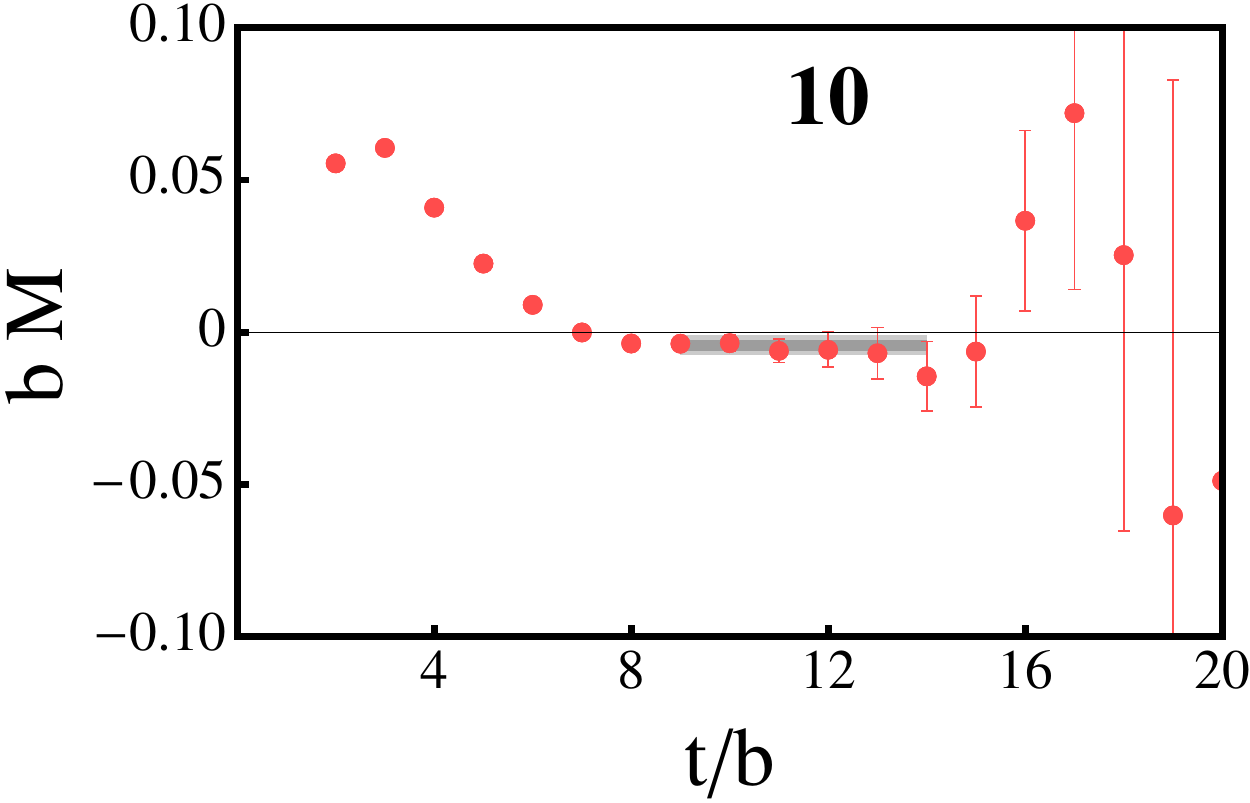}
  \includegraphics[width=0.32\textwidth]{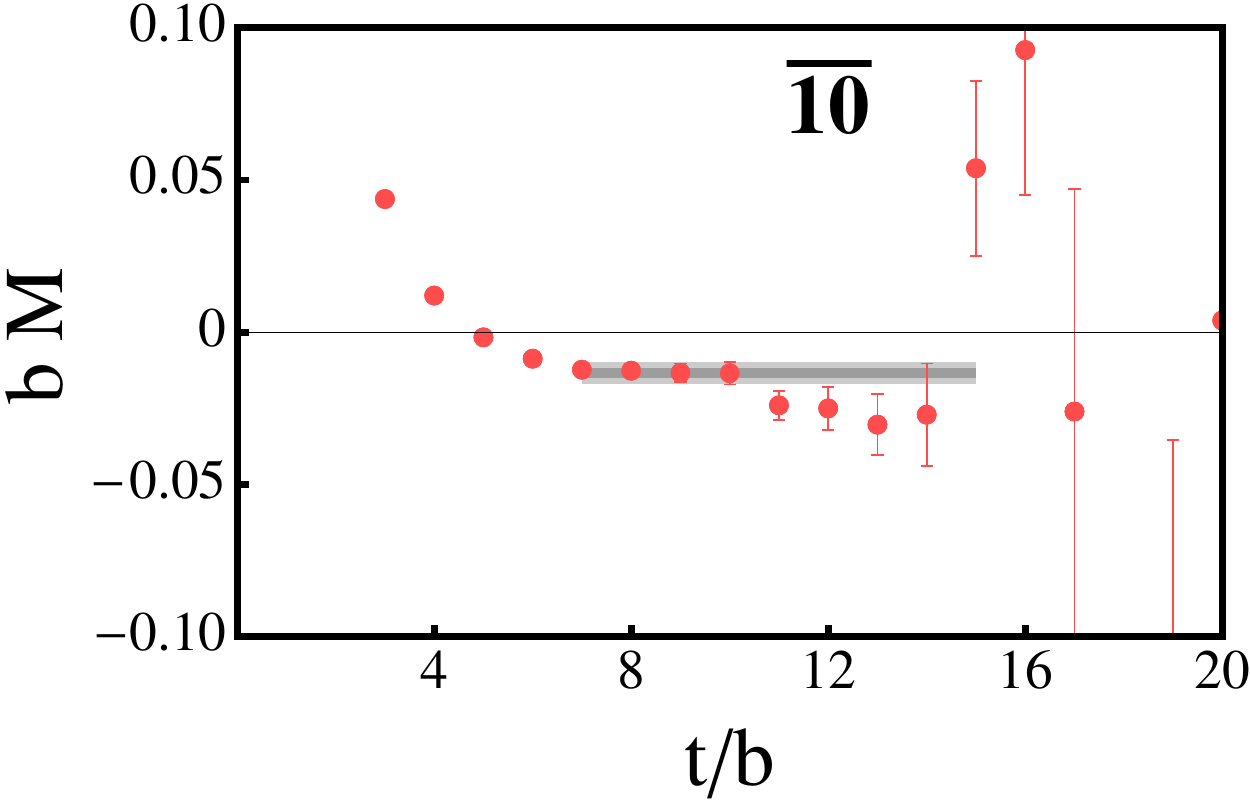}\ \ \ \ 
  \includegraphics[width=0.32\textwidth]{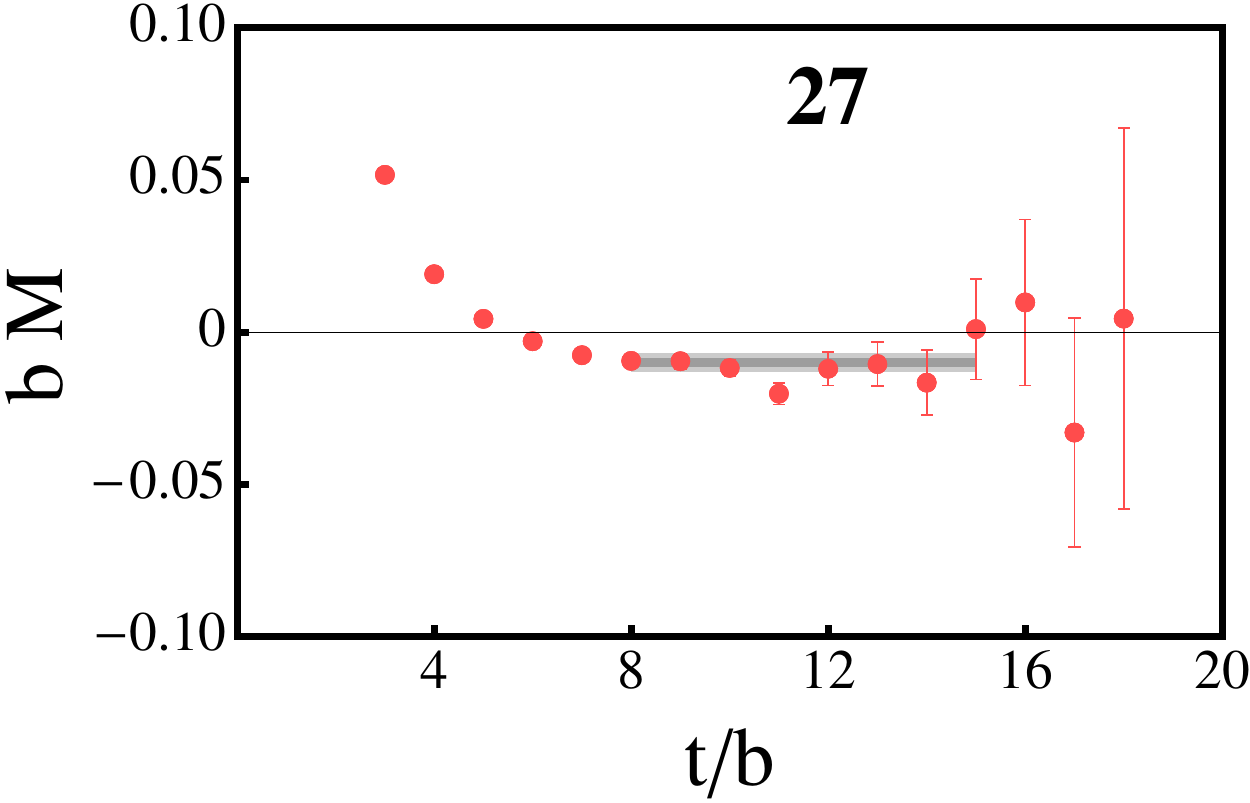}
  \caption{
EMPs associated with $|{\bf P}|=0$ two-baryon
    correlation functions computed with the
    \cfgc\ ensemble.  
 \textcolor{\revcolor}{
The inner (darker) shaded region corresponds to the statistical uncertainty
of the extracted energy, while  the outer (lighter) shaded region
corresponds to the
statistical and fitting systematic uncertainties combined in quadrature.
}
The time-extent of each band corresponds to the choice  of the fitting interval
for each correlation function.
From left to
    right, the top row corresponds to the {\bf 1}, ${\bf 8}_A$,
    {\bf 10} SU(3)  irreps 
 \textcolor{\revcolor}{
(corresponding to the H-dibaryon, $I=0$ $N\Xi$ in the $\siii-\diii$ coupled channels
and $n\Sigma^-$ in the  $\siii-\diii$ coupled channels, respectively)}, 
and the bottom row corresponds to the $\overline{\bf 10}$ and {\bf 27}
 \textcolor{\revcolor}{
(corresponding to the deuteron and di-neutron, respectively)}.
 }
  \label{fig:A2EMPsL32}
\end{figure}

\begin{figure}[!t]
  \centering
  \includegraphics[width=0.99\textwidth]{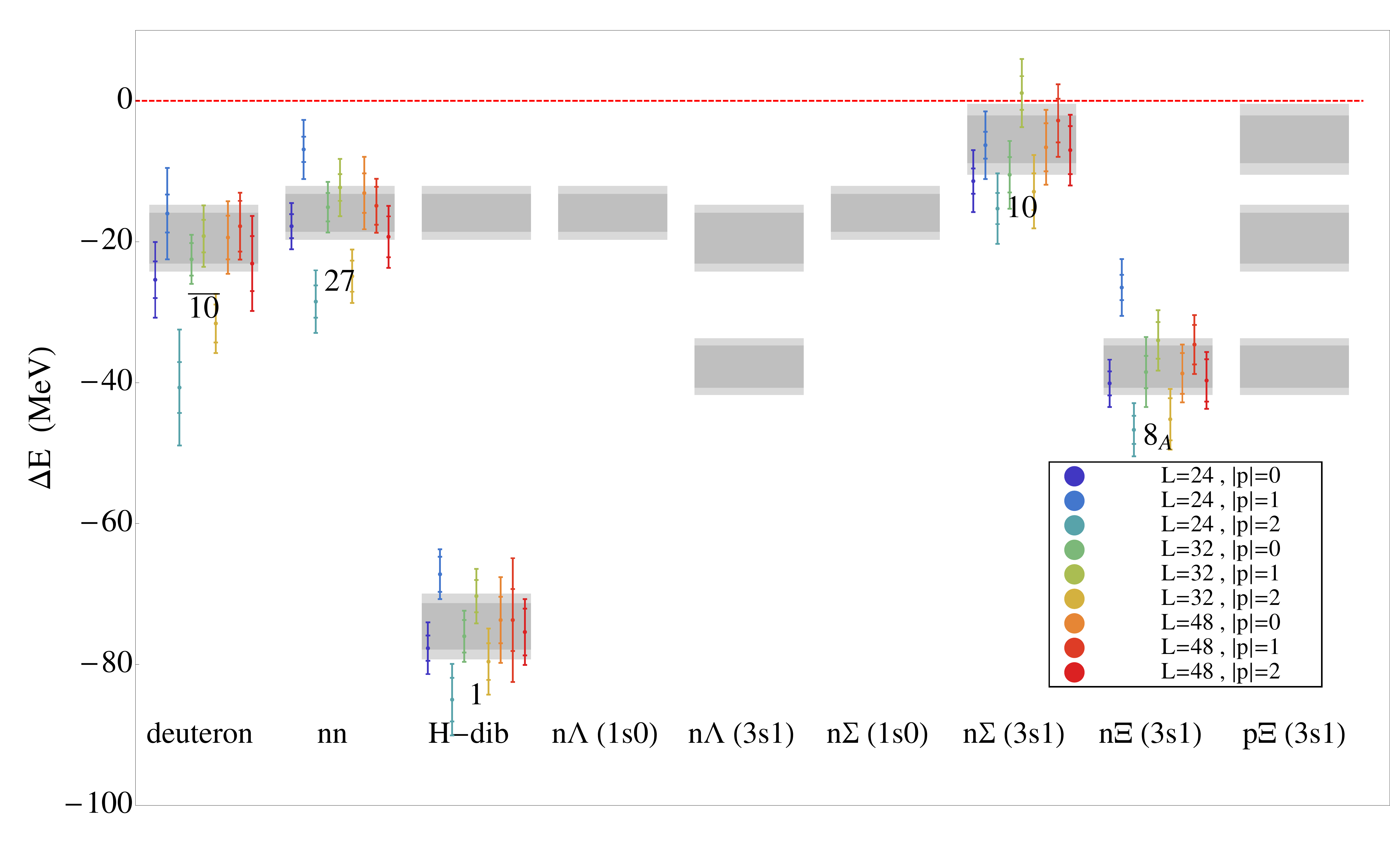}
  \caption{Binding energies in the $A=2$ systems relative to
    two non-interacting baryons ($B=-\Delta E$).  
The points and associated
    uncertainties are the results of the LQCD calculations given in
    Table~\protect\ref{tab:A2energies24},
    Table~\protect\ref{tab:A2energies32} and
    Table~\protect\ref{tab:A2energies48}. 
The  dark (statistical uncertainty)
and light (statistical and systematic uncertainties combined in quadrature)
    horizontal bands denote the average of the bindings calculated on
    the \cfgc\ ensemble, which are taken as the infinite-volume
    estimate.  Where only bands are shown, SU(3) symmetry
    has been used to determine the spectrum.  }
  \label{fig:A2data}
\end{figure}

\noindent
In general, the two-body states can be classified by isospin,
strangeness, parity and angular momentum.  In the limit of SU(3)-flavor symmetry, the energy
eigenstates can also be classified by SU(3) quantum numbers.  The
lowest-lying baryons transform as ${\bf 8}$ under SU(3), and, therefore,
the two-body states have degeneracies determined by the dimensionality
of the irreps in the product
\begin{eqnarray} {\bf 8}\otimes {\bf 8}
  & = & 
  {\bf 27}\oplus{\bf 10}\oplus\overline{\bf 10}\oplus{\bf
    8}_S\oplus{\bf 8}_A\oplus{\bf 1}
  \ \ \ .
  \label{eightsquared}
\end{eqnarray}
As the wavefunction of such systems is totally antisymmetric,
the s-wave $\si$ channels transform under SU(3) as ${\bf 27}\oplus{\bf
  8}_S\oplus{\bf 1}$, while the $\siii$-$\diii$ coupled channels
transform as ${\bf 10}\oplus\overline{\bf 10}\oplus{\bf 8}_A$.  
The source structures we have employed, in which the quark-level
operators reside at one point in the spatial volume, have vanishing overlap
with the ${\bf 8}_S$ irrep, and as a result, we are unable to determine the
energy of this two-body irrep.
Correlation functions are not constructed directly in terms of their SU(3)
transformation properties, but the contributing  
SU(3) irreps can be deduced from their structure: 
$\overline{\bf 10}$ from the deuteron,
${\bf 27}$ from the di-nucleon,
${\bf 1}\oplus {\bf 27}$ from the H-dibaryon (the ${\bf 8}_S$ is absent),
${\bf 10}$ from $n\Sigma^-$ in the 
$\siii$-$\diii$ coupled channels,
and
${\bf 8}_A$ from $I=0$ $N\Xi$ in the $\siii$-$\diii$ coupled channels.
EMPs extracted from the two-body correlation functions 
for systems at rest
calculated with the \cfgc\ ensemble 
are shown in fig.~\ref{fig:A2EMPsL32}.  
The energies of states that are negatively
shifted relative to two free baryons  
are presented in Table~\ref{tab:A2energies24},
Table~\ref{tab:A2energies32} and Table~\ref{tab:A2energies48},
respectively, and displayed in fig.~\ref{fig:A2data}.

The energies of the states that are presented in this work, along with their
statistical uncertainties,
are 
determined from a single-parameter
correlated $\chi^2$-minimization procedure performed over a specific time
interval of EMPs 
and from exponential fits to the correlation functions directly,
with covariance matrices determined with either Jackknife or Bootstrap.
The systematic uncertainty that is assigned to these energies is determined by
varying the fit interval over a range of values consistent with the identified
plateau region.

A number of
scattering states with positive energy-shifts relative to two free
baryons have also been identified using different correlation functions, 
but their uncertainties are large
enough to preclude clean extraction of scattering phase-shifts using
L\"uschers method~\cite{Luscher:1986pf,Luscher:1990ux}, 
and we 
defer analysis of these states to a later time when adequate statistics have
been
accumulated.  
\begin{table}
\begin{center}
\begin{minipage}[!ht]{16.5 cm}
  \caption{Two-body binding energies (MeV) calculated with the
    \cfga\  ensemble. 
The first uncertainty is statistical, the second is the fitting systematic and
the third is due to the lattice spacing.
  }
\label{tab:A2energies24}
\end{minipage}
\setlength{\tabcolsep}{1em}
\begin{tabular}{c|ccc}
\hline
      SU(3) irrep  & $|{\bf n}|=0$& $|{\bf n}|=1$ & $|{\bf  n}|=2$\\
\hline
      {\bf 1}             
      & \textcolor{\revcolor}{77.7(1.8)(3.2)(0.8)} 
      & \textcolor{\revcolor}{67.2(2.5)(2.5)(0.8)}
      & \textcolor{\revcolor}{85.0(3.1)(4.0)(0.9)}      \\
      ${\bf 8}_A$         
      &  \textcolor{\revcolor}{40.1(1.7)(2.9)(0.4) }
      &  \textcolor{\revcolor}{26.5(1.8)(3.6)(0.3) }
      &  \textcolor{\revcolor}{46.7(2.0)(3.2)(0.5) } \\
      ${\bf 10}$          
      &  \textcolor{\revcolor}{11.4(1.8)(4.0)(0.1)}
      &  \textcolor{\revcolor}{6.3(1.9)(4.4)(0.1) }
      &  \textcolor{\revcolor}{15.3(2.2)(4.5)(0.1) }\\
      $\overline{\bf 10}$ 
      &  \textcolor{\revcolor}{25.4(2.6)(4.7)(0.3)}
      &  \textcolor{\revcolor}{16.0(2.7)(5.9)(0.2) }
      &  \textcolor{\revcolor}{40.7(3.6)(7.4)(0.5) } \\
      ${\bf 27}$          
      &  \textcolor{\revcolor}{17.8(1.7)(2.8)(0.2)}
      &  \textcolor{\revcolor}{6.9(1.8)(3.8)(0.1) }
      &  \textcolor{\revcolor}{28.5(2.3)(3.8)(0.3) }\\
\hline
\end{tabular}
%noalign{\smallskip\hrule}\cr}
\begin{minipage}[t]{16.5 cm}
\vskip 0.0cm
\noindent
\end{minipage}
\end{center}
\end{table}     
\begin{table}
\begin{center}
\begin{minipage}[!ht]{16.5 cm}
  \caption{Two-body binding energies (MeV) calculated with the
    \cfgb\  ensemble. 
The first uncertainty is statistical, the second is the fitting systematic and
the third is due to the lattice spacing.
  }
\label{tab:A2energies32}
\end{minipage}
\setlength{\tabcolsep}{1em}
\begin{tabular}{c|ccc}
\hline
      SU(3) irrep  & $|{\bf n}|=0$ & $|{\bf n}|=1$ & $|{\bf  n}|=2$ \\
\hline
      {\bf 1}             
      & \textcolor{\revcolor}{  76.0(2.3)(2.8)(0.8)} 
      & \textcolor{\revcolor}{ 70.3(2.3)(3.1)(0.7)} 
      & \textcolor{\revcolor}{79.6(2.6)(3.9)(0.9)}\\
      ${\bf 8}_A$         
      & \textcolor{\revcolor}{ 38.5(2.3)(4.4)(0.4)} 
      & \textcolor{\revcolor}{ 34.0(2.6)(3.4)(0.4)} 
      &  \textcolor{\revcolor}{45.2(3.0)(3.1)(0.5)}\\
      ${\bf 10}$          
      & \textcolor{\revcolor}{ 10.5(2.5)(4.1)(0.1)}  
      & \textcolor{\revcolor}{ 1.1(2.4)(4.2)(0.0)}  
      &  \textcolor{\revcolor}{12.9(2.6)(4.5)(0.1)} \\
      $\overline{\bf 10}$ 
      &  \textcolor{\revcolor}{ 22.5(2.3)(2.6)(0.2)} 
      & \textcolor{\revcolor}{ 19.2(2.3)(3.7)(0.2)} 
      & \textcolor{\revcolor}{31.6(2.7)(3.2)(0.3)}\\
      ${\bf 27}$          
      &  \textcolor{\revcolor}{ 15.1(2.0)(2.0)(0.2)} 
      & \textcolor{\revcolor}{12.3(1.9)(3.6)(0.1)}  
      &  \textcolor{\revcolor}{24.9(2.2)(3.1)(0.3)} \\
\hline
\end{tabular}
%noalign{\smallskip\hrule}\cr}
\begin{minipage}[t]{16.5 cm}
\vskip 0.0cm
\noindent
\end{minipage}
\end{center}
\end{table}     
\begin{table}
\begin{center}
\begin{minipage}[!ht]{16.5 cm}
  \caption{Two-body binding energies (MeV) calculated with the
    \cfgc\  ensemble.  
The first uncertainty is statistical, the second is the fitting systematic and
the third is due to the lattice spacing.
The second to last column corresponds to an average of 
the $|{\bf  n}|=0,1,2$ calculations, which is taken to be the infinite volume
value.
The last column gives the value of $\kappa_0$ times the spatial lattice size
for $L=48$.
  }
\label{tab:A2energies48}
\end{minipage}
\setlength{\tabcolsep}{0.3em}
\begin{tabular}{c|ccc|cc}
\hline
      SU(3) irrep  & $|{\bf n}|=0$ & $|{\bf n}|=1$ & $|{\bf  n}|=2$  &
   \    $L=\infty$   &  $\kappa_0 L$ \\
\hline
      {\bf 1}             
      &  \textcolor{\revcolor}{73.7(3.3)(5.1)(0.8)}  
      &  \textcolor{\revcolor}{73.7(4.4)(7.6)(0.8)}
      &  \textcolor{\revcolor}{75.4(3.3)(3.3)(0.8)}  
      &  \BHgs
      &  \textcolor{\revcolor}{12.3} \\
      ${\bf 8}_A$         
      & \textcolor{\revcolor}{38.7(2.9)(2.9)(0.4) } 
      & \textcolor{\revcolor}{34.6(2.8)(3.1)(0.4) } 
      & \textcolor{\revcolor}{39.7(3.0)(2.7)(0.4) } 
      & \BnXJone
      & \textcolor{\revcolor}{8.8} \\
      ${\bf 10}$          
      & \textcolor{\revcolor}{6.6(3.4)(4.1)(0.0)}  
      & \textcolor{\revcolor}{2.8(3.1)(4.1)(0.0) }
      & \textcolor{\revcolor}{7.0(3.4)(3.7)(0.0) } 
      &  \BnSJone
      &  \textcolor{\revcolor}{ 3.3 } \\
      $\overline{\bf 10}$ 
      &  \textcolor{\revcolor}{19.7(3.1)(4.1)(0.2) } 
      & \textcolor{\revcolor}{17.8(3.6)(3.1)(0.2) } 
      & \textcolor{\revcolor}{23.1(3.9)(5.5)(0.2) } 
      & \Bd
      &  \textcolor{\revcolor}{ 6.3 } \\
      ${\bf 27}$          
      & \textcolor{\revcolor}{13.1(2.8)(4.3)(0.2) } 
      &  \textcolor{\revcolor}{14.9(2.7)(2.7)(0.2) } 
      &  \textcolor{\revcolor}{19.3(2.9)(3.3)(0.2)} 
      & \Bnn
      &  \textcolor{\revcolor}{5.7} \\
\hline
\end{tabular}
%noalign{\smallskip\hrule}\cr}
\begin{minipage}[t]{16.5 cm}
\vskip 0.0cm
\noindent
\end{minipage}
\end{center}
\end{table}     

In sufficiently large  volumes, the binding momentum associated with a two-body bound state 
at rest in the lattice volume will scale as 
\begin{eqnarray}
  \kappa(L) & = & \kappa_0\ +\ {6 Z_\psi^2\over L} e^{-\kappa_0 L}
\ +\ ...
  \ \ \ ,
  \label{eq:extrapform}
\end{eqnarray}
where $\kappa_0$ is the infinite-volume binding momentum,
$\kappa_0=\sqrt{ M_B B}$, where $B$ is the binding energy and 
$Z_\psi$ is the residue of the
bound-state pole~\cite{Luscher:1986pf,Luscher:1990ux,Beane:2003da}.
Analogous FV scaling formulas
for systems moving in the lattice volume are
known~\cite{Davoudi:2011md},
but at this order in the expansion differ from the relation in 
eq.~(\ref{eq:extrapform}) only by the coefficient of the second term.
In 
the \cfgb and \cfgc
lattice volumes, 
the energies of the two-body bound states do not exhibit
statistically significant volume dependence.  
Consequently, using  eq.~(\ref{eq:extrapform}) to
determine the infinite-volume binding energies does not provide a refinement
over simply taking the binding energies determined in the \cfgc\ ensemble, and
the latter is used as the best estimate of the infinite-volume binding energies,
the results of which are shown in Table~\ref{tab:A2energies48}.
The expected differences between the infinite-volume bindings and those in the 
\cfgc\ ensemble can be estimated from the values of $\kappa_0 L$ given in  
Table~\ref{tab:A2energies48}.  
With the exception of the state in the {\bf 10} irrep, 
the states are small enough compared to the lattice volume to make the
finite-volume effects negligible.

There are a few important results that should be highlighted.  The
deuteron is found to be substantially more deeply bound in the present
calculations, $B_d^{n_f=3}=\BdMeV$, than in the quenched
calculations~\cite{Yamazaki:2011nd} in which a binding energy of
 \textcolor{\revcolor}{$B_d^{n_f=0}=9.1(1.1)(0.5)~{\rm MeV}$} at a similar pion mass is found.  
The
H-dibaryon is found to be deeply bound with $B_H=\BHgsMeV$, approximately
twice as bound as the result found by 
HALQCD~\cite{Inoue:2011ai} at a similar quark mass.
In recent work we reported that the $n\Sigma^-$
interaction in the $\siii-\diii$ channel was extremely repulsive at a
pion mass of $m_\pi\sim 390~{\rm MeV}$\cite{Beane:2012ey}, consistent
with the phase-shift analysis of experimental data at the physical
pion mass.  At the SU(3) symmetric point, we find that this state has
moved close to threshold and is even consistent with being bound,
indicating that there is significant light quark mass dependence in
this channel at the heavier quark masses (beyond the regime of applicability of
the relevant EFT).

As the calculations have been performed at the SU(3) symmetric point, the states discussed above
provide a nearly complete set of two-baryon ground states,
with only the ${\bf 8}_S$ irrep being absent. 
Furthermore, since the
determinations of the various energy levels in the two-body sector are
correlated, their differences can be determined more precisely than
their individual values. In Table \ref{tab:A2split} we present the
splittings between the various irreps.
\begin{table}
\begin{center}
\begin{minipage}[!ht]{16.5 cm}
  \caption{Two-body energy splittings, $E_{I_1}-E_{I_2}$ (MeV) between
    different multiplets calculated with the \textcolor{\revcolor}{ \cfgb}  ensemble.  
    The column refers to representation $I_1$ and the row to
    representation $I_2$.
The first uncertainty is statistical, the second is the fitting systematic and
the third is due to the lattice spacing.
  }
\label{tab:A2split}
\end{minipage}
\setlength{\tabcolsep}{0.7em}
\begin{tabular}{c|cccc}
\hline
     $I_2$\textbackslash $I_1$ &  ${\bf 8}_A$ &   ${\bf 10}$ & $\overline{\bf 10}$ & ${\bf 27}$ \\
\hline
      {\bf 1}  
      &  \textcolor{\revcolor}{34.3(0.7)(1.2)(0.4)} 
      & \textcolor{\revcolor}{ 65.9(0.4)(0.9)(0.7)} 
      & \textcolor{\revcolor}{49.3(1.4)(1.7)(0.5)} 
      &  \textcolor{\revcolor}{ 55.4(1.2)(1.8)(0.6) }  \\          
      ${\bf 8}_A$          
      & --- 
      &  \textcolor{\revcolor}{  31.0(0.8)(1.6)(0.3)  }
      &  \textcolor{\revcolor}{ 14.2(1.1)(2.0)(0.2) } 
      &  \textcolor{\revcolor}{ 20.5(1.2)(2.3)(0.2) } \\          
      ${\bf 10}$           
      & --- 
      & --- 
      &  \textcolor{\revcolor}{-17.7(1.4)(2.2)(0.2) }  
      & \textcolor{\revcolor}{-11.0(1.4)(2.4)(0.1) }  \\          
      $\overline{\bf 10}$   
      & --- 
      & --- 
      & --- 
      &  \textcolor{\revcolor}{ 5.8(1.0)(1.0)(0.1) } \\          
\hline
\end{tabular}
%noalign{\smallskip\hrule}\cr}
\begin{minipage}[t]{16.5 cm}
\vskip 0.0cm
\noindent
\end{minipage}
\end{center}
\end{table}     
The energy difference between the $\overline{\bf 10}$ and ${\bf 27}$
corresponds to the deuteron-di-nucleon mass difference.  This splitting is
found to be small, and consistent with zero within the uncertainties of the
calculation.  Theoretically, it has been established from SU(2) that these states become
degenerate in the large-$N_c$ limit of QCD~\cite{Kaplan:1995yg},
with a fractional splitting (and violation of Wigner's SU(4) symmetry) that scales as $1/N_c^2$.
Extending the argument to the strange sectors shows that 
the other  splittings are only
$1/N_c$ suppressed, and not $1/N_c^2$ suppressed~\cite{Kaplan:1995yg}.  Such scalings are
consistent with what we have found, but verification of the scaling will
require significantly higher statistics in the calculations.

%%%%%%%%%%%%%%%%%%%%%%%%%%%%%%%%%%%%%%%%%%%%%%%%%%%%
\section{Three-Body Systems}
\label{sec:BBB}
\noindent
The correlation functions for the three-body systems are generated
using the procedure described previously.  As is the case for two-body
systems, the states in the spectrum for each system can be classified
by their SU(3) quantum numbers in the limit of SU(3)-flavor symmetry.
The three-body states can be assigned to the SU(3) irreps in ${\bf
  8}\otimes {\bf 8}\otimes{\bf 8}$, which can be straightforwardly
constructed as
\begin{eqnarray} {\bf 8}\otimes {\bf 8}\otimes{\bf 8}
  & = & 
  {\bf 64}\oplus 2\  {\bf 35} \oplus 2\ \overline{\bf 35} \oplus 6\  {\bf 27}
  \oplus 4\  {\bf 10} \oplus 4\  \overline{\bf 10} \oplus 8\  {\bf 8}\oplus 2\  {\bf 1}
  \ \ \ .
  \label{eq:eightcubed}
\end{eqnarray}
However, the local sources constructed from only the
upper components of the quark fields produce correlation functions containing
a subset of these irreps,
\begin{eqnarray} {\bf 8}\otimes {\bf 8}\otimes{\bf 8}
  & \rightarrow & 
  {\bf 35} \oplus  \overline{\bf 35} \oplus 2\  {\bf 27}
  \oplus {\bf 10} \oplus   \overline{\bf 10} \oplus 
2\  {\bf 8}\oplus \
  {\bf 1}
  \ \ \ ,
  \label{eq:eightcubedlocal}
\end{eqnarray}
and further decomposition into states with $J^\pi={1\over 2}^+$ and
$J^\pi={3\over 2}^+$ gives
\begin{eqnarray} 
\left(\ {\bf 8}\otimes {\bf 8}\otimes{\bf 8}\ \right)_{J^\pi=1/2^+}
  & \rightarrow & 
  {\bf 35} \oplus  \overline{\bf 35} \oplus  {\bf 27}
  \oplus {\bf 8}
\nonumber\\
\left(\ {\bf 8}\otimes {\bf 8}\otimes{\bf 8}\ \right)_{J^\pi=3/2^+}
  & \rightarrow & 
{\bf 27} \oplus  {\bf 10} \oplus   \overline{\bf 10} \oplus  {\bf 8} 
\oplus \  {\bf 1}
  \ \ \ .
  \label{eq:eightcubedlocalJdecompo}
\end{eqnarray}
It is clear from the SU(3) irreps contributing to the three-body systems  
that, with our source structure, a given correlation function 
contains contributions from multiple SU(3) irreps.  
With a relatively small number of states identified with the
present set of correlation functions, the SU(3) classification of states 
is difficult to establish from the spectra alone.  
More generally, it is expected that the spectrum of states in any given correlation
function becomes increasingly complicated with increasing numbers of baryons
even when constrained by SU(3)-flavor symmetry.
As the focus of this work is systems containing only a  small number of strange
quarks, we
have chosen to use the same notation as in hypernuclear spectroscopy.
States in \hetppn\ (same as \htpnn\ by
isospin symmetry), \hetL\ (same as \htL\ and \nnL\ by isospin symmetry),
the isosinglet \htL, 
and the isotriplet 
\hetS\ have been identified in the three-body sector.

Correlation functions 
calculated with LQCD
will not only contain contributions from the ground state and excited states of
the bound nuclei but also continuum states that consist of all
possible sub-clusterings of the baryons.  For instance, the
correlation functions used to extract the \hetppn\  nuclear states will
also contain contributions from the deuteron-proton and  di-proton-neutron
in addition to the  proton-proton-neutron continuum states.  With sufficient precision
in the calculation, one will be able to use these levels to extract, for
instance, the deuteron-proton scattering phase-shift~\cite{Bour:2011ef}.  
Given that the two-body sector is well-established, the spectrum
of such continuum states can be approximately constructed.  
Clearly,
states of the \hetppn\  nucleus can only be cleanly identified when they are not
close in energy to 
the expected location of non-interacting continuum
states.  
The generalization of this discussion applies to
other systems comprised of three or more baryons. 
In Appendix \ref{sec:scatt_states}, 
an example of 
the expected FV scattering-state spectrum is constructed for
each of the volumes used in this analysis,
demonstrating the extent of this problem in large volumes.

%%%%%%%%%%%%%%%%%%%%%%%%%%%%%%%%%%%%%%%%%%%%%%%%%%%%
\subsection{$I={1\over 2}$,  $J^\pi={1\over 2}^+$ : \htpnn\ and \hetppn }
\label{sec:3Hand3He}
\begin{figure}[!ht]
  \centering
  \includegraphics[width=0.32\textwidth]{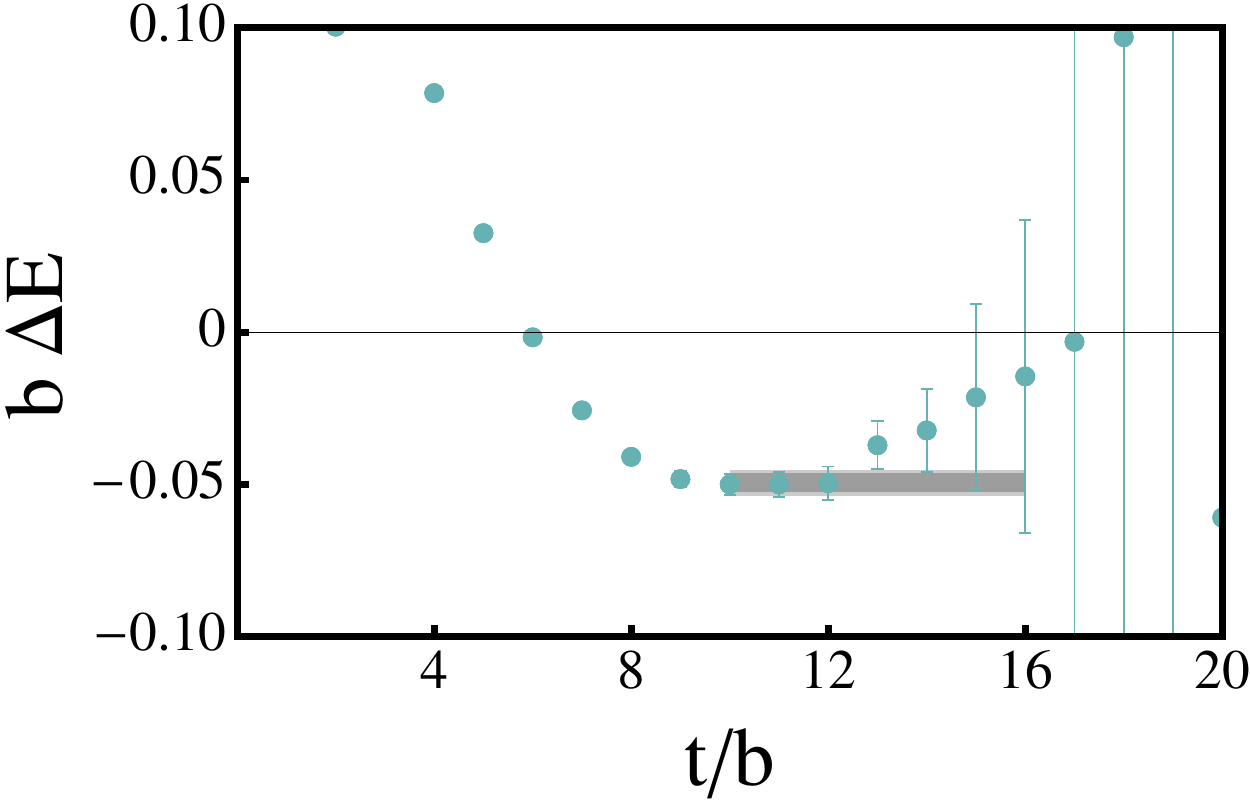}\ 
  \includegraphics[width=0.32\textwidth]{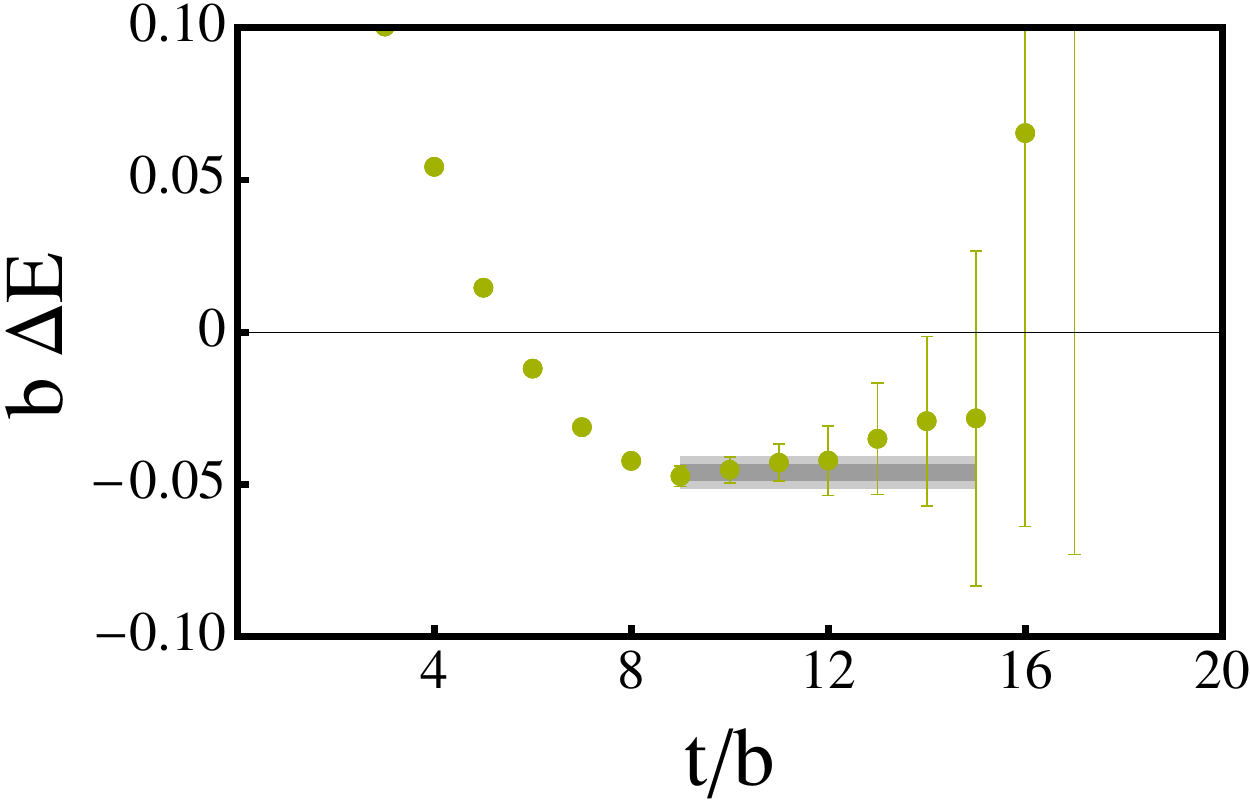}\ 
  \includegraphics[width=0.32\textwidth]{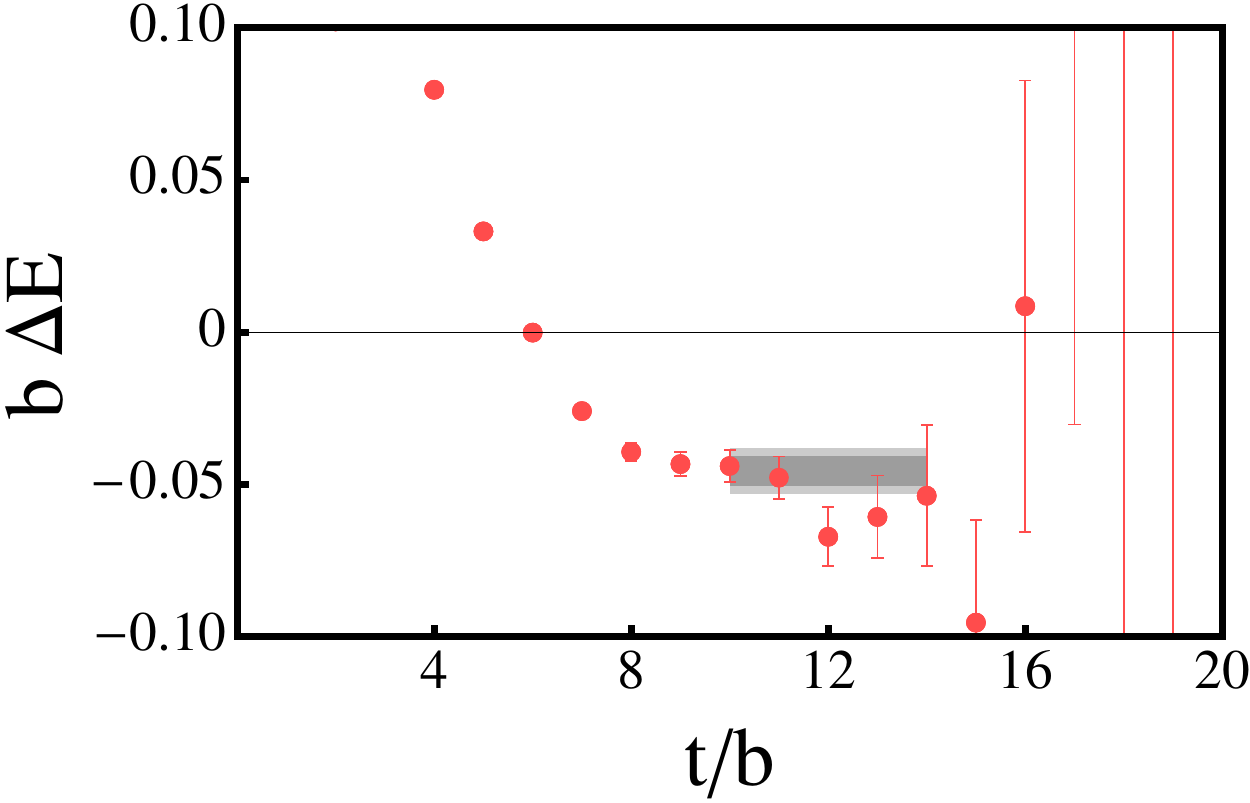}
  \caption{EMPs associated with
    $J^\pi={1\over 2}^+$ \hetppn\ (\htpnn) $|{\bf P}|=0$ correlation functions
    computed with the \cfga\ (left), \cfgb\ (center) and \cfgc\
    (right) ensembles.  
 \textcolor{\revcolor}{
The inner (darker) shaded region corresponds to the statistical uncertainty
of the extracted energy, while  the outer (lighter) shaded region
corresponds to the
statistical and fitting systematic uncertainties combined in quadrature.
}
 }
  \label{fig:He3EMPs}
\end{figure}

\noindent
In nature, the $I={1\over 2}$, $J^\pi={1\over 2}^+$ ground state of
the \hetppn\ nucleus is the only bound state of two protons and a
neutron, and it is known to be dominantly composed of two protons in a
$\si$ state coupled to an s-wave neutron.  Four \hetppn\ correlation
functions, resulting from different source structures defined by
$s=0$, $I={1\over 2}$ and $J^\pi={1\over 2}^+$ quantum numbers
transforming as a
$\overline{\bf 35}$ of SU(3), have
been constructed.\footnote{The only possible SU(3) irrep with these quantum
  numbers is the $\overline{\bf 35}$.}
EMPs obtained from correlation functions in each of the three ensembles, 
from which the energy of the lowest-lying \hetppn\ states have been
determined, are shown in fig.~\ref{fig:He3EMPs}.
\begin{table}
\begin{center}
\begin{minipage}[!ht]{16.5 cm}
  \caption{
    The calculated \textcolor{\revcolor}{$J^\pi={1\over 2}^+$} binding energy of \hetppn\  (\htpnn) in the
    \textcolor{\revcolor}{ \cfga } ensemble.
    ``g.s.''  denotes the ground state.
The first uncertainty is statistical, the second is the fitting systematic and
the third is due to the lattice spacing.
  }
\label{tab:He3energies24}
\end{minipage}
\setlength{\tabcolsep}{1em}
\resizebox{\linewidth}{!}{%
\begin{tabular}{c|ccc}
\hline
      \hetppn  &   \multicolumn{3}{c}{\cfga}  \\
\hline
      & $|{\bf n}|^2=0$  & $|{\bf n}|^2=1$ & $|{\bf n}|^2=2$ 
\\
      \hline
      g.s.  (MeV) 
      &   \textcolor{\revcolor}{65.4(5.1)(4.4)(0.7) }
      &  \textcolor{\revcolor}{42.8(3.8)(8.9)(0.4) }
      &  \textcolor{\revcolor}{46.3(5.3)(6.7)(0.5) }
\\
\hline
\end{tabular}
}
%noalign{\smallskip\hrule}\cr}
\begin{minipage}[t]{16.5 cm}
\vskip 0.0cm
\noindent
\end{minipage}
\end{center}
\end{table}  
\begin{table}
\begin{center}
\begin{minipage}[!ht]{16.5 cm}
  \caption{
    The calculated  \textcolor{\revcolor}{$J^\pi={1\over 2}^+$} binding energy of \hetppn\  (\htpnn) in the
    \textcolor{\revcolor}{ \cfgb } ensemble.
    ``g.s.''  denotes the ground state.
The first uncertainty is statistical, the second is the fitting systematic and
the third is due to the lattice spacing.
  }
\label{tab:He3energies32}
\end{minipage}
\setlength{\tabcolsep}{1em}
\resizebox{\linewidth}{!}{%
\begin{tabular}{c|ccc}
\hline
      \hetppn  &  \multicolumn{3}{c}{\cfgb}\\
\hline
      & $|{\bf n}|^2=0$ & $|{\bf n}|^2=1$ & $|{\bf n}|^2=2$  \\
      \hline
      g.s.  (MeV) 
      &  \textcolor{\revcolor}{63.2(3.9)(7.0)(0.7)} 
      &  \textcolor{\revcolor}{52.9(5.7)(9.9)(0.6) } 
      &  \textcolor{\revcolor}{55.7(6.4)(10.1)(0.6)} 
\\
\hline
\end{tabular}
}
%noalign{\smallskip\hrule}\cr}
\begin{minipage}[t]{16.5 cm}
\vskip 0.0cm
\noindent
\end{minipage}
\end{center}
\end{table}  
\begin{table}
\begin{center}
\begin{minipage}[!ht]{16.5 cm}
  \caption{
    The calculated \textcolor{\revcolor}{$J^\pi={1\over 2}^+$} binding energy of \hetppn\  (\htpnn) in the
    \textcolor{\revcolor}{ \cfgc } ensemble.
    ``g.s.''  denotes the ground state.
The first uncertainty is statistical, the second is the fitting systematic and
the third is due to the lattice spacing.
  }
\label{tab:He3energies48}
\end{minipage}
\setlength{\tabcolsep}{1em}
\resizebox{\linewidth}{!}{%
\begin{tabular}{c|ccc}
\hline
      \hetppn 
      & \multicolumn{3}{c}{\cfgc} \\
\hline
      & $|{\bf n}|^2=0$ & $|{\bf n}|^2=1$ & $|{\bf n}|^2=2$  \\
      \hline
      g.s.  (MeV) 
      &  \textcolor{\revcolor}{61.9(8.9)(10.9)(0.7)} 
      & \textcolor{\revcolor}{53.0(7.1)(8.0)(0.6) } 
      & \textcolor{\revcolor}{ 50.0(6.1)(9.2)(0.6) }\\
\hline
\end{tabular}
}
%noalign{\smallskip\hrule}\cr}
\begin{minipage}[t]{16.5 cm}
\vskip 0.0cm
\noindent
\end{minipage}
\end{center}
\end{table}  
\begin{figure}[!ht]
  \centering
  \includegraphics[width=0.99\textwidth]{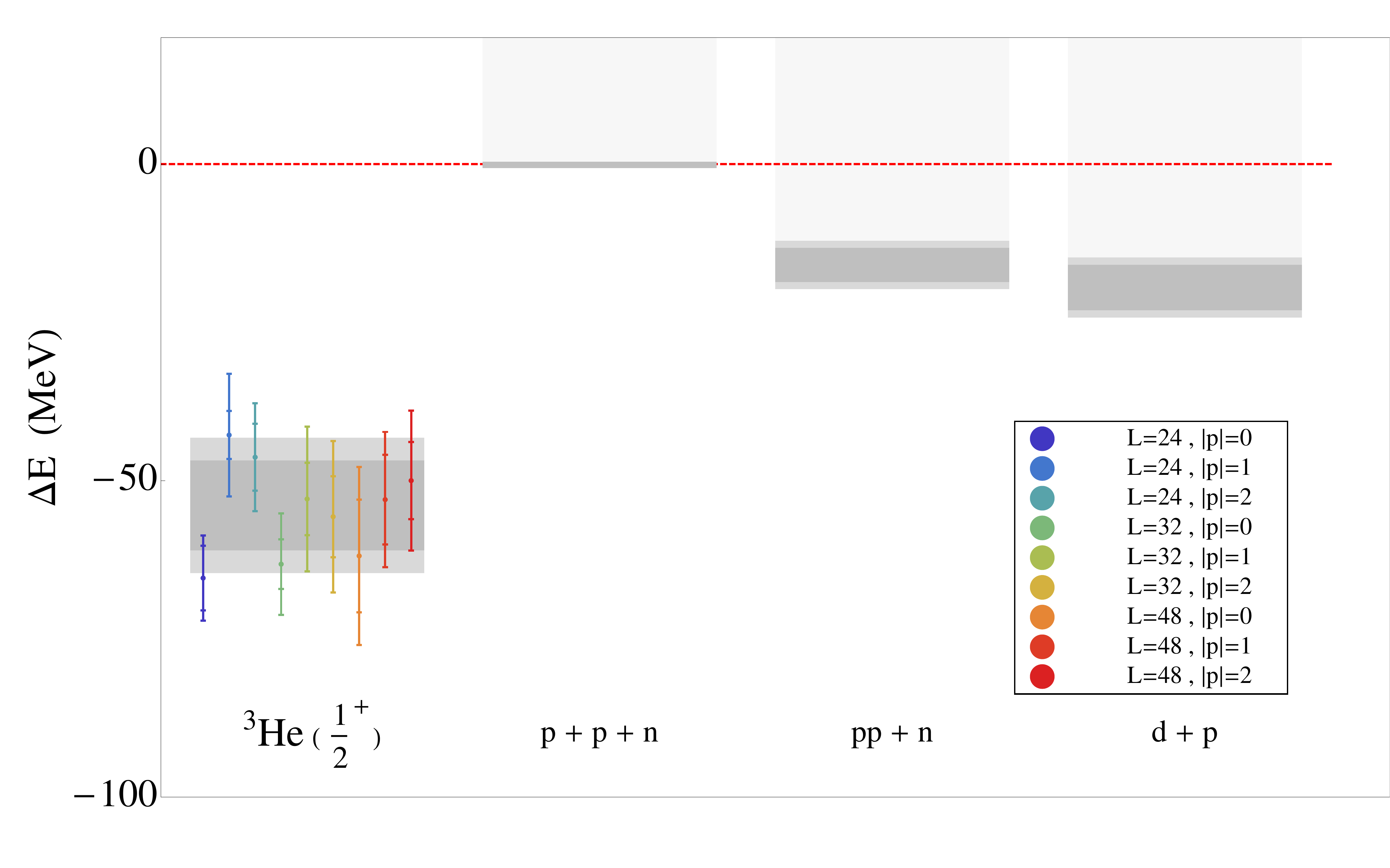}
\caption{The bound-state energy levels in the  $J^\pi={1\over 2}^+$ \hetppn\
  (\htpnn) sector.
The points and their associated uncertainties correspond to the energies of the
states extracted from the correlation functions with the quantum numbers of the
ground state of \hetppn.
The locations of the scattering thresholds associated with non-interacting  
deuteron-proton, di-proton-neutron and proton-proton-neutron continuum
states, determined from the  single-hadron spectrum and the  two-body binding
energies given in Table~\protect\ref{tab:A2energies48} 
are shown.
\label{fig:A3levels}}
\end{figure}
The \hetppn\ bound-state energies on the ensembles 
are given in Tables~\ref{tab:He3energies24}, \ref{tab:He3energies32} and
\ref{tab:He3energies48} 
and are shown in fig.~\ref{fig:A3levels}
along with the thresholds for non-interacting continuum
states.\footnote{Finite-volume 
effects will lead to small shifts in these thresholds.}  
The exact form of
infinite-volume extrapolation of three- and higher-body bound-state
energies is as yet unknown, though expected to be exponential (see
Refs.~\cite{Kreuzer:2010ti,Polejaeva:2012ut,Kreuzer:2012sr} for related
discussions). For the current study, we simply average the results
obtained from the system at rest and from the boosted systems on the
\cfgc\ ensemble to provide an estimate of the infinite-volume binding
energy of
\begin{eqnarray}
  B^{(\infty)}(^3{\rm He}) & = & \BhetMeV
  \ \ \ .
  \label{eq:he3states}
\end{eqnarray}
The energy of this state is 
significantly lower than any of the expected continuum
states, based upon where they would lie in the spectrum in the absence
of interactions.  
Therefore, we conclude that this is the ground state of \hetppn.

While it is tempting to compare these results with the experimental
spectrum of \hetppn, one should refrain at present, since
these calculations are performed in the SU(3) limit of QCD 
and without electromagnetism.  
The ground-state binding energy
will receive a shift due to the electromagnetic interaction between
the two protons.  On the other-hand, the exact isospin symmetry
directly relates this spectrum to that of the triton. In nature the
triton binding energy per nucleon is $B/A\sim 2.83~{\rm MeV}$, while
at the SU(3) symmetric point we find that $B/A\sim \BhetGSperAapprox$,
more than an order of magnitude larger.

The \hetppn\  ground-state energy that we have calculated in this
$n_f=3$ calculation is substantially different from that obtained with
quenched calculations at a comparable pion
mass~\cite{Yamazaki:2009ua}, which find an infinite-volume
extrapolated value of $B^{(\infty)}_{n_f=0}(^3{\rm
  He})=18.2(3.5)(2.9)~{\rm MeV}$. 
A likely explanation for
the difference is quenching artifacts, which are unlikely to cancel between the
bound system and the threshold states. The difference in the
total energy (not the binding energy) of the 
\hetppn\ 
ground state between the two calculations is of ${\cal O}(1 \%)$, smaller
than the differences observed between single-hadron masses in quenched
and unquenched calculations~\cite{Ali  Khan:2001tx}. 
Additionally, the contributions from continuum states
that must be present in both calculations at some level 
(see  Appendix \ref{sec:scatt_states})
may pollute
the extraction of the \hetppn\ ground state, particularly in large volumes.

%%%%%%%%%%%%%%%%%%%%%%%%%%%%%%%%%%%%%%%%%%%%%%%%%%%%
\subsection{$I=0$,  $J^\pi={1\over 2}^+$  and   $J^\pi={3\over 2}^+$ : \htL - The Hypertriton}
\label{sec:3HLam}
\begin{figure}[!ht]
  \centering
  \includegraphics[width=0.32\textwidth]{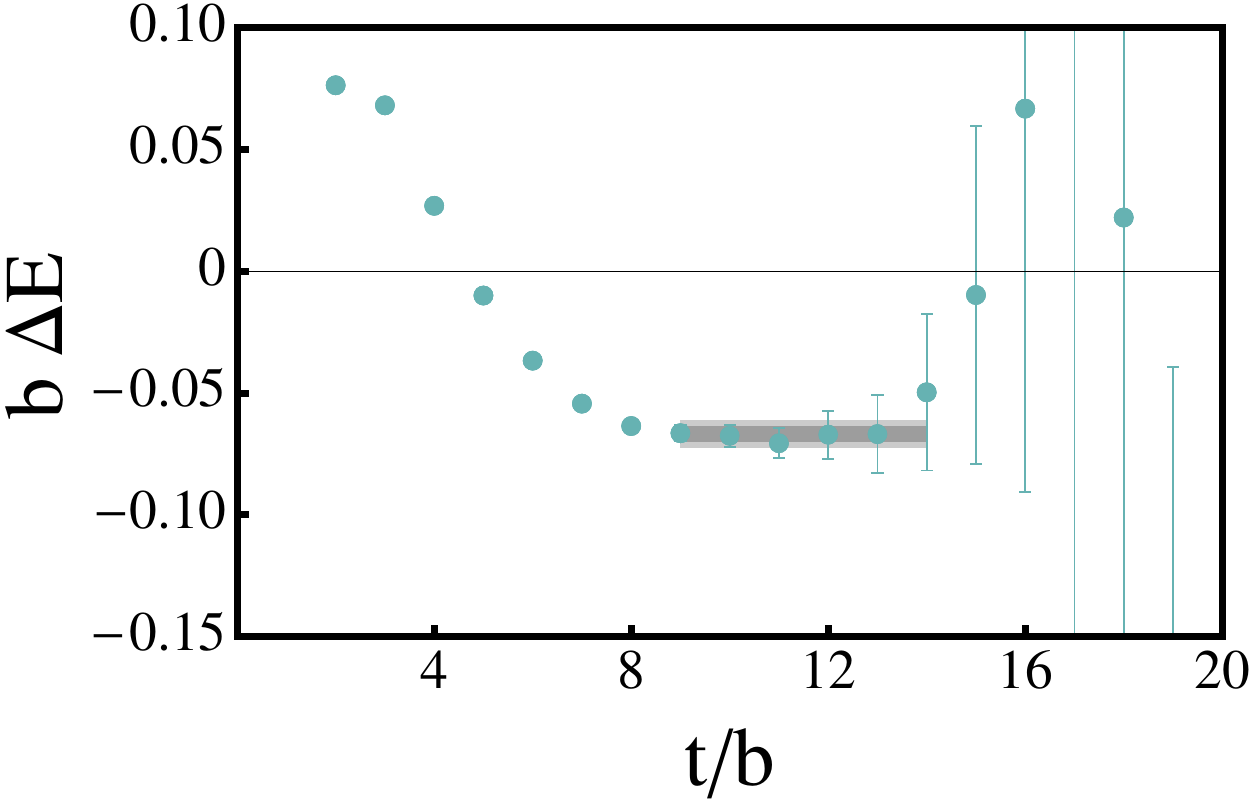}
  \includegraphics[width=0.32\textwidth]{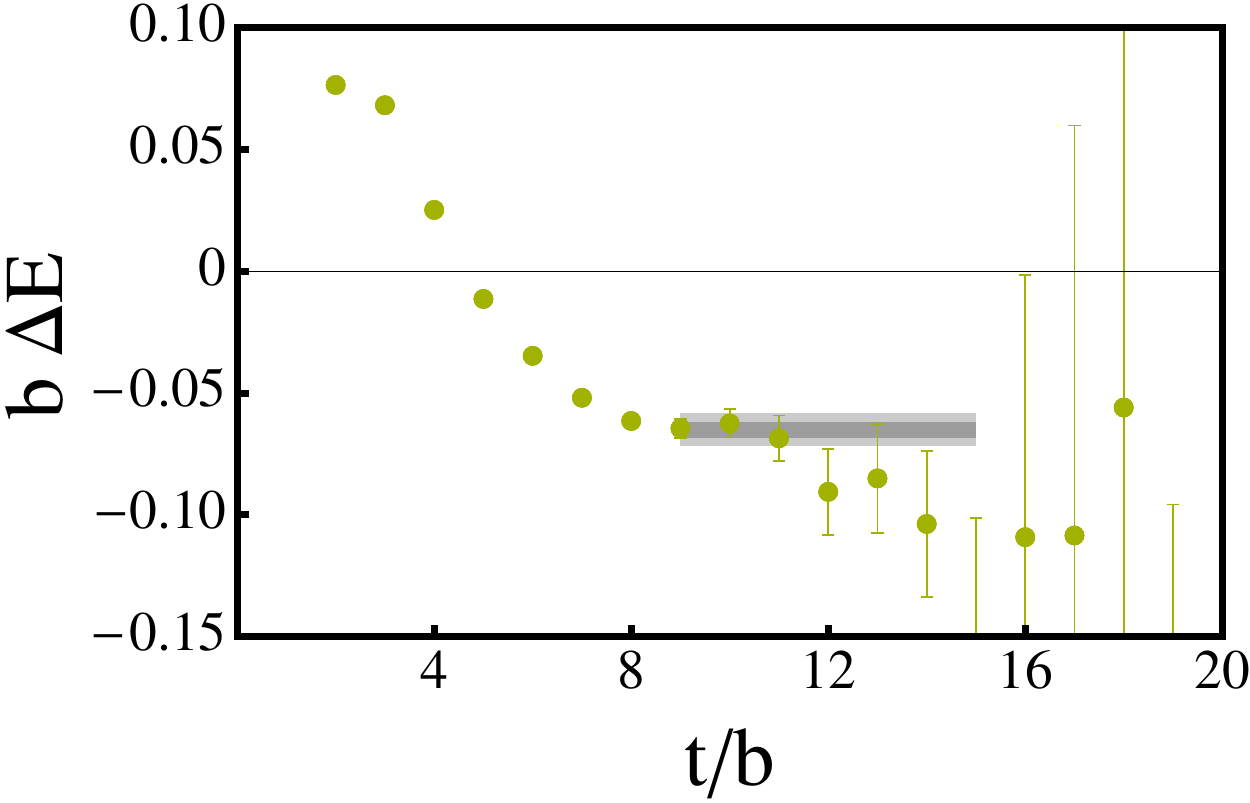}
  \includegraphics[width=0.32\textwidth]{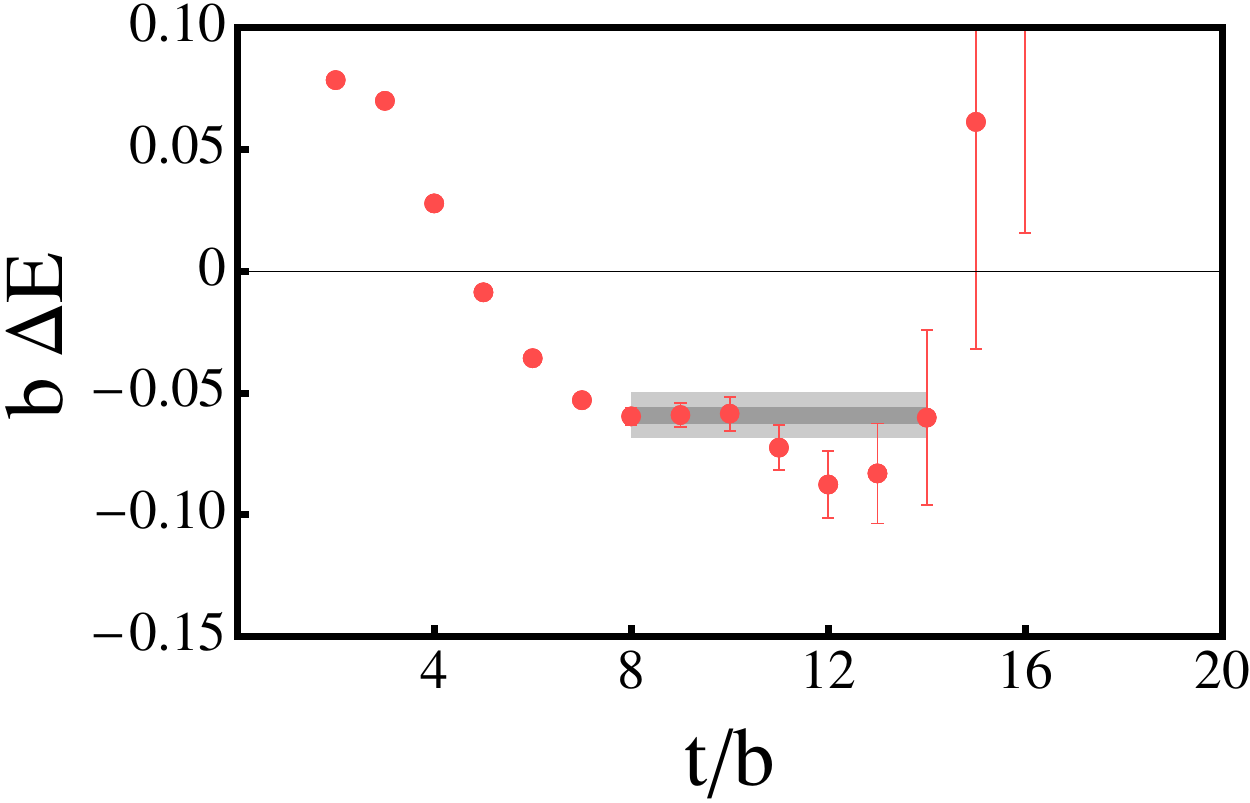}
  \caption{The EMPs associated with a $J^\pi={3\over 2}^+$ hypertriton (\htL)
correlation function computed with the \cfga\ (left),
    \cfgb\ (center) and \cfgc\ (right) ensembles, with momentum $|{\bf
      P}|=0$.  
 \textcolor{\revcolor}{
The inner (darker) shaded region corresponds to the statistical uncertainty
of the extracted energy, while  the outer (lighter) shaded region
corresponds to the
statistical and fitting systematic uncertainties combined in quadrature.
}
 }
  \label{fig:H3LEMPs}
\end{figure}

\begin{figure}[!ht]
  \centering
  \includegraphics[width=0.99\textwidth]{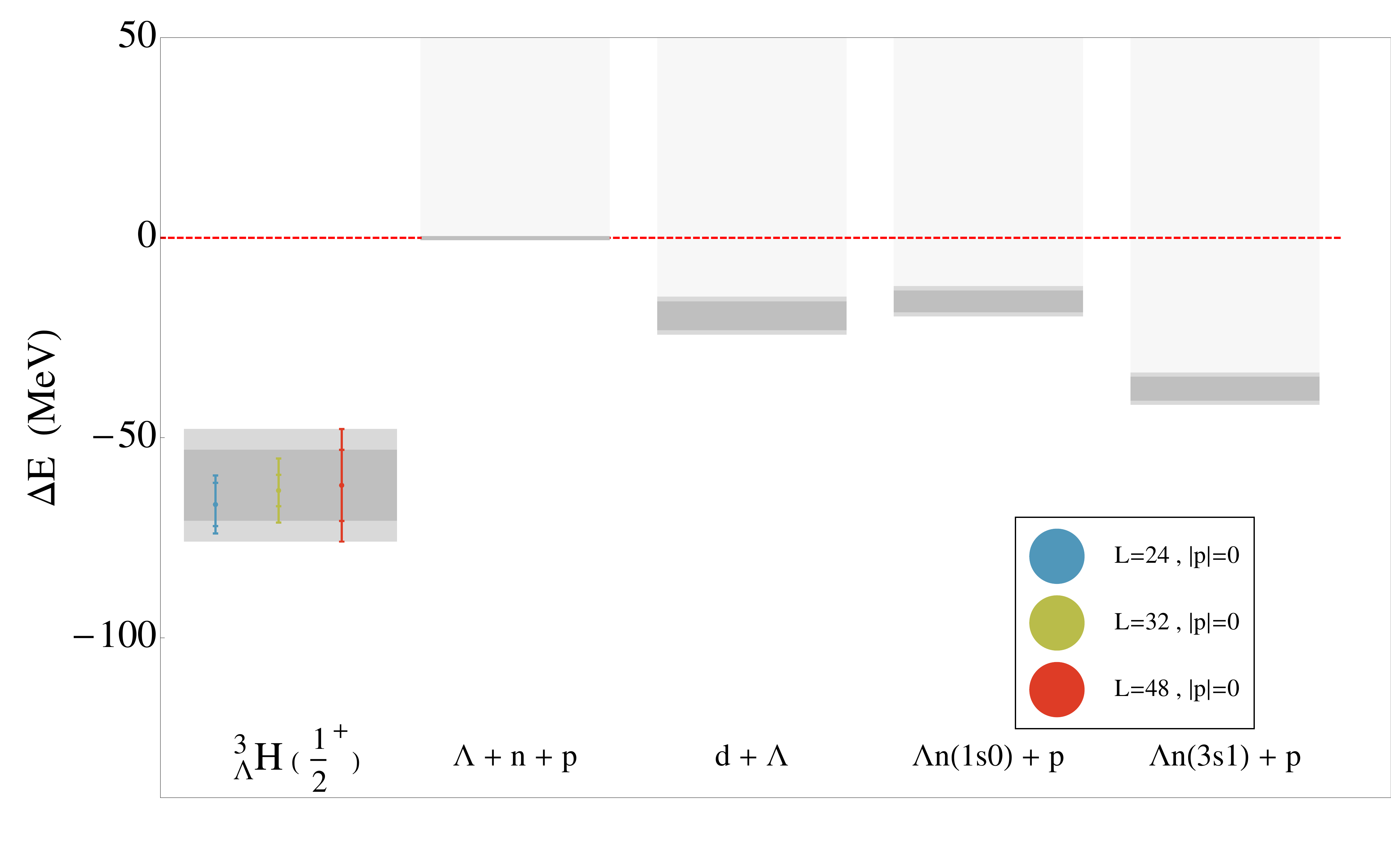}
  \includegraphics[width=0.99\textwidth]{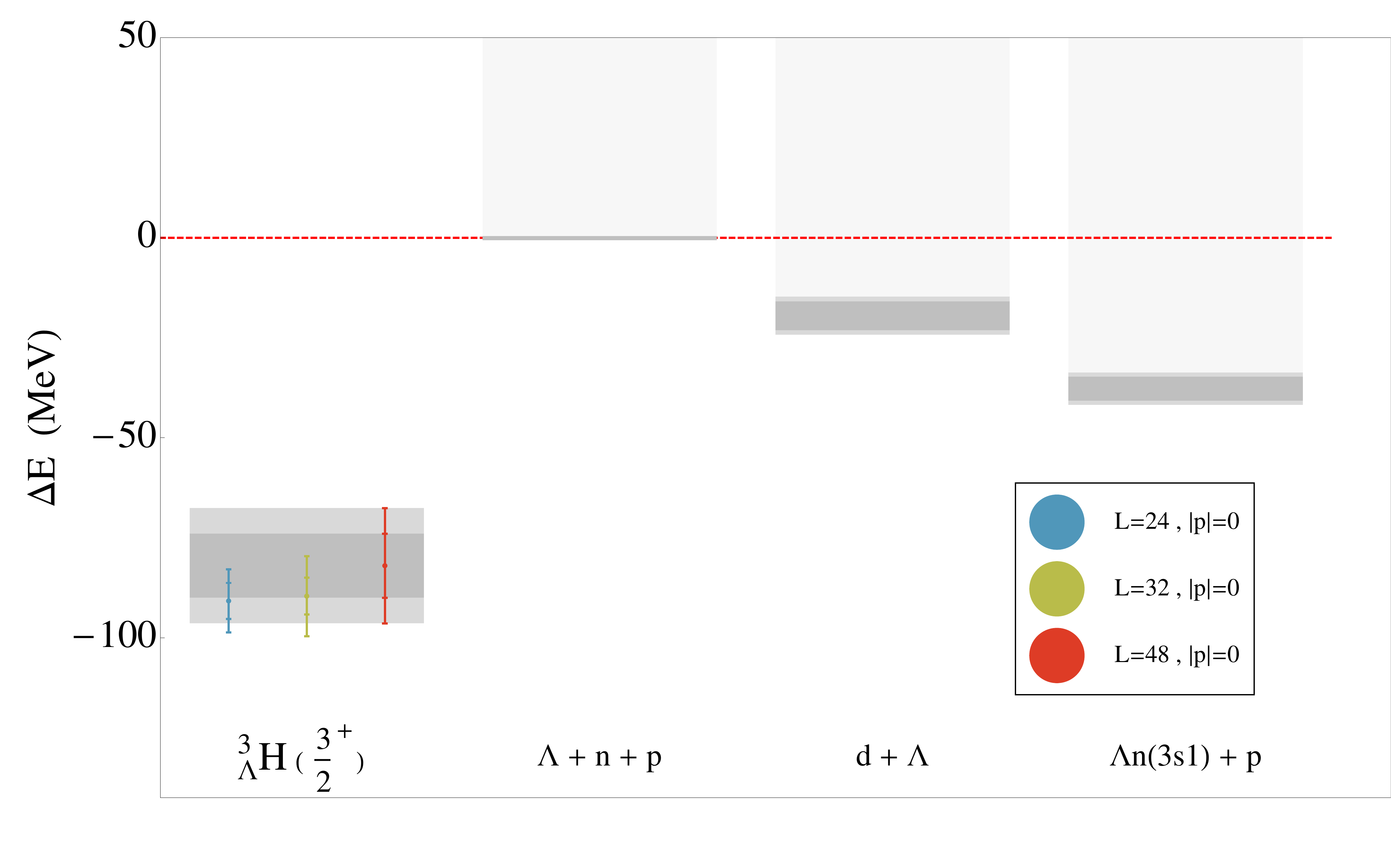}
  \caption{The bound-state energy levels in the $J^\pi={1\over 2}^+$
    (upper panel) and $J^\pi={3\over 2}^+$ (lower panel) hypertriton
(\htL) sector.
    The points and their associated uncertainties correspond to the
    energies of the states extracted from the correlation functions
    with the quantum numbers of the ground state of 
$J^\pi={1\over 2}^+$ and 
$J^\pi={3\over 2}^+$ \htL.  The locations of the energy-levels associated with
    non-interacting 
    continuum states, determined from
    the two-body binding energies given in
    Table~\protect\ref{tab:A2energies48}, are shown.
    }
  \label{fig:H3Llevels}
\end{figure}

\noindent
The hypertriton, \htL, with the quantum numbers of $np\Lambda$ and
$I=0$ is the simplest hypernucleus produced in the laboratory, having a
total binding energy of $B\sim 2.35~{\rm MeV}$.
With a
$\Lambda$-separation energy of just $B^\Lambda\sim 0.13~{\rm MeV}$,
it is consistent with a $\Lambda$ weakly bound to a deuteron.  The
ground state has $J^\pi={1\over 2}^+$ and has been identified as a
member of the $\overline{\bf 35}$ of flavor SU(3)~\cite{Ram:1966}.  It
continues to be the focus of experimental efforts, for instance, in
heavy-ion collisions at RHIC~\cite{Chen:2009ku}
and the HypHI project at GSI,
where in the latter it is being used as a ``phase-zero'' calibration
nucleus for the production and detection systems~\cite{Saito:2012zz}.  
We have
calculated correlation functions in both the $J^\pi={1\over 2}^+$ and
$J^\pi={3\over 2}^+$ channels and have identified the lowest-lying state in
each.
Two of the correlation functions associated with the  $J^\pi={1\over 2}^+$ channel  are
pure $\overline{\bf 35}$ and are in the same irrep as \hetppn, and 
hence the
energy of the identified  states are the same.
Further, the $J^\pi={3\over 2}^+$ channel is pure  $\overline{\bf 10}$.  
EMPs in the $J^\pi={3\over 2}^+$ channel from these correlation functions are shown
in fig.~\ref{fig:H3LEMPs}, from which the energies of the lowest lying
states have been determined, and are given in Table~\ref{tab:H3Lenergies}.
The EMPs in the  $J^\pi={1\over 2}^+$ channel are not shown, as they are
identical to those of \hetppn, shown in fig.~\ref{fig:He3EMPs}.
The extracted spectra of bound states are  shown in
fig.~\ref{fig:H3Llevels}.
\begin{table}
\begin{center}
\begin{minipage}[!ht]{16.5 cm}
  \caption{
    The calculated   \textcolor{\revcolor}{ $J^\pi={3\over 2}^+$} binding energies in \htL.
    ``g.s.''  denotes the ground state.
The energies in the $J^\pi={1\over 2}^+$ channel are the same as those of
\hetppn\ by SU(3) symmetry, see Tables~\protect\ref{tab:He3energies24},
\protect\ref{tab:He3energies32} and \protect\ref{tab:He3energies48}.
The first uncertainty is statistical, the second is the fitting systematic and
the third is due to the lattice spacing.
  }
\label{tab:H3Lenergies}
\end{minipage}
\setlength{\tabcolsep}{1em}
\begin{tabular}{c|ccc}
\hline
 \htL        & \cfga & \cfgb & \cfgc \\
      \hline
      $J^\pi={3\over 2}^+$ g.s.  (MeV) 
      &  \textcolor{\revcolor}{90.8(4.5)(6.5)(1.0) } 
      &  \textcolor{\revcolor}{  89.6(4.6)(8.9)(1.0) } 
      & \BhtLJthreehalf \\
\hline
\end{tabular}
%noalign{\smallskip\hrule}\cr}
\begin{minipage}[t]{16.5 cm}
\vskip 0.0cm
\noindent
\end{minipage}
\end{center}
\end{table}     
Taking the results obtained in the \cfgc\ ensemble to be the best
estimate of the \htL\ infinite-volume binding energies gives,
\begin{eqnarray}
  B^{(\infty)}(\htLJhalf ) & = & \BhtLJhalfMeV
  \nonumber\\
  B^{(\infty)}(\htLJthreehalf) & = & \BhtLJthreehalfMeV
  \ \ \ ,
  \label{eq:h3Lstates}
\end{eqnarray}
where we have used the \hetppn\ result for the $J^\pi={1\over 2}^+$ binding
energy, which includes calculations of boosted systems.

The observed states are significantly below the scattering thresholds
and are consistent with a bound \htL\  nucleus at these values of the
quark masses in the absence of electromagnetism. Interestingly, the
lowest energy state is in the $J^\pi=\frac{3}{2}^+$ spin channel. As the
measurements of the two spin states are correlated, the spin splitting
can be extracted with high precision, resulting in
\begin{eqnarray}
  B^{(\infty)}(\htLJthreehalf) -
  B^{(\infty)}(\htLJhalf) & = & \BhtLJthreehalfminusonehalf
  \ \ \ .
  \label{eq:h3Ldiff}
\end{eqnarray}

%%%%%%%%%%%%%%%%%%%%%%%%%%%%%%%%%%%%%%%%%%%%%%%%%%%%
\subsection{$I=1$, $J^\pi={1\over 2}^+$: \hetL , \htLtilde\  and \nnL }
\label{sec:3HeLamANDnnL}
\begin{figure}[!ht]
  \centering
  \includegraphics[width=0.32\textwidth]{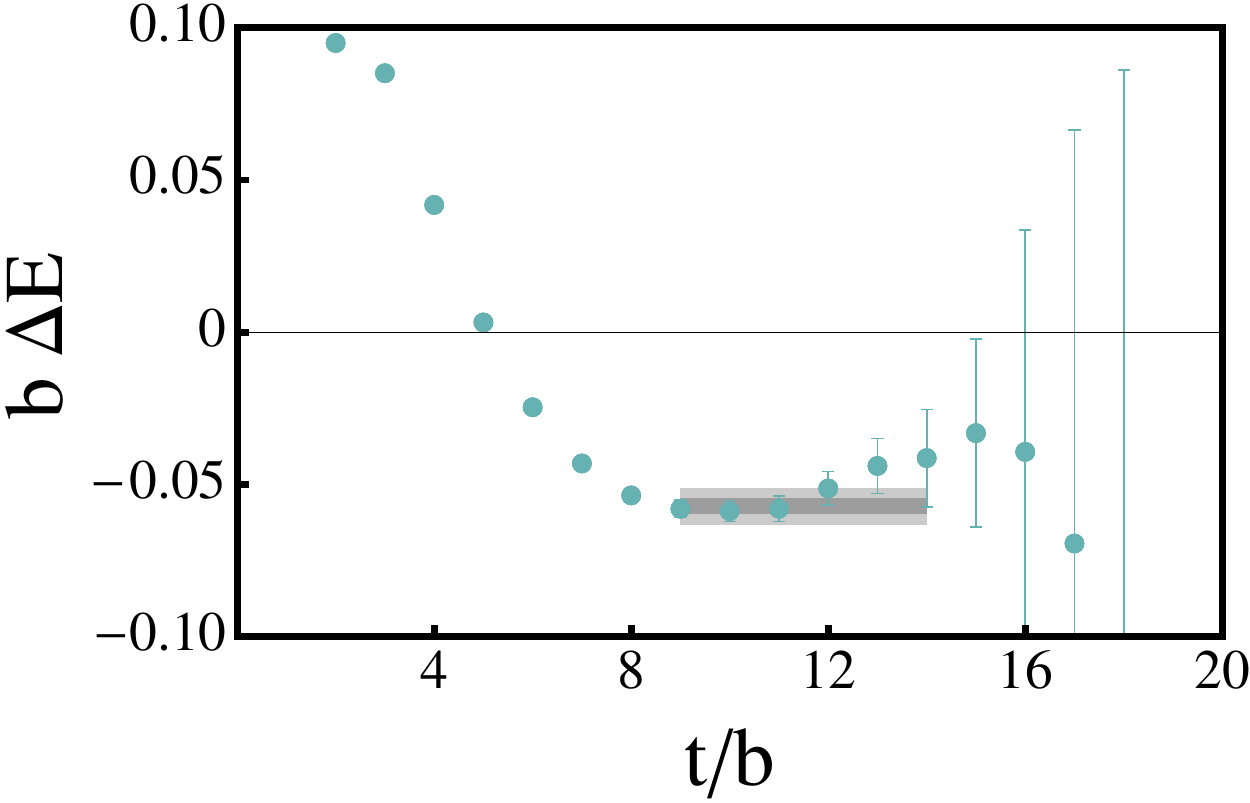}
  \includegraphics[width=0.32\textwidth]{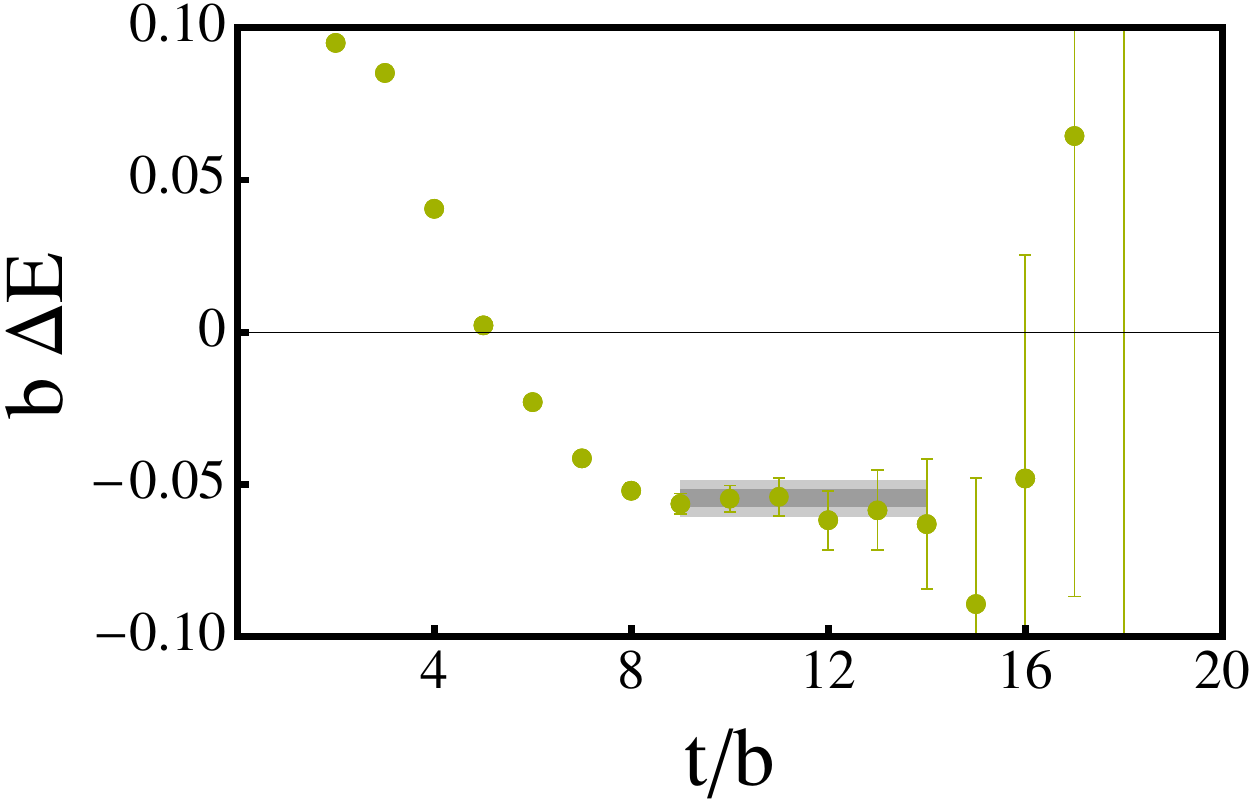}
  \includegraphics[width=0.32\textwidth]{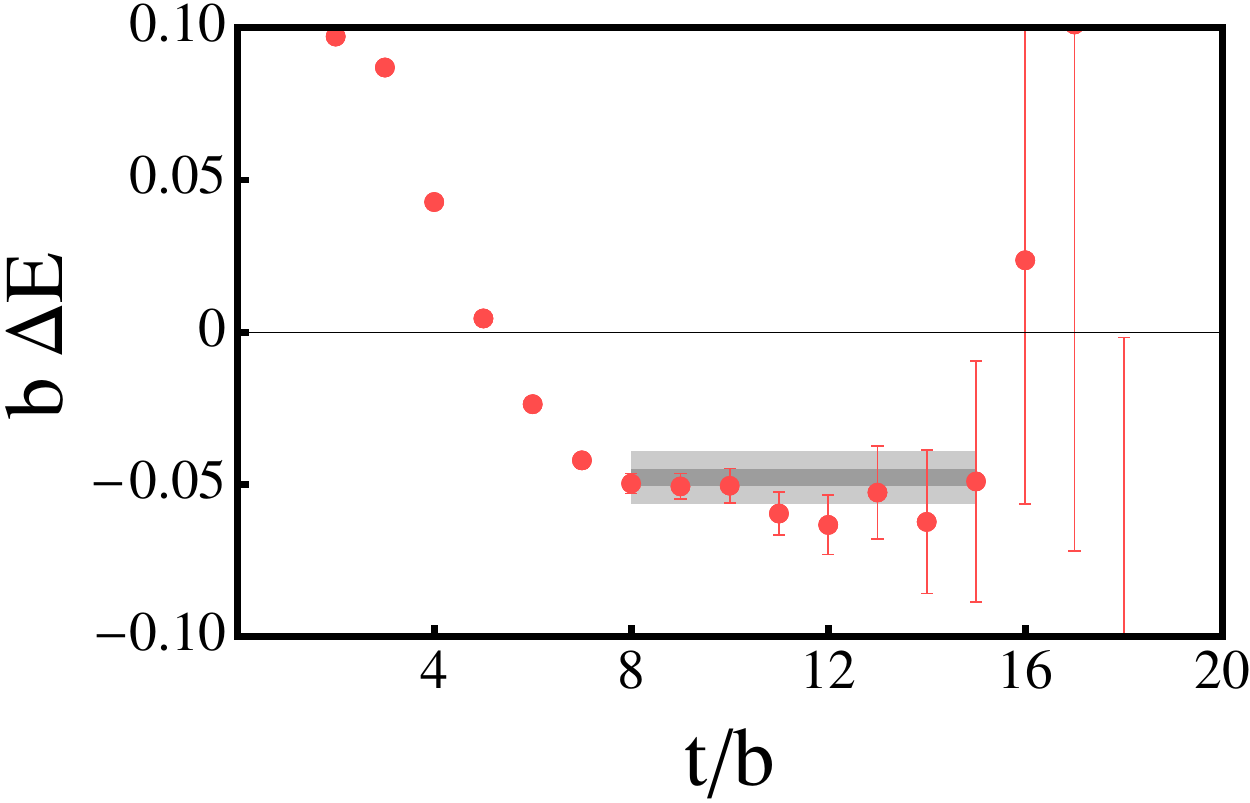}
  \caption{The EMPs associated with one 
    $J^\pi={1\over 2}^+$ \hetL\ (\htL\  and \nnL) correlation function
    computed with the \cfga\ (left), \cfgb\ (center) and \cfgc\
    (right) ensembles, with momentum $|{\bf P}|=0$.  
 \textcolor{\revcolor}{
The inner (darker) shaded region corresponds to the statistical uncertainty
of the extracted energy, while  the outer (lighter) shaded region
corresponds to the
statistical and fitting systematic uncertainties combined in quadrature.
}
}
  \label{fig:He3LEMPs}
\end{figure}

\noindent
The isotriplet of states\footnote{We refer to the $np\Lambda$ state
  with the $np$ coupled to $I=1$ as \htLtilde\ to differentiate it from
  the \htL\ state in which the $np$ couple to $I=0$.}, \hetL,
\htLtilde\ and \nnL, are degenerate in the absence of electromagnetism
and in the limit of exact isospin symmetry, and can have
$J^\pi={1\over 2}^+$ and $J^\pi={3\over 2}^+$.  The $J^\pi={1\over
  2}^+$ is expected to be the lowest-lying state, with a significant
component of the wavefunction having the two nucleons in the $\si$
channel coupled to the $\Lambda$.  The $J^\pi={3\over 2}^+$ state
cannot have such a $NN\Lambda$ configuration in its wavefunction by
the Pauli principle without placing the baryons in orbital
excitations but will have configurations of the form of two nucleons
in the $\siii$-$\diii$ channel coupled to a $\Sigma^+$. In the SU(3)
limit, this can be nearby in energy, but when SU(3) breaking is
included, the energy for the $J^\pi={3\over 2}^+$ will increase,
largely dictated by the $\Sigma$-$\Lambda$ mass splitting, and become
less  phenomenologically interesting.  Consequently, we will focus
first on the $J^\pi={1\over 2}^+$ channel.  The EMPs from one of the
eight correlation functions of these quantum numbers are shown in
fig.~\ref{fig:He3LEMPs}, from which the energies of the lowest-lying
states have been determined.
\begin{table}
\begin{center}
\begin{minipage}[!ht]{16.5 cm}
  \caption{
    The calculated binding energies in \hetL\ (\htL\ and \nnL).
    ``g.s.''  denotes the ground state.
The first uncertainty is statistical, the second is the fitting systematic and
the third is due to the lattice spacing.
}
\label{tab:He3Lenergies}
\end{minipage}
\setlength{\tabcolsep}{1em}
\begin{tabular}{c|ccc}
\hline
 \hetL        & \cfga & \cfgb & \cfgc \\
      \hline
      g.s.   (MeV) 
      & \textcolor{\revcolor}{77.6(3.6)(7.5)(0.8)} 
      &  \textcolor{\revcolor}{ 74.1(3.9)(7.3)(0.8) } 
      & \BhetLJhalf \\
\hline
\end{tabular}
%noalign{\smallskip\hrule}\cr}
\begin{minipage}[t]{16.5 cm}
\vskip 0.0cm
\noindent
\end{minipage}
\end{center}
\end{table}     
\begin{figure}[!ht]
  \centering
  \includegraphics[width=0.99\textwidth]{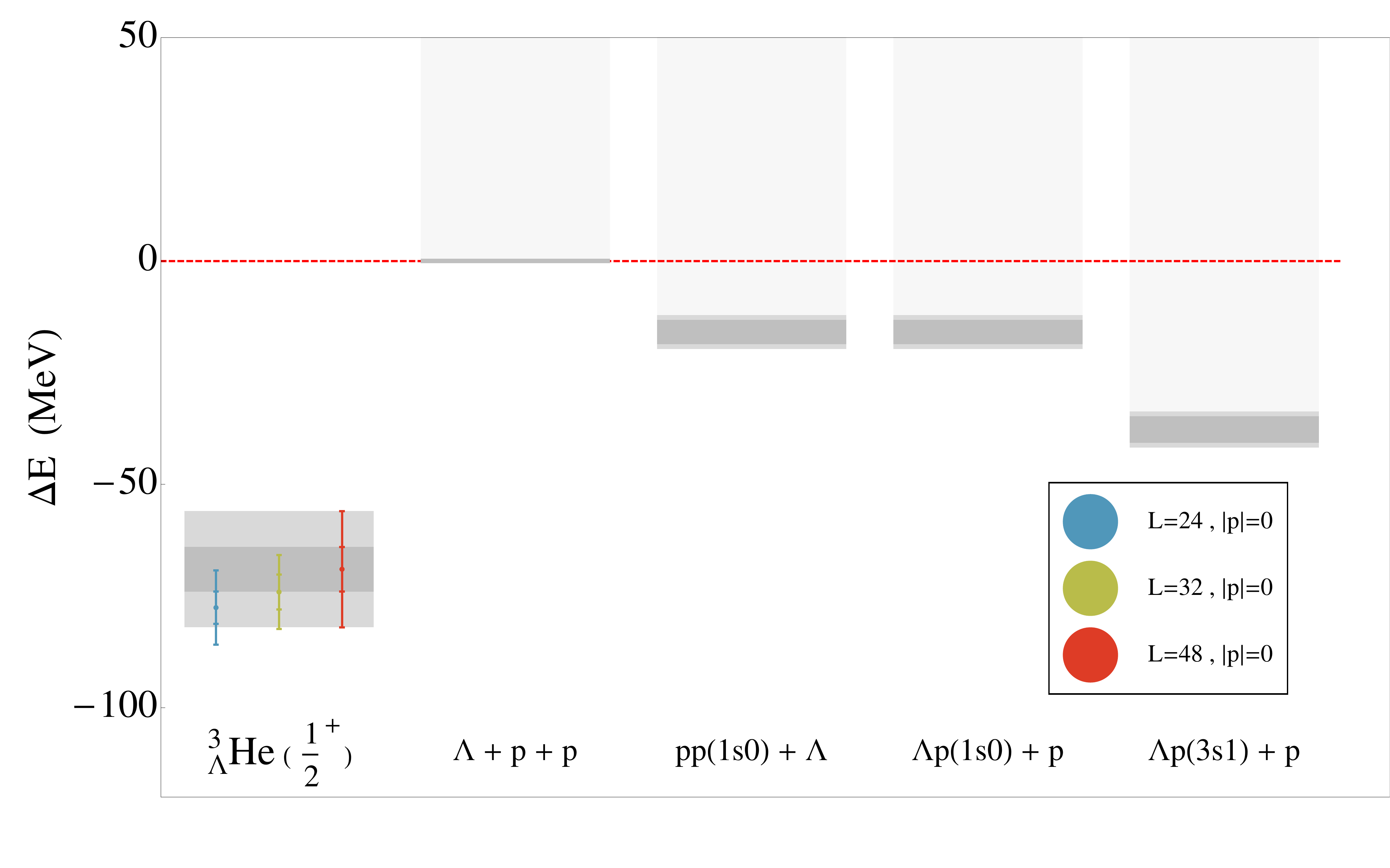}
  \caption{The bound-state energy levels in the $J^\pi={1\over 2}^+$
    \hetL\ (\htL\ and \nnL) sector.  The points and their associated
    uncertainties correspond to the energies of the states extracted
    from the correlation functions with the quantum numbers of the
    ground state of \hetL.  The locations of the energy-levels
    associated with non-interacting di-proton-$\Lambda$, $\Lambda$N-N
    and $\Lambda$-N-N continuum states, determined from the
   two-body binding energies given in
    Table~\protect\ref{tab:A2energies48}, are shown.
 }
  \label{fig:He3Llevels}
\end{figure}
The extracted spectrum of bound states is given in
Table~\ref{tab:He3Lenergies}, and shown in fig.~\ref{fig:He3Llevels}.
Taking the result obtained on the \cfgc\ ensemble as the estimate of
the infinite-volume binding energy, we find
\begin{eqnarray}
  B^{(\infty)}(_{\Lambda} ^3{\rm He}(1/2^+)) & = & \BhetLJhalfMeV
  \ \ \ .
  \label{eq:he3Lstates}
\end{eqnarray}
The ground state is significantly more deeply bound than any of the
continuum states, and we identify this as the ground state of the
\hetL\ nucleus (and hence also bound \htLtilde\ and \nnL\ due to isospin symmetry).
The correlation function from which this ground-state energy was extracted is
a superposition of 
$\overline{\bf 35}$ and ${\bf 27}$ SU(3) irreps.
Another element of the  ${\bf 27}$ irrep is in the $s=-3$ sector, with 
$I=1$, $J^\pi={1\over 2}^+$ and with the baryon structure of $N\Xi\Lambda$.
One of the  correlation functions associated with this state is pure  ${\bf 27}$, and the
energy of the lowest-lying state in this correlation function is 
found to be the same as that in the 
\hetL\ correlation function within the uncertainties of the calculations,
suggesting that the {\bf 27} state is 
lower in energy than
\textcolor{\revcolor}{
or nearly degenerate with
}
the $\overline{\bf 35}$.

Experimentally, there is no evidence for a bound \hetL\ nucleus as the
$\Lambda$-nucleon interactions are not sufficient to overcome the Coulomb
repulsion between the protons.  Further, the small binding of the
hypertriton, with a significant deuteron-$\Lambda$ component, 
strongly suggests that the corresponding $I=1$ state will be unbound, and it is
likely, but yet to be verified, that the \nnL\ electrically
neutral nucleus is also unbound.  However, our calculations 
provide compelling evidence for a bound state in this channel in the
limit of SU(3)-flavor symmetry, and we expect that the bound state
persists over a range of light-quark masses.

%%%%%%%%%%%%%%%%%%%%%%%%%%%%%%%%%%%%%%%%%%%%%%%%%%%%
\subsection{ $I=1$,  $J^\pi={3\over 2}^+$ : \hetS }
\label{sec:3HeSig}
\begin{figure}[!ht]
  \centering
  \includegraphics[width=0.32\textwidth]{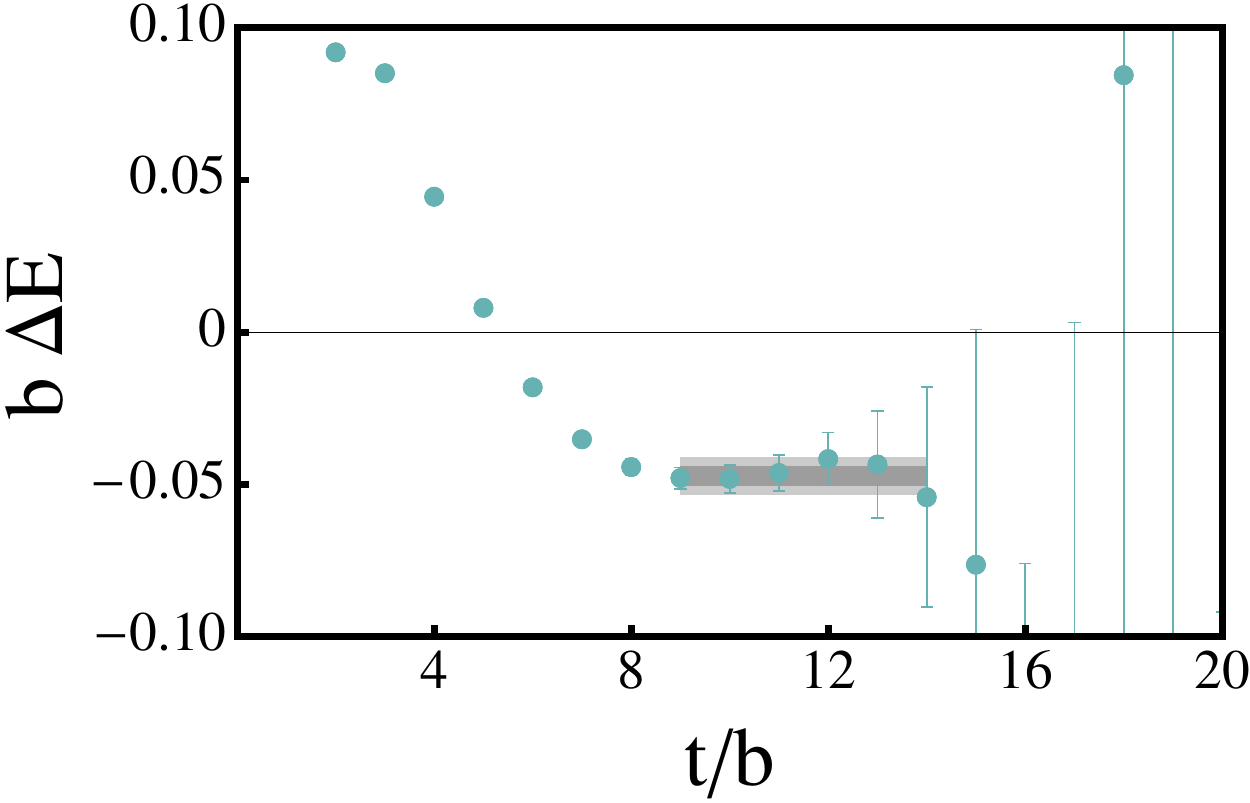}
  \includegraphics[width=0.32\textwidth]{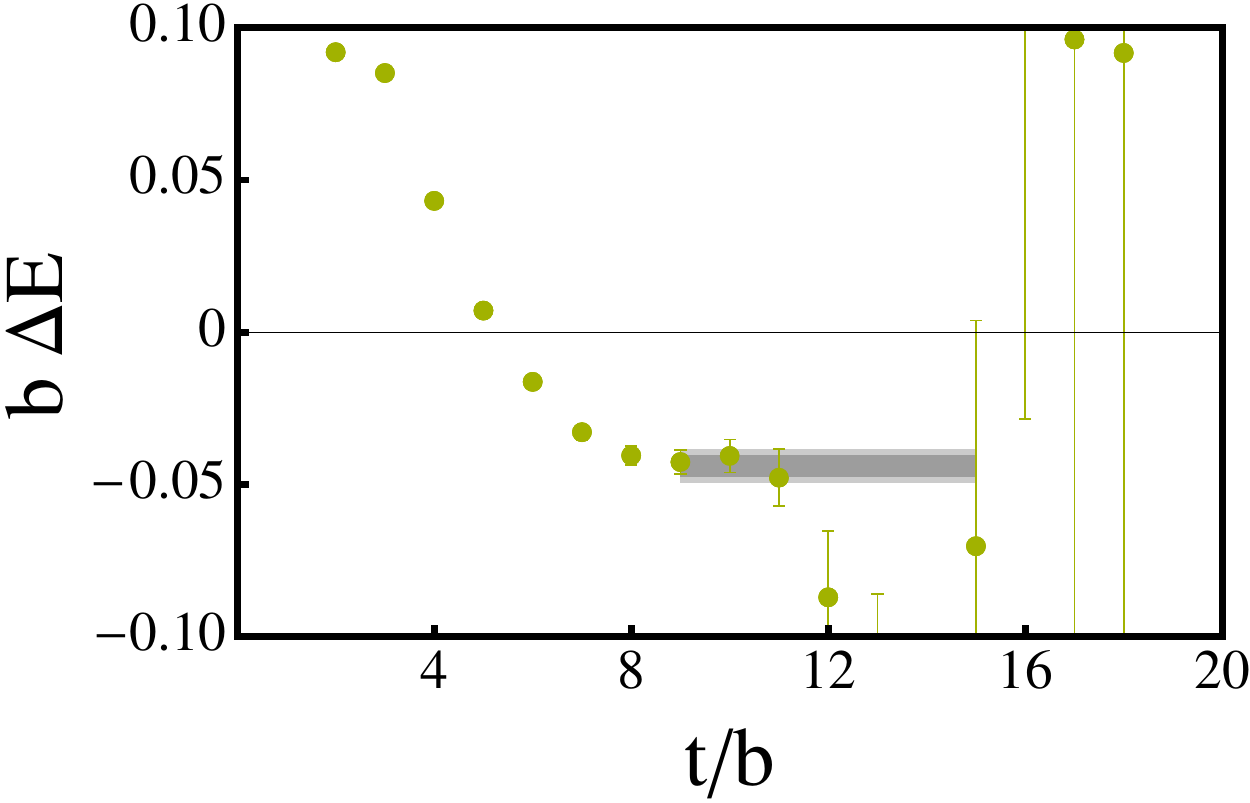}
  \includegraphics[width=0.32\textwidth]{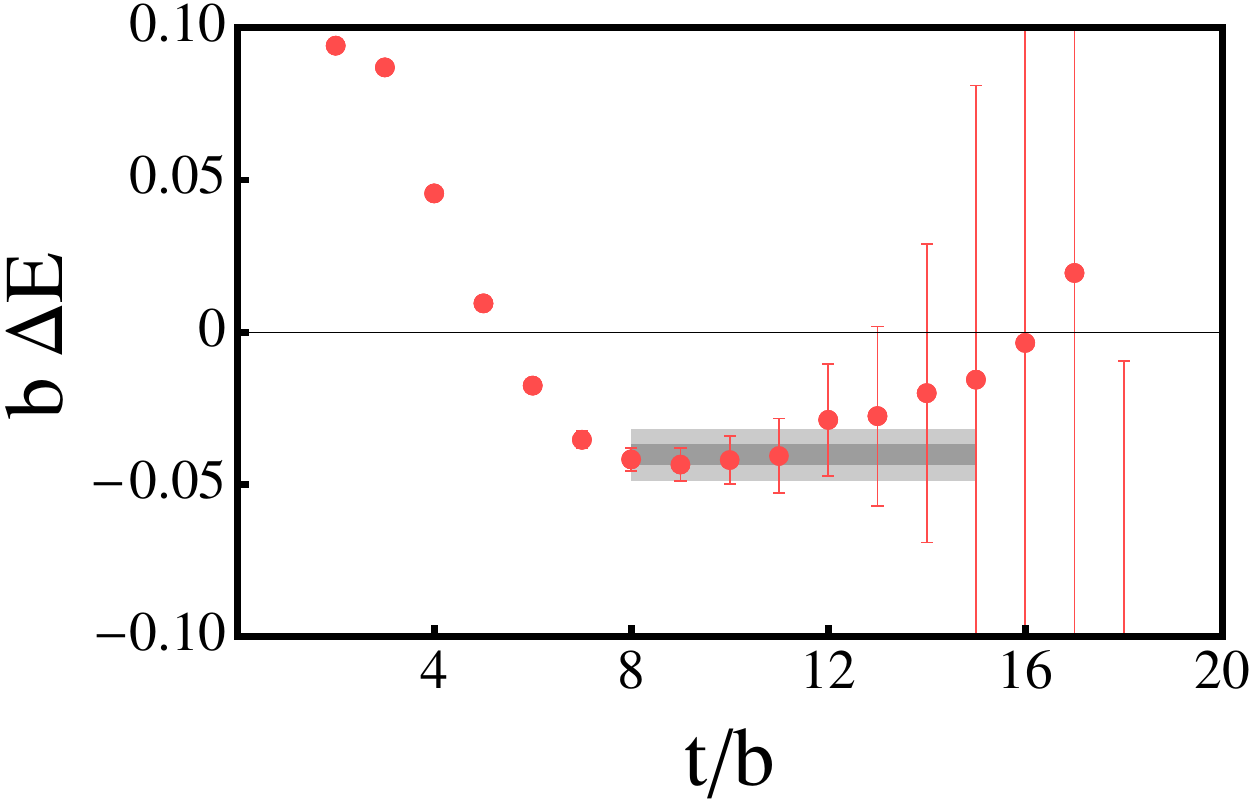}
  \caption{The EMPs associated with one 
    $J^\pi={3\over 2}^+$ \hetS\ correlation function
    computed with the \cfga\ (left), \cfgb\ (center) and \cfgc\
    (right) ensembles, with momentum $|{\bf P}|=0$.  
 \textcolor{\revcolor}{
The inner (darker) shaded region corresponds to the statistical uncertainty
of the extracted energy, while  the outer (lighter) shaded region
corresponds to the
statistical and fitting systematic uncertainties combined in quadrature.
}
}
  \label{fig:He3LJ1p5EMPs}
\end{figure}

\noindent
As discussed above, for the $I=1$, $s=-1$, $J^\pi={3\over 2}^+$, three body state,
an NN$\Lambda$ component is forbidden (for all baryons in a relative s-wave) 
and one important contribution to the ground-state wavefunction
is $pn\Sigma$, where the nucleons couple to
$I=0$, $J=1$,  as in \htL. As yet, the only observed $\Sigma$ hypernucleus is
$^4_\Sigma {\rm He}$ ($ppn\Sigma^0$)~\cite{Hayano:1988pn,he4Sig}, but at the SU(3)
point it is possible that this three-body system binds.
The sources used to generate this correlation function transform as
${\bf 27}$ under SU(3)\footnote{This {\bf 27} irrep is different from that
  in the $J^\pi={1\over 2}^+$ channel.  In principle the ground state of the
  system could reside in the {\bf 64} irrep, but this  is not accessible with our present
  operator structure.},
and result in EMPs that exhibit clear plateaus.
The ground-state energies extracted from the three ensembles are given in
Table~\ref{tab:He3LJ1p5energies},
and the associated EMPs are shown in fig.~\ref{fig:He3LJ1p5EMPs}.
\begin{table}
\begin{center}
\begin{minipage}[!ht]{16.5 cm}
  \caption{
    The calculated binding energies in $J^\pi={3\over 2}^+$ 
\hetS.
    ``g.s.''  denotes the ground state.
The first uncertainty is statistical, the second is the fitting systematic and
the third is due to the lattice spacing.
  }
\label{tab:He3LJ1p5energies}
\end{minipage}
\setlength{\tabcolsep}{1em}
\begin{tabular}{c|ccc}
\hline
 \hetS        & \cfga & \cfgb & \cfgc \\
      \hline
      g.s.   (MeV) 
      & \textcolor{\revcolor}{64.3(4.5)(7.9)(0.7) }  
      &  \textcolor{\revcolor}{ 58.2(5.2)(7.7)(0.6)}
      & \BhetS \\
\hline
\end{tabular}
%noalign{\smallskip\hrule}\cr}
\begin{minipage}[t]{16.5 cm}
\vskip 0.0cm
\noindent
\end{minipage}
\end{center}
\end{table}     
\begin{figure}[!ht]
  \centering
  \includegraphics[width=0.99\textwidth]{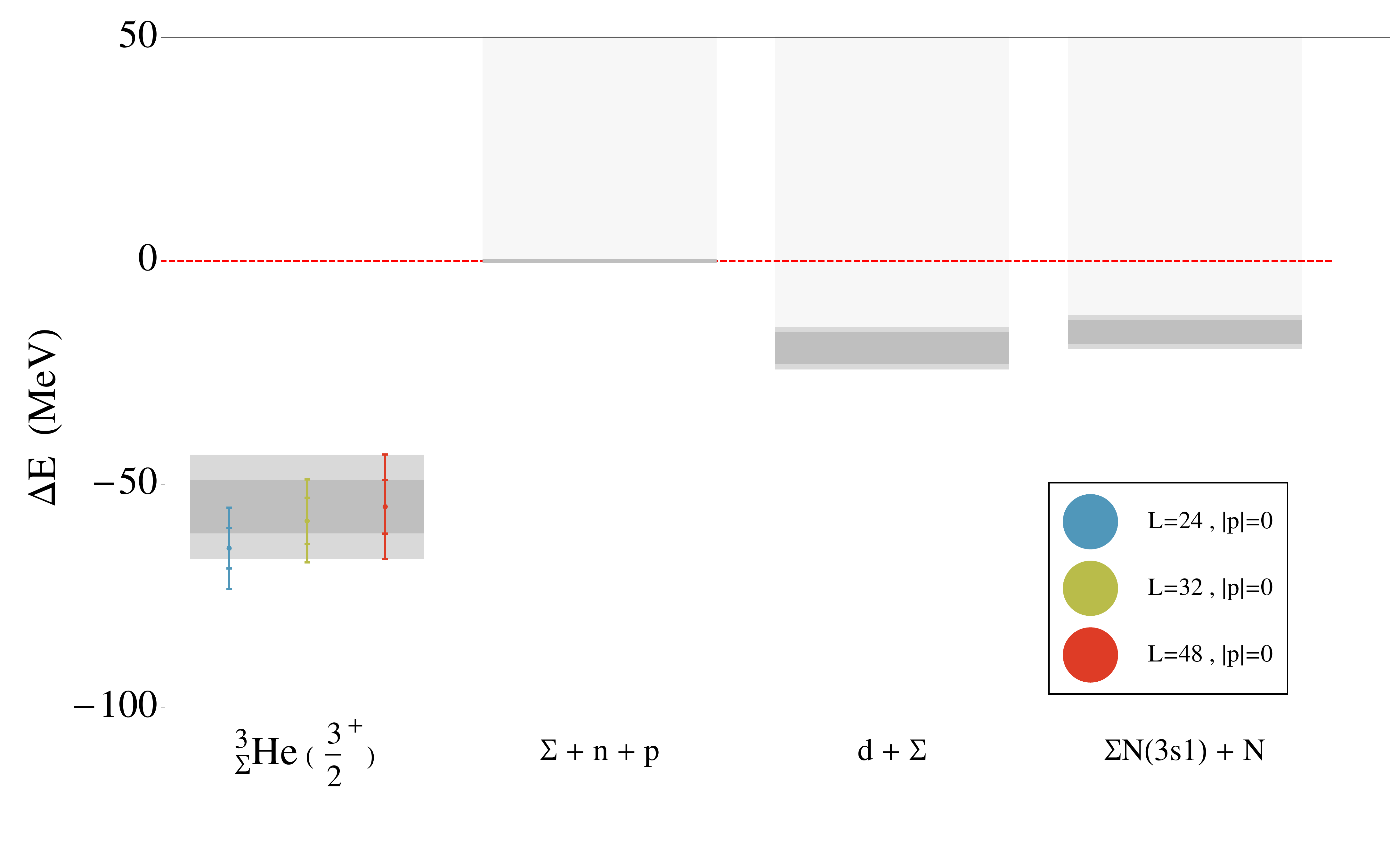}
  \caption{The bound-state energy levels in the $J^\pi={3\over 2}^+$
    \hetS\  sector.  The points and their associated
    uncertainties correspond to the energies of the states extracted
    from the correlation functions with the quantum numbers of the
    ground state of \hetS.  The locations of the energy-levels
    associated with non-interacting  continuum states, determined from the
   two-body binding energies given in
    Table~\protect\ref{tab:A2energies48}, are shown.
 }
  \label{fig:He3LJ1p5levels}
\end{figure}
The ground-state energy and the anticipated continuum thresholds based upon the
non-interacting two-body energies
are shown in fig.~\ref{fig:He3LJ1p5levels}.

%%%%%%%%%%%%%%%%%%%%%%%%%%%%%%%%%%%%%%%%%%%%%%%%%%%%
\section{Four-Body Systems}
\label{sec:BBBB}
\noindent
There are a large number of four-body systems and states that could be
explored theoretically with LQCD at the SU(3) symmetric point,
dictated by the product of four {\bf 8}'s,
\begin{eqnarray} {\bf 8} \otimes {\bf 8}\otimes {\bf 8}\otimes {\bf 8}
  & = & 8\ {\bf 1} \oplus 32\ {\bf 8} \oplus 20\ {\bf 10} \oplus 20\
  {\bf \overline{10}} \oplus 33\ {\bf 27} \oplus 2\ {\bf 28} \oplus 2\
  {\bf \overline{28}} \oplus 15\ {\bf 35} \oplus 15\ {\bf
    \overline{35}}
  \nonumber\\
  && \oplus\: 12\ {\bf 64} \oplus 3\ {\bf 81} \oplus 3\ {\bf
    \overline{81} } \oplus {\bf 125} \ \ \ ,
  \label{eq:fouroctets}
\end{eqnarray}
giving a total of 166 lowest-lying states (one per distinct irrep)
with distinguishable quantum numbers.
The local sources that have been used in this work to generate correlation
functions project onto a subset of
the irreps,
\begin{eqnarray} 
\left( \ {\bf 8} \otimes {\bf 8}\otimes {\bf 8}\otimes {\bf 8}\ \right)_{J^\pi=0^+}
  & \rightarrow & 
{\bf 1}\oplus {\bf 27}\oplus\overline{\bf 28}
\nonumber\\
\left( \ {\bf 8} \otimes {\bf 8}\otimes {\bf 8}\otimes {\bf 8}\ \right)_{J^\pi=1^+}
  & \rightarrow & 
{\bf 8}\oplus {\bf 10}\oplus\overline{\bf 10}\oplus\overline{\bf 35}
\nonumber\\
\left( \ {\bf 8} \otimes {\bf 8}\otimes {\bf 8}\otimes {\bf 8}\ \right)_{J^\pi=2^+}
  & \rightarrow & 
{\bf 8}\oplus {\bf 27}
\ \ \ ,
\label{eq:fouroctetssubset}
\end{eqnarray}
which greatly reduces the complexity of individual correlation functions.
In order to restrict ourselves to systems that are currently of
phenomenological importance, we explore systems containing up to two
strange quarks only, the isosinglet \hef, the isodoublet \hfL\ and
\hefL, the isosinglet \hfLL\ and the isotriplet \hefLL, \hfLL, and
\nnLL.

%%%%%%%%%%%%%%%%%%%%%%%%%%%%%%%%%%%%%%%%%%%%%%%%%%%%
\subsection{$I=0$,  $J^\pi=0^+$ : \hef}
\label{sec:4He}
\begin{figure}[!ht]
  \centering
  \includegraphics[width=0.32\textwidth]{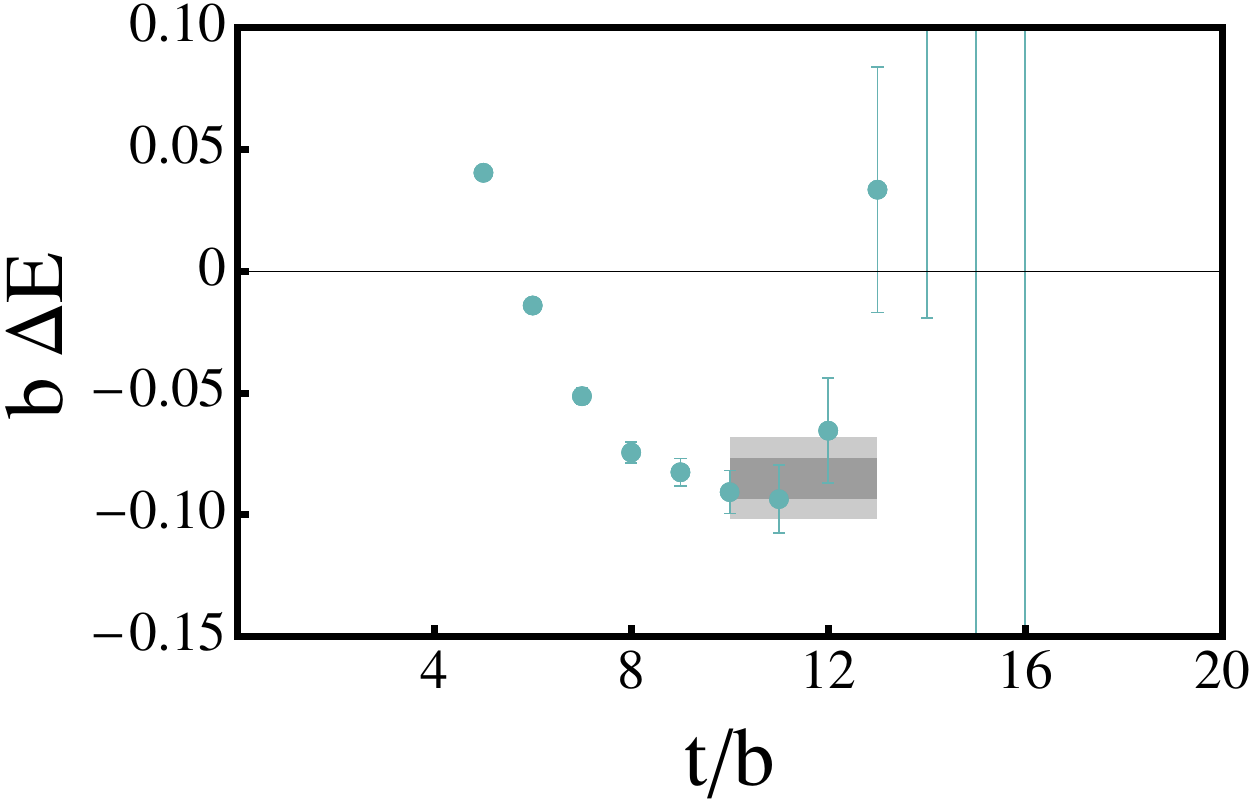}
  \includegraphics[width=0.32\textwidth]{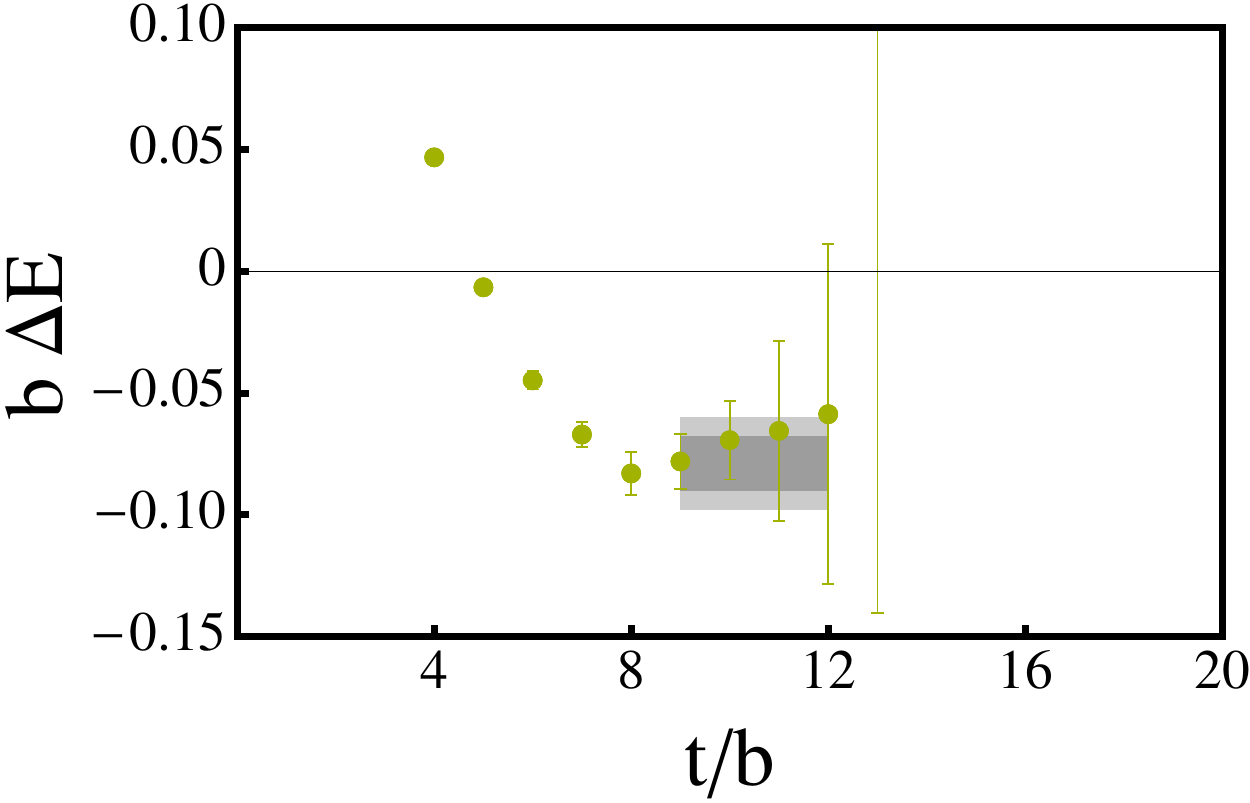}
  \includegraphics[width=0.32\textwidth]{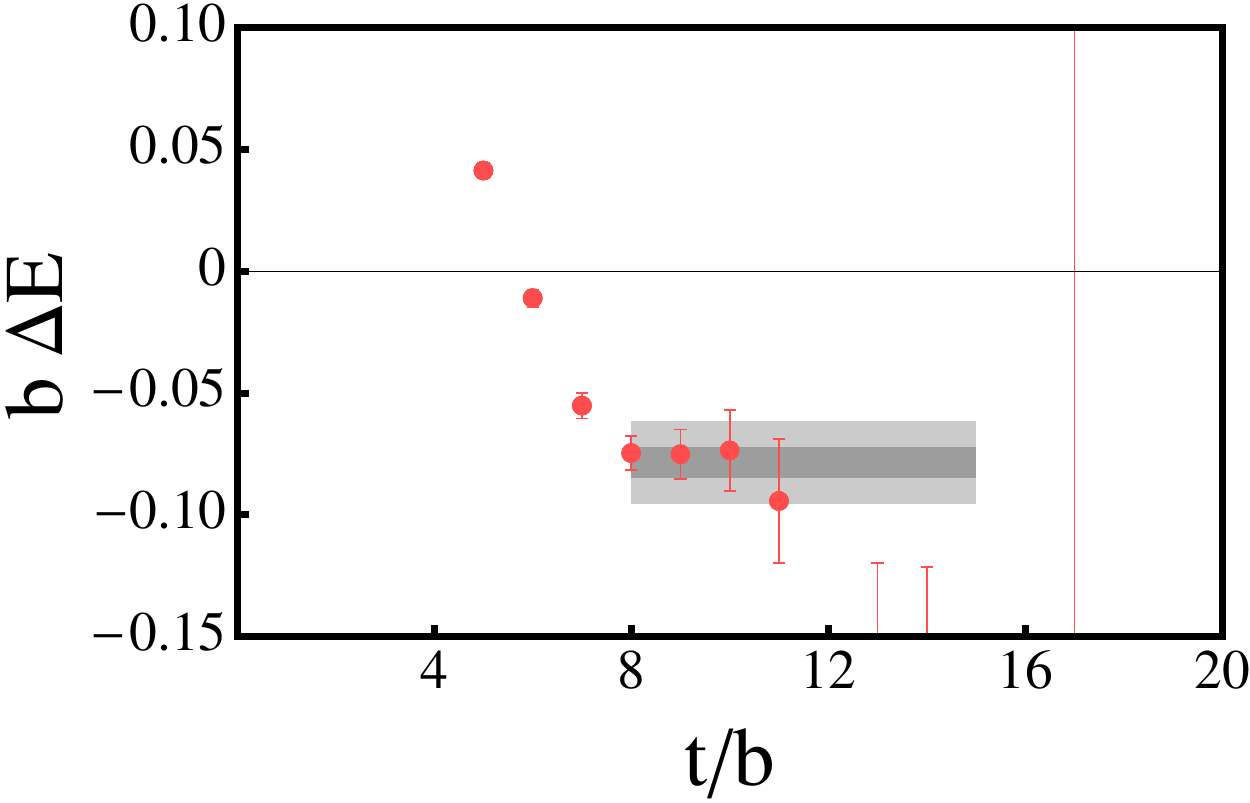}
  \caption{EMPs associated with a $|{\bf P}|=0$
    $J^\pi=0^+$ \hef\   correlation function computed with the \cfga\
    (left), \cfgb\ (center) and \cfgc\ (right) ensembles.  
 \textcolor{\revcolor}{
The inner (darker) shaded region corresponds to the statistical uncertainty
of the extracted energy, while  the outer (lighter) shaded region
corresponds to the
statistical and fitting systematic uncertainties combined in quadrature.
}
  }
  \label{fig:He4EMPs}
\end{figure}

\begin{figure}[!ht]
  \centering
  \includegraphics[width=0.99\textwidth]{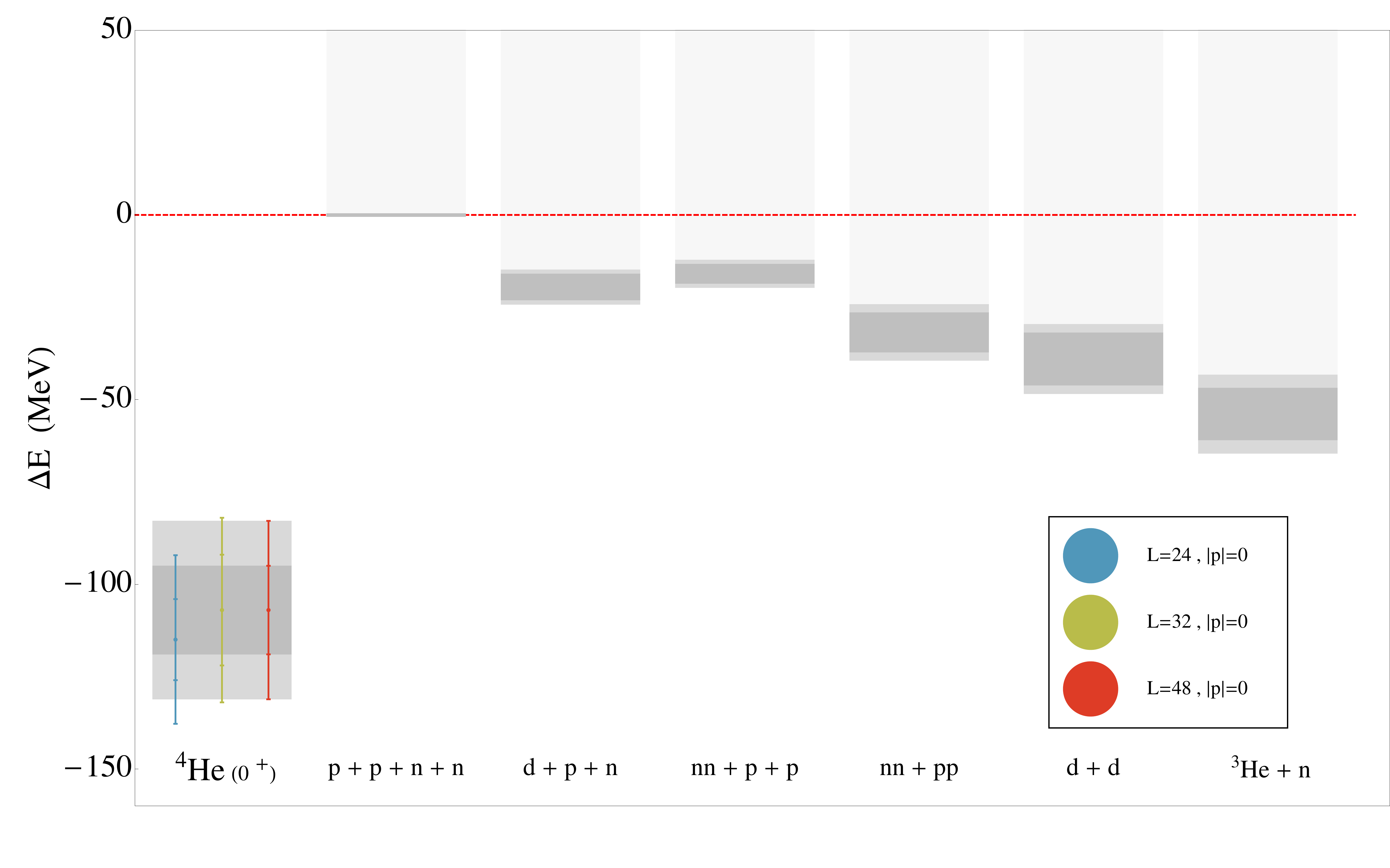}
  \caption{The bound-state energy levels in the $J^\pi=0^+$ \hef\
    sector.  The points and their associated uncertainties correspond
    to the energies of the states extracted from the correlation
    functions with the quantum numbers of the ground state of \hef.
    The locations of the energy-levels associated with non-interacting
    N-\hetppn, d-d, di-nucleon-di-nucleon, di-nucleon-N-N, d-N-N and
    N-N-N-N continuum states, determined from the two-body binding energies given in
    Table~\protect\ref{tab:A2energies48} and the three-body
    energies given in eq.~(\ref{eq:he3states}), are shown.
    }
  \label{fig:He4levels}
\end{figure}

\noindent
In nature, the \hef\ nucleus is anomalously deeply bound when compared
to nuclei nearby in the periodic table, due to its closed shell structure, 
with a total binding energy of
$B\sim 28~{\rm MeV}$, or a binding energy per nucleon of
$B/A\sim 7~{\rm MeV}$.  We anticipate that at the SU(3) symmetric
point, the binding energy of \hef\ will be even deeper given the
bindings of the deuteron and di-neutron found in the two-body sector.
Two of the \hef\ correlation functions, resulting from different source
structures defined by $s=0$, $I=0$ and $J^\pi=0^+$ quantum numbers,
transform as an element of the $\overline{\bf 28}$ irrep of SU(3)\footnote{The $\overline{\bf 28}$ is the only
  allowed $I=0$, $s=0$, A=4 irrep.}.
EMPs of one of these correlation functions are shown in
fig.~\ref{fig:He4EMPs}, from which the
energies of the lowest lying states have been determined.
The extracted spectrum of bound states, only calculated for the system
at rest in the lattice volume, is given in
Table~\ref{tab:He4energies} and shown in fig.~\ref{fig:He4levels}.
\begin{table}
\begin{center}
\begin{minipage}[!ht]{16.5 cm}
  \caption{
    The calculated binding energies in \hef.
    ``g.s.''  denotes the ground state.
The first uncertainty is statistical, the second is the fitting systematic and
the third is due to the lattice spacing.
  }
\label{tab:He4energies}
\end{minipage}
\setlength{\tabcolsep}{1em}
\begin{tabular}{c|ccc}
\hline
 \hef        & \cfga & \cfgb & \cfgc \\
      \hline
      g.s.   (MeV) 
      &  \textcolor{\revcolor}{115(11)(20)(1)}
      &  \textcolor{\revcolor}{ 107(15)(20)(1) } 
      & \Bhef \\
\hline
\end{tabular}
%noalign{\smallskip\hrule}\cr}
\begin{minipage}[t]{16.5 cm}
\vskip 0.0cm
\noindent
\end{minipage}
\end{center}
\end{table}     
Also shown in fig.~\ref{fig:He4levels} are the thresholds of non-interacting
continuum states, based upon the two-body and three-body bound-state
spectra.  Using the result obtained on the \cfgc\ ensemble as an
estimate of the binding energy in infinite volume, we find
\begin{eqnarray}
  B^{(\infty)}(^4{\rm He}) & = & \BhefMeV
  \ \ \ .
  \label{eq:he4states}
\end{eqnarray}
While this state is somewhat more deeply bound than any continuum
state, the precision of the calculation is not sufficient to
unambiguously  distinguish the state from the n +\hetppn\ continuum.  
In order to eliminate this ambiguity in state
identification, further calculations are required, and additional
source structure should be used to increase the size of the basis of
correlation functions.

The \hef\  ground-state energy that we have calculated in this $n_f=3$
calculation is substantially different from that obtained with
quenched calculations at a comparable pion
mass~\cite{Yamazaki:2009ua}, which find an infinite-volume
extrapolated value of $B^{(\infty)}_{n_f=0}(^4{\rm
  He})=27.7(7.8)(5.5)~{\rm MeV}$, close to the experimental value.

%%%%%%%%%%%%%%%%%%%%%%%%%%%%%%%%%%%%%%%%%%%%%%%%%%%%
\subsection{$I={1\over 2}$,  $J^\pi=0^+$  : \hefL\  and \hfL}
\label{sec:4HLamAND4HeLam}
\noindent
In nature, the \hefL\ hypernucleus has been well studied
experimentally and theoretically.  The $\Lambda$-separation energy of
the \hefL\ $J^\pi=0^+$ ground state is measured to be
$S_\Lambda=2.39(0.03)~{\rm MeV}$, and for the $J^\pi=1^+$ first
excited state is $S_\Lambda=1.24(0.05)~{\rm MeV}$.  These two
lowest-lying states are consistent with the $\Lambda$ coupled to a
\hetppn\ $J^\pi={1\over 2}^+$ core. 
A recent review of this system can be found in Ref.~\cite{Gal:2010xn}.

We have calculated correlation functions in
the $J^\pi=0^+$ channel, which should provide the ground state, but not the
nearby $J^\pi=1^+$ first excited state.  
The sources employed to produce the correlation functions are elements of 
the same
$\overline{\bf  28}$ irrep of SU(3) as those of \hef,
and hence the extracted  states have the same energy\footnote{
The $s=-1$, $I={1\over 2}$ systems 
of various spin configurations
have components transforming in the 
$\overline{\bf 81}$ and {\bf 125} irreps that are inaccessible to our operator
construction, but that may in principle contain the ground state of this system.
}.
The EMPs from these  correlation functions are the same as those shown in
fig.~\ref{fig:He4EMPs}, from which the energies of the lowest lying
states have been determined, and are the same as those 
in Table~\ref{tab:He4energies}.  
The spectrum 
in this channel, and a subset of associated continuum states, are the same as those in
fig.~\ref{fig:He4levels}.  There are no continuum states from other SU(3)
irreps lying lower than those
associated with the \hef\ spectrum (assuming that we have correctly identified
the ground states in the three-body sector).  
However, due to the presence of
different SU(3) irreps in this channel, the spectrum of excited states of 
the nucleus, and the 
continuum states,
is expected to be different from that in the \hef\ channel.

As is the case for \hef, while the lowest-lying state extracted from
the correlation functions has a central value that is lower than any
of the non-interacting continuum states, the precision of the
calculation is not sufficient to completely exclude the possibility that it is 
a continuum state, e.g. \hetppn+$\Lambda$, or \hetL+N.  The
extrapolated binding energy is given in eq.~(\ref{eq:he4states}).

%%%%%%%%%%%%%%%%%%%%%%%%%%%%%%%%%%%%%%%%%%%%%%%%%%%%
\subsection{$I=1$,  $J^\pi=0^+$  : \hefLL, \hfLL, and \nnLL}
\label{sec:4HeLamLam}
\begin{figure}[!ht]
  \centering
  \includegraphics[width=0.32\textwidth]{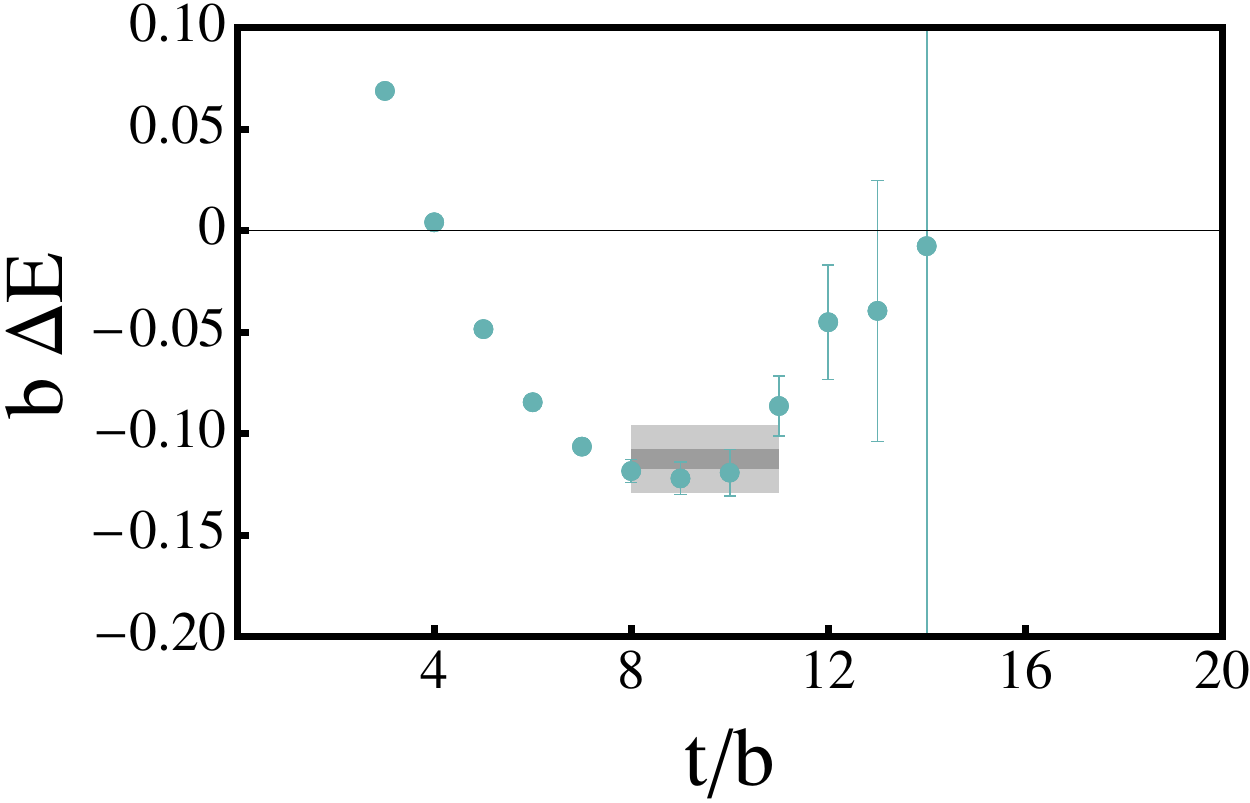}
  \includegraphics[width=0.32\textwidth]{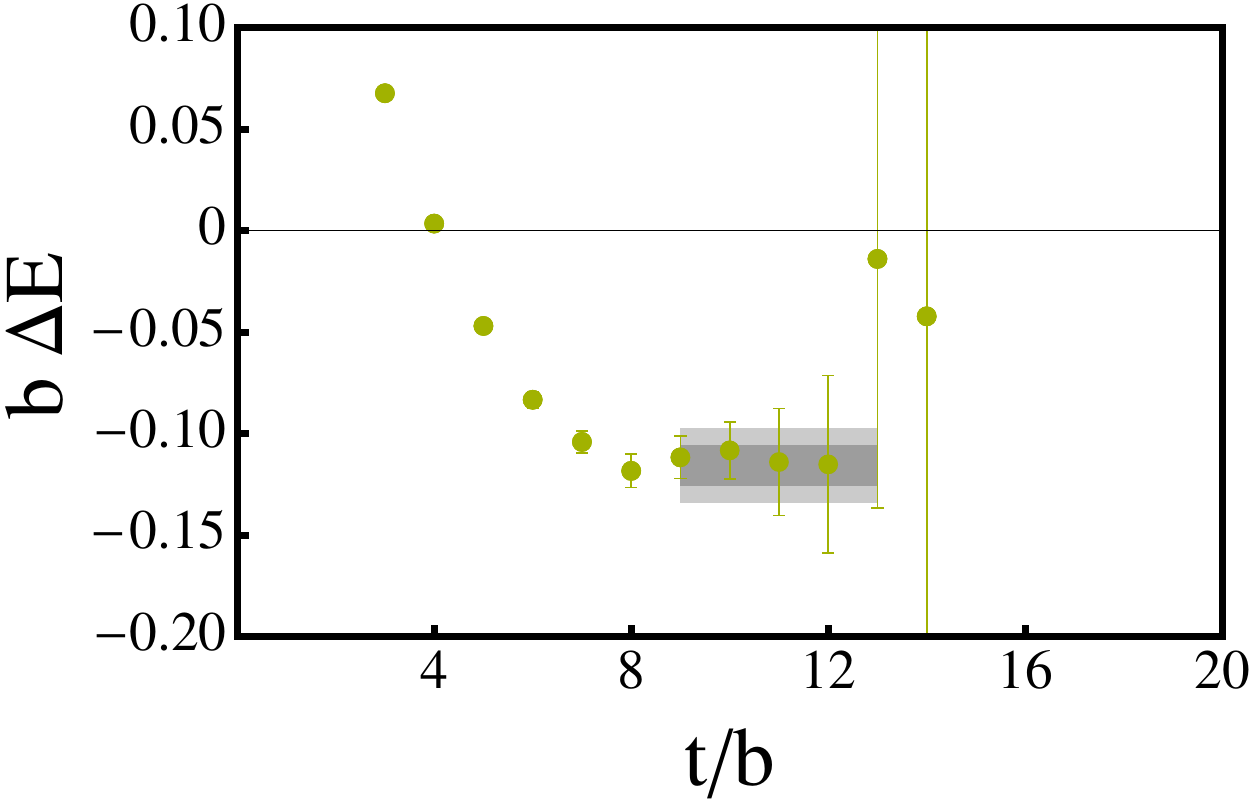}
  \includegraphics[width=0.32\textwidth]{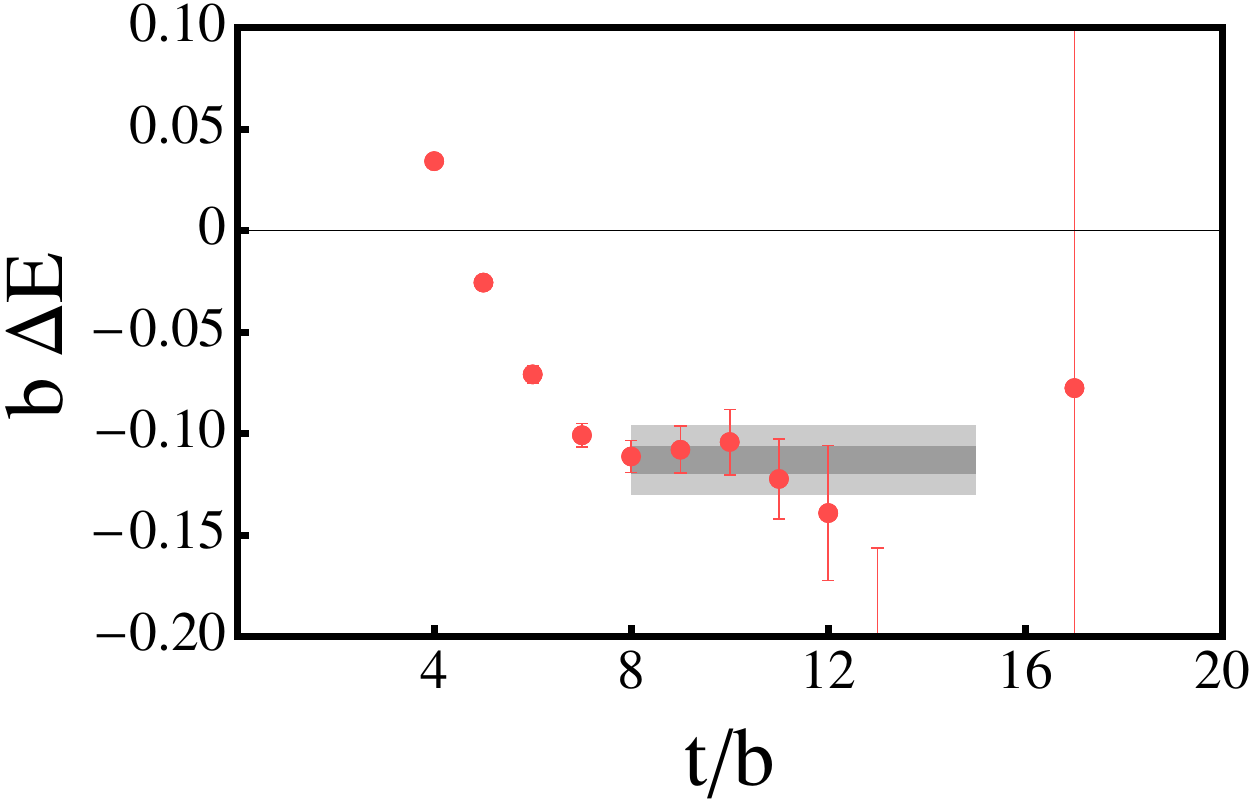}
  \caption{The EMPs associated with one of the eight
    $J^\pi=0^+$ \hefLL\  correlation functions computed with the \cfga\
    (left), \cfgb\ (center) and \cfgc\ (right) ensembles, with
    momentum $|{\bf P}|=0$.  
 \textcolor{\revcolor}{
The inner (darker) shaded region corresponds to the statistical uncertainty
of the extracted energy, while  the outer (lighter) shaded region
corresponds to the
statistical and fitting systematic uncertainties combined in quadrature.
}
  }
  \label{fig:He4LLEMPs}
\end{figure}

\begin{figure}[!ht]
  \centering
  \includegraphics[width=0.99\textwidth]{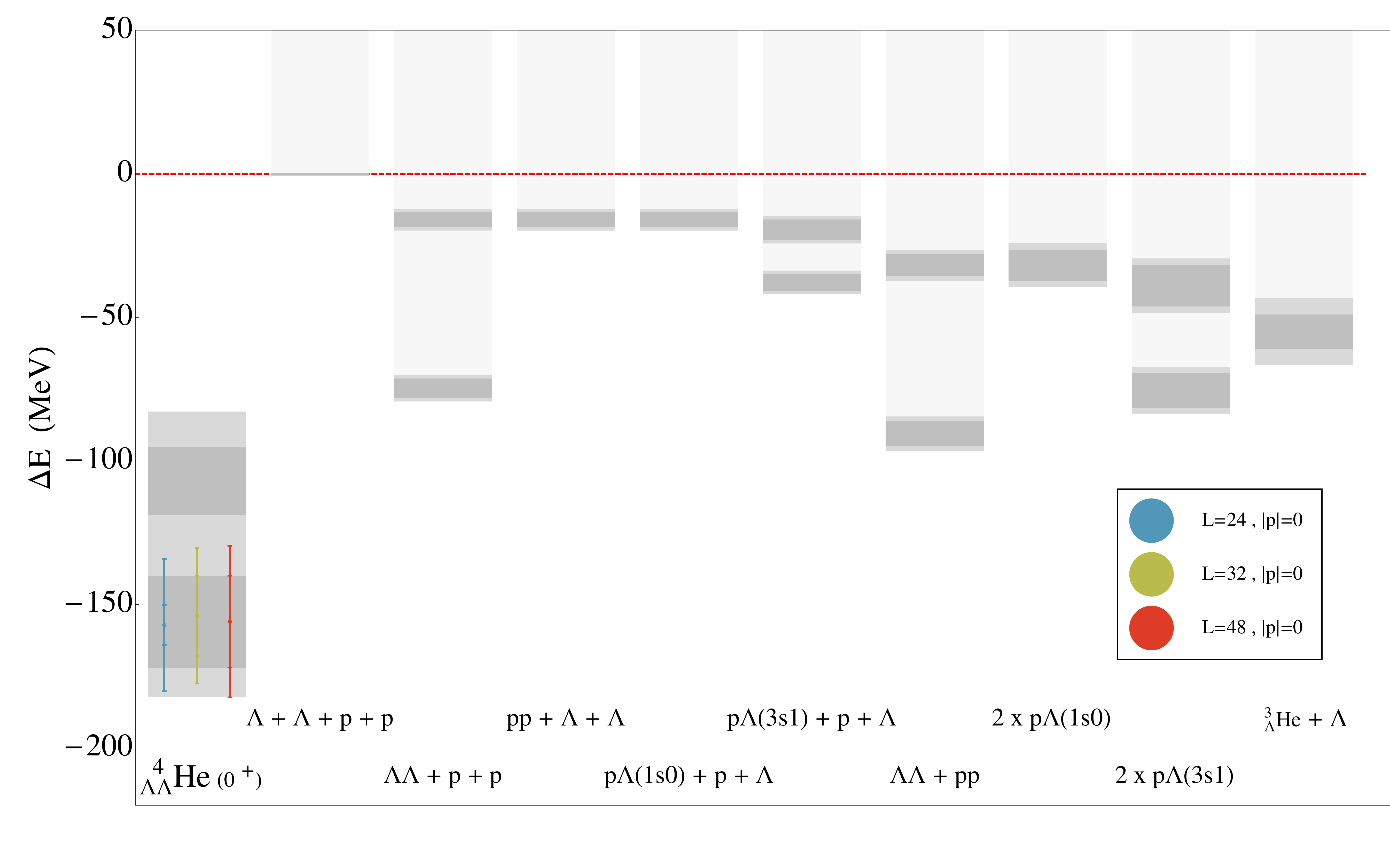}
  \caption{The bound-state energy levels in the $J^\pi=0^+$ \hefLL\
    (\hfLL\ and \nnLL) sector.  The points and their associated
    uncertainties correspond to the energies of the states extracted
    from the correlation functions with the quantum numbers of the
    ground state of \hefLL.  
The excited state of the \hefLL\ , in the $\overline{\bf 28}$, has the same
energy as the ground state of \hef.
The locations of the energy-levels
    associated with non-interacting $\Lambda$-\hetL,
    N$\Lambda$-N$\Lambda$, H-dibaryon-di-nucleon,
    N$\Lambda$-N-$\Lambda$, di-nucleon-$\Lambda$-$\Lambda$,
    H-dibaryon-N-N, and $\Lambda$-$\Lambda$-N-N continuum states,
    determined from the two-body binding
    energies given in Table~\protect\ref{tab:A2energies48} and
    the three-body energies given in eq.~(\ref{eq:he3states}) and 
    eq.~(\ref{eq:he3Lstates}), are
    shown. 
}
  \label{fig:He4LLlevels}
\end{figure}

\noindent
At the SU(3) symmetric point, with a deeply
bound H-dibaryon, bound di-neutron and attractive $\Lambda n$
interaction, we naively expect to find 
that \hefLL\ and its isospin partners are bound.
This is in contrast to the situation at the physical point, where a doubly
strange hypernucleus that is stable against strong decay has not been
conclusively observed
(for recent reviews of the status of experimental investigations into
doubly strange hypernuclei see, 
for example, Ref.~\cite{Nagae:2010zz,Pochodzalla:2010zz,Gal:2010xn}).
The states in \hefLL\ (with  $s=-2$ and $I=1$)
and its isospin partners can reside in the 
{\bf 27},  $\overline{\bf 28}$, $\overline{\bf 35}$, $\overline{\bf 81}$, 
 {\bf 64} and {\bf 125} irreps of SU(3).
However, the sources employed in this work produce correlation functions in
the $\overline{\bf 28}$ and {\bf 27} irreps only, and
therefore the complete spectrum cannot be definitively determined.
EMPs from one of  the correlation functions are shown in fig.~\ref{fig:He4LLEMPs},
from which the energies of the lowest-lying states have been
determined.
The extracted ground-state energies, only calculated for the system
at rest in the lattice volume, are given in
Table~\ref{tab:He4LLenergies}, and shown in
fig.~\ref{fig:He4LLlevels}.
The energy of the lowest state in the  correlation function 
with contributions from the $\overline{\bf 28}$ and {\bf 27}
is
found to be the same within uncertainties with that from a pure {\bf 27}
correlation function.  
The energy of the lowest state in the $\overline{\bf 28}$ is that of the ground
state of \hef\ by SU(3) symmetry, and is significantly larger than that of the 
{\bf 27}, and clearly the {\bf 27} is dominating the large-time behavior of the
mixed correlation function.
\begin{table}
\begin{center}
\begin{minipage}[!ht]{16.5 cm}
  \caption{
    The calculated binding energies in \hefLL.
    ``g.s.''  denotes the ground state.
The first uncertainty is statistical, the second is the fitting systematic and
the third is due to the lattice spacing.
  }
\label{tab:He4LLenergies}
\end{minipage}
\setlength{\tabcolsep}{1em}
\begin{tabular}{c|ccc}
\hline
 \hefLL        & \cfga & \cfgb & \cfgc \\
      \hline
      g.s.   (MeV) 
      &  \textcolor{\revcolor}{ 157(7)(22)(2) } 
      &  \textcolor{\revcolor}{ 154(14)(19)(2) } 
      &  \BhefLL \\
\hline
\end{tabular}
%noalign{\smallskip\hrule}\cr}
\begin{minipage}[t]{16.5 cm}
\vskip 0.0cm
\noindent
\end{minipage}
\end{center}
\end{table}     
Using the result obtained on the \cfgc\ ensemble as an estimate of the
binding energy in infinite volume, we find that
\begin{eqnarray}
  B^{(\infty)}(^{~\, 4}_{\Lambda\Lambda}{\rm He}) & = & \BhefLL ~{\rm MeV}
  \ \ \ .
  \label{eq:he4LLstates}
\end{eqnarray}
The ground state is more bound than any continuum state 
(although we have been
unable to cleanly isolate the ground state of the doubly strange three-body hypernuclei)
and we identify this as the ground state of the \hefLL, \hfLL, \nnLL\ 
isotriplet.   However, it is possible that this is an excited state of the
nucleus, with irreps other than the $\overline{\bf 28}$ and {\bf 27} containing a
lower-energy state.  Further, it is also possible that this state is a
continuum scattering state associated with N+\htLL.
Clearly, further calculations are also required to unambiguously distinguish the
energy of the {\bf 27} ground state from that of the $\overline{\bf 28}$ excited state.

%%%%%%%%%%%%%%%%%%%%%%%%%%%%%%%%%%%%%%%%%%%%%%%%%%%%
\section{Five-Body Systems}
\label{sec:BBBBB}
\noindent
There are a plethora of five-body systems that
can be explored theoretically at the SU(3) symmetric
point, dictated, in part,  by the product of five {\bf 8}'s, 
\begin{eqnarray} {\bf 8} \otimes {\bf 8}\otimes {\bf 8}\otimes {\bf 8}
  \otimes {\bf 8} & = & 32\ {\bf 1} \oplus 145\ {\bf 8} \oplus 100\
  {\bf 10} \oplus 100\ {\bf \overline{10}} \oplus 180\ {\bf 27} \oplus
  20\ {\bf 28} \oplus 20\ {\bf \overline{28}}
  \nonumber\\
  && \oplus \: 100\ {\bf 35} \oplus 100\ {\bf \overline{35}} \oplus
  94\ {\bf 64} \oplus 5\ {\bf 80} \oplus 5\ {\bf \overline{80} }
  \oplus 36\ {\bf 81} \oplus 36\ {\bf \overline{81} }
  \nonumber\\
  && \oplus \: 20\ {\bf 125} \oplus 4\ {\bf 154} \oplus 4\ {\bf
    \overline{154} } \oplus {\bf 216} \ \ \ .
  \label{eq:fiveoctets}
\end{eqnarray}
In this work, we explore one five-body state that can be produced by local
quark-level operators, involving only their upper components, with all
five baryons in a relative s-wave.
Unfortunately, this system, with $s=-3$, has not been experimentally observed.

%%%%%%%%%%%%%%%%%%%%%%%%%
\subsection{$I=0$,  $J^\pi={3\over 2}^+$ : \hymag}
\begin{figure}[htbp]
  \begin{center}
    \includegraphics[width=0.32\textwidth]{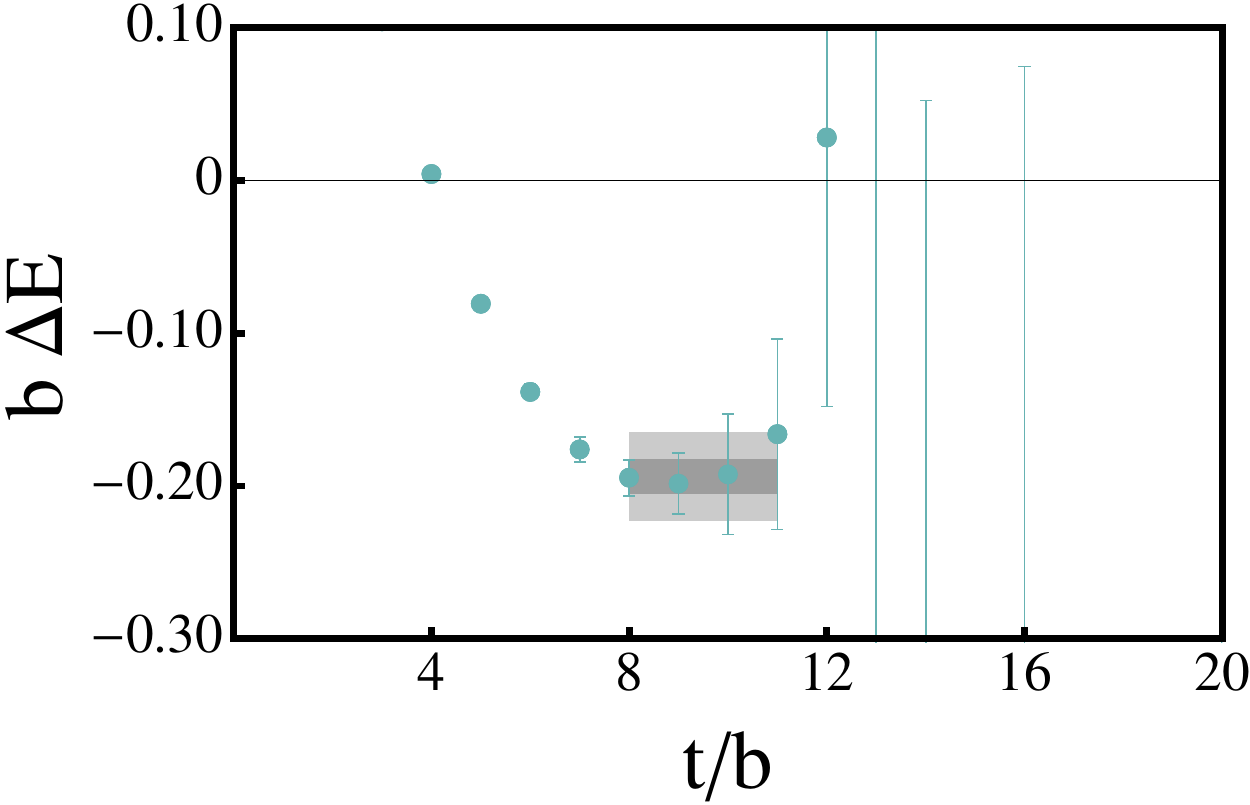}
    \includegraphics[width=0.32\textwidth]{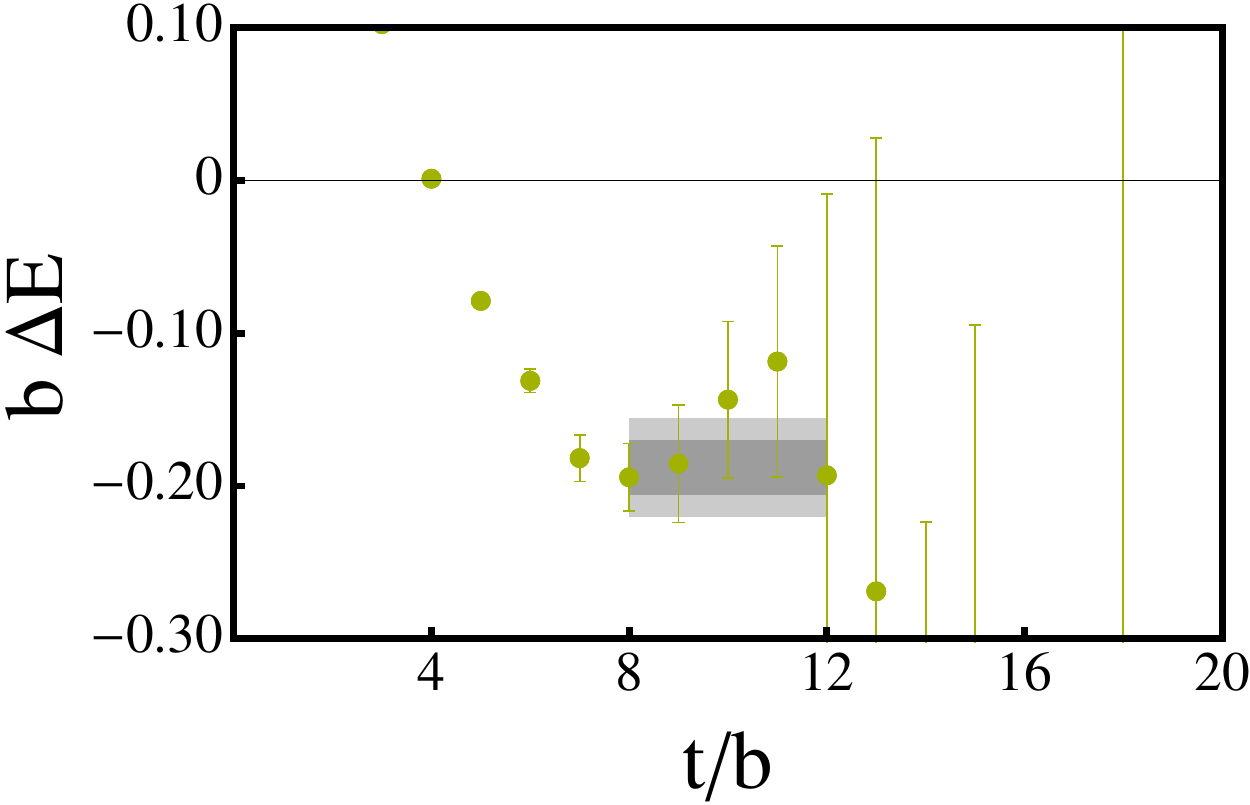}
    \includegraphics[width=0.32\textwidth]{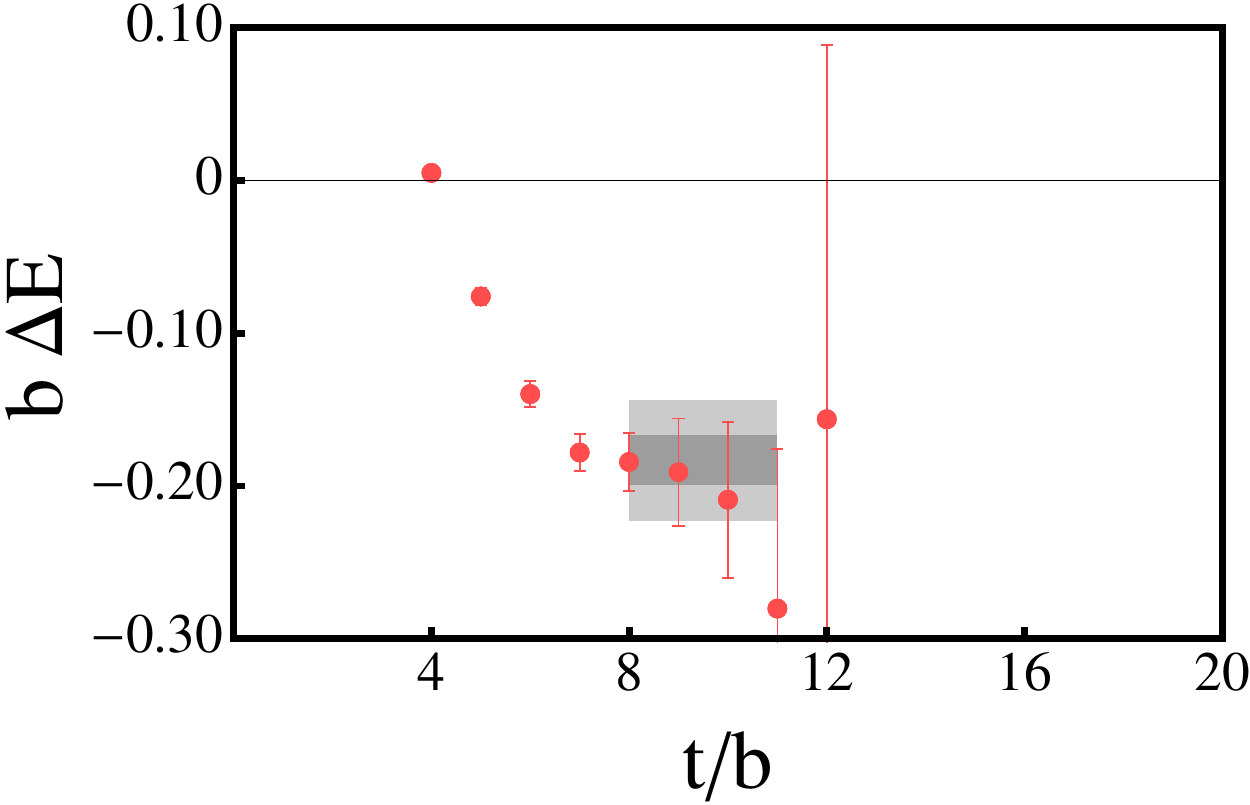}
    \caption{The EMPs of the single correlation function for the
      \hymag\ state.
 \textcolor{\revcolor}{
The inner (darker) shaded region corresponds to the statistical uncertainty
of the extracted energy, while  the outer (lighter) shaded region
corresponds to the
statistical and fitting systematic uncertainties combined in quadrature.
}}
    \label{fig:hymagEMPs}
  \end{center}
\end{figure}

\noindent
The \hymag\  state has  $I=0$, $s=-3$, $J^\pi=3/2^+$, and belongs to a
$\overline{\bf 10}$ irrep of SU(3).
Extending the standard hypernuclear nomenclature, it may be referred to as
$_{\Lambda\Xi^0}^{~\:\:\:5}{\rm H}$. 
Experimentally, it is not clear
how such a state could be produced and, given the two-body
interactions, it is not expected to be bound at the physical values of
the light-quark masses.
\begin{table}
\begin{center}
\begin{minipage}[!ht]{16.5 cm}
  \caption{
    The calculated binding energies in $_{\Lambda\Xi^0}^{~\:\:\:5}{\rm H}$.
    ``g.s.''  denotes the ground state.
The first uncertainty is statistical, the second is the fitting systematic and
the third is due to the lattice spacing.
  }
\label{tab:hymagenergies}
\end{minipage}
\setlength{\tabcolsep}{1em}
\begin{tabular}{c|ccc}
\hline
 \hef        & \cfga & \cfgb & \cfgc \\
      \hline
      g.s.   (MeV) 
      &  \textcolor{\revcolor}{273(19)(39)(3) }
      &  \textcolor{\revcolor}{255(25)(37)(3) } 
      &  \Bhymag  \\
\hline
\end{tabular}
%noalign{\smallskip\hrule}\cr}
\begin{minipage}[t]{16.5 cm}
\vskip 0.0cm
\noindent
\end{minipage}
\end{center}
\end{table}     
 \textcolor{\revcolor}{
The EMPs for this system in each of the lattice volumes are shown in 
fig.~\ref{fig:hymagEMPs}, from which it is clear that the lowest state is
negatively shifted with the energies given in Table~\ref{tab:hymagenergies}.
}
 \textcolor{\revcolor}{
It is not clear that the $\overline{\bf 10}$ contains the ground
  state of the system, or if it corresponds to a continuum state.
}

While it is interesting to study this state for algorithmic reasons, 
the states of more importance are those that can be accessed experimentally,
those with $s=0,-1,-2$.
These more interesting systems have baryons in a relative
$p$-wave, i.e. $p$-shell nuclei and hypernuclei, and require retaining the
lower components
of the quark fields in the local operators by parity considerations.
Unfortunately, we find that such operators have poor overlap onto such systems,
and produce noisy correlation functions.
These nuclei can be accessed with nonlocal operators and are the subject of future work.

%%%%%%%%%%%%%%%%%%%%%%%%%%%%%%%%%%%%%
\section{Summary and Conclusions}
\label{sec:Conclusions}
\begin{figure}[!ht]
  \centering
  \includegraphics[width=0.99\textwidth]{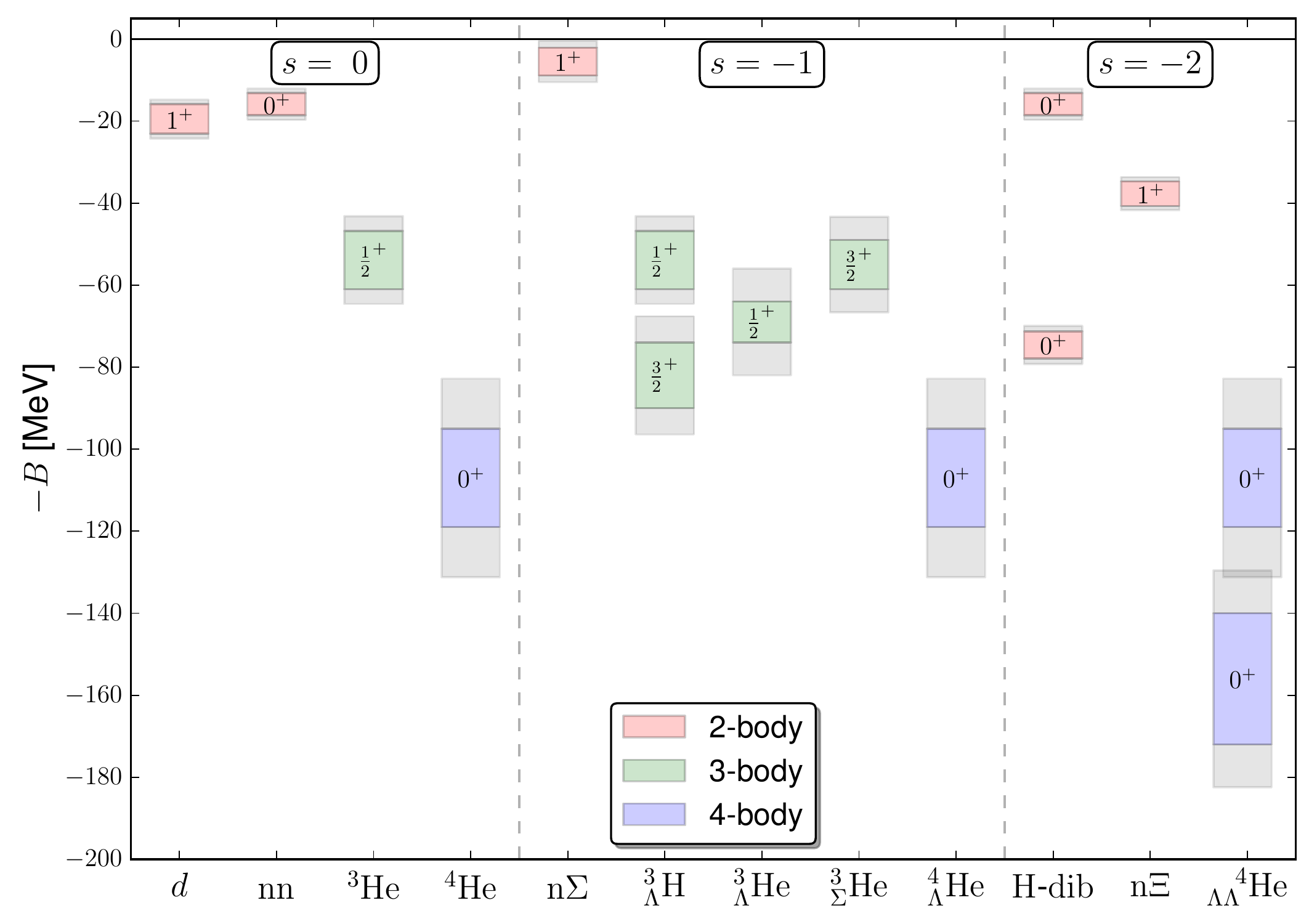}
  \caption{A compilation of the nuclear energy levels,
with spin and parity  $J^\pi$, 
 determined in this work.
}
  \label{fig:NuclearSummary}
\end{figure}

\noindent
We have presented the results of Lattice QCD calculations of various of the
lightest nuclei and hypernuclei with $A\leq5$
and with light-quark masses at the
(unphysical) SU(3)-flavor symmetric point equal to the
physical strange quark mass.  
These calculations
were
performed in three lattice volumes with spatial extent \Lafm, \Lbfm\ and
\Lcfm, and with one lattice spacing of $b\sim\blattapprox$.  
Using a new algorithm to perform the Wick contractions,
ground-state energies of a number of nuclear states
were determined from one or more correlation function(s) 
generated from local quark-level operators 
for systems at rest or  moving in the lattice volumes.
A summary of the binding energies determined in this work can be found in
Table~\ref{tab:bindingsummary},
and is shown  in fig.~\ref{fig:NuclearSummary}.
\begin{table}
\begin{center}
\begin{minipage}[!ht]{16.5 cm}
  \caption{
    Summary of the extracted ground-state binding energies of the nuclei and hypernuclei
    studied in this work.}
\label{tab:bindingsummary}
\end{minipage}
\setlength{\tabcolsep}{1em}
\resizebox{\linewidth}{!}{%
\begin{tabular}{c|ccccccc}
\hline
      State & $A$ & $s$ & $I$ & $J^\pi$ & SU(3) irrep & Binding Energy
      (MeV) & $\sim B/A$ (MeV) \\
      \hline
      $d$ (deuteron) & 2 & 0 & 0 & $1^+$ & $\overline{\bf 10}$ & \Bd & 10 \\
      $nn$ (di-neutron) & 2 & 0 & 1 & $0^+$ & ${\bf 27}$ & \Bnn & 8 \\
      $n\Sigma$ & 2 & -1 & $\frac{3}{2}$ & $1^+$ & ${\bf 10}$ & \BnSJone & 3 \\
      $H$ (H-dibaryon) & 2 & -2 & 0 & $0^+$ & {\bf 1} & \BHgs & 37 \\
      $n\Xi$ & 2 & -2 & 0 & $1^+$ & ${\bf 8}_A$ & \BnXJone & 19 \\
      \hline
      \hetppn ,  \htpnn  & 3 & 0 & ${1\over 2}$ & ${1\over 2}^+$ &
      $\overline{\bf 35}$ & \Bhet & 18 \\
      \htL (hypertriton) & 3 & -1 & 0 & ${1\over 2}^+$ & $\overline{\bf 35}$ &
      \BhtLJhalf & 18 \\
      \htL (hypertriton) & 3 & -1 & 0 & ${3\over 2}^+$ & $\overline{\bf 10}$ &
      \BhtLJthreehalf & 27\\
      \hetL ,  \htLtilde  , \nnL & 3 & -1 & 1 &  ${1\over 2}^+$  & {\bf 27} &
      \BhetLJhalf & 23\\
      \hetS & 3 & -1 & 1 & $\frac{3}{2}^+$ & {\bf 27} & \BhetS & 18\\
      \hline
      \hef & 4 & 0 & 0 & $0^+$ & $\overline{\bf 28}$ & \Bhef & 27\\
      \hefL , \hfL & 4 & 0 & 0 & $0^+$  & $\overline{\bf 28}$ & \BhefL & 27 \\
      \hefLL , \hfLL , \nnLL & 4 & 0 & 0 & $0^+$ & ${\bf 27}$  & \BhefLLtwentyseven & 39\\
\hline
\end{tabular}
}
%noalign{\smallskip\hrule}\cr}
\begin{minipage}[t]{16.5 cm}
\vskip 0.0cm
\noindent
\end{minipage}
\end{center}
\end{table}     
The approximate binding energy per baryon, which is seen to
be significantly larger than found in nature,
is also shown in  Table~\ref{tab:bindingsummary}.

In contrast to QCD with the light-quark masses at their physical
values, at the SU(3) symmetric point all two-body channels except possibly
$N\Sigma(^3S_1)$ contain a bound state in their spectrum.  The SU(3)
{\bf 1} H-dibaryon is the most deeply bound two-body state, and its
excitation, transforming as a {\bf 27} of SU(3), is also
bound. 
The nature of the sources used in this work, each derived from the
same light-quark propagator, are such that states in the symmetric
${\bf 8}_S$ of SU(3) are not produced in the correlation functions, and
as such, we are unable to locate these states in the two-body
spectrum.
The energy splitting between the deuteron and the di-nucleon is found to be
smaller than the splittings to the other SU(3) irreps, consistent with what is
found in nature, and the result of a large-$N_c$ analysis.
\textcolor{\revcolor}{
It is interesting to note that the deuteron remains a finely tuned system even
at this heavy pion mass. In nature, the ratio of the deuteron  binding momentum to the
pion mass (which defines the range of the nuclear force) is $\sqrt{M_N
  B_d}/m_\pi\sim 0.33$, where $M_N$ is the nucleon mass and $B_d$ is the deuteron
binding energy.  This quantity is exploited as an expansion parameter in
the low-energy effective field theory description of nuclear
interactions~\cite{Chen:1999tn}.
Our calculations reveal that $\sqrt{M_N B_d}/m_\pi\sim 0.24$ at $m_\pi\sim
800~{\rm MeV}$, which, by this measure, is even more finely tuned than at the physical
light-quark masses.
}

In the three-body sector, we are able to cleanly identify the
$J^\pi={1\over 2}^+$ ground state of \hetppn\ and its isospin partner
\htpnn, and the total binding energy is determined to be \BhetMeV.  
In the case of the hypertriton, \htL, the states in both the $J^\pi={1\over
  2}^+$ and $J^\pi={3\over 2}^+$ channels are consistent with being bound
nuclear states and not continuum states.  They are both found to be deeply
bound, with the $J^\pi={3\over 2}^+$ state being somewhat more bound than the 
$J^\pi={1\over 2}^+$ state. This is in contrast to the situation in nature,
where the $J^\pi={1\over 2}^+$ hypertriton is found to be very weakly bound.
The $J^\pi={1\over 2}^+$ ground state of
\hetL, and its isospin partners \htL\ and \nnL, are cleanly identified,
with a binding energy of \BhetLJhalfMeV, which is substantially lower
than the corresponding continuum states.
Further, the $J^\pi={3\over 2}^+$ \hetS\  ground state is observed to be more
bound than continuum states but is somewhat less phenomenologically
interesting, as it does not contain an NN$\Lambda$ component.

In the case of \hef, a bound $J^\pi=0^+$ ground
state has been identified, which, while lower in energy than any of the continuum states,
cannot be unambiguously identified as a bound \hef\ nucleus because of the
precision of the calculations.  
As the sources employed for \hefL\ and
\hef\ are in the same SU(3) irrep, their spectra are
identical  in the present calculations, and as such, this 
ambiguity
is present for \hefL\ also.  
The ground state of \hefLL\ and its isospin partners \hfLL\ and \nnLL\ can be clearly identified, with a
binding energy of 
\BhefLL~{\rm MeV}.

Finally,
we have calculated correlation functions in an exotic five-baryon channel, with
$s=-3$.  
Significantly more calculations will need to be performed in order to cleanly
identify a ground state in this system, but this calculation has demonstrated
that the contractions for five-body systems can now be performed.

It is now clear, but hardly a surprise,  that the spectrum of nuclei and hypernuclei change
dramatically from light-quark masses at the SU(3) symmetric point to
the physical point.  
While we had already learned this from the recent
work on the H-dibaryon, and  nucleon-nucleon scattering lengths,
this has now been demonstrated to be true for even larger systems.  
\textcolor{\revcolor}{
While the binding energy per nucleon of the deuteron (and di-neutron) is 
about $10~{\rm MeV}$, for \hetppn\  and \hef\  it is near $25~{\rm MeV}$.
These values are significantly larger than the $1.1~{\rm MeV}$, $ 2.6~{\rm
  MeV}$ 
and $7.0 ~{\rm MeV}$, respectively,  at the physical
pion mass.
}
It will be
interesting to learn how the various thresholds for binding
evolve with the light-quark masses.  
Providing accurate binding energies for any given light-quark masses will require the inclusion of
electromagnetic effects, the leading contributions of which can be
determined at the classical level and simply added to the results of
the LQCD calculations.
\textcolor{\revcolor}{
A deeper understanding of the origin of the binding energies calculated in
this work will
require a series of nuclear few-body calculations, which are beyond the scope
of the present work.  In particular, it is important to understand the relative
contribution from the two-body, three-body, and higher-body
contributions to the $A\ge 3$ nuclei and hypernuclei, which can only be
accomplished using modern few-body techniques.
}

\textcolor{\revcolor}{
Our results suggest  that quenching in LQCD calculations produces
significantly larger errors in the binding of nuclei than it does in the hadron
masses. 
This is not too surprising given the modifications to the long-range
component of the nucleon-nucleon interaction due to quenching.
It was shown in Ref.~\cite{Beane:2002nu} that the hairpin interactions that
arise in quenched and partially-quenched theories generate exponential
contributions to the nucleon-nucleon interaction in addition to the usual
Yukawa interactions at long distances.  Therefore, one anticipates significant
modifications to the binding of nuclei, especially for finely-tuned systems.
}

By diversifying and refining the source structure used to generate the
correlation functions, the continuum states in each
channel can be explored.
In the case of two-body continuum states, such as n+\hetppn\ in the \hef\ 
channel, the  established scattering formalism of L\"uscher will allow for the
scattering phase shifts in  n+\hetppn\  to be rigorously
determined from QCD below the inelastic threshold.  
For the three-body and higher-body continuum states, further formal
developments are required in order to rigorously determine multibody S-matrix elements.

Lattice QCD has evolved to the point where first-principles
calculations of  light nuclei are now possible, as
demonstrated by the calculations at unphysically heavy light-quark
masses presented in this work.  
The experimental program
in hypernuclear physics, and the difficulties encountered in
accurately determining rates for low energy nuclear reactions, 
warrant continued effort in, and development of, the application of
LQCD to  nuclear physics.
Clearly, calculations at smaller lattice spacings at the SU(3)
symmetric point are required in order to remove the systematic
uncertainties in the nuclear binding energies at these quark masses.
\textcolor{\revcolor}{While not providing quantities that
  can be directly compared with experiment, these calculations provide
  valuable information about the quark-mass dependence of spectrum of the
  lightest nuclei, and hence the nuclear forces, 
and will shed light on the fine-tunings that are  present in nuclear physics.
}
In order to impact directly the experimental program in
nuclear and hypernuclear physics, analogous calculations must be performed at
lighter quark masses, ideally at their physical values.

\vskip0.2in

\acknowledgments
We thank R. Edwards and B. Jo\'{o} for help with QDP++ and
Chroma~\cite{Edwards:2004sx}.  We acknowledge computational support
from the USQCD SciDAC project, the National Energy Research Scientific
Computing Center (NERSC, Office of Science of the US DOE,
DE-AC02-05CH11231), the UW HYAK facility, LLNL, the PRACE Research Infrastructure
resource CURIE based in France at the Tr\`{e}s Grand Centre de Calcul, TGCC, 
and the NSF through XSEDE resources under grant number TG-MCA06N025.
SRB was supported in part
by the NSF CAREER grant PHY-0645570.  The work of EC and AP is
supported by the contract FIS2008-01661 from MEC (Spain) and FEDER.
H-WL and MJS were supported in part by the DOE grant DE-FG03-97ER4014,
and the NSF MRI grant PHY-0922770 (HYAK).
WD and KO were supported in part by DOE grants DE-AC05-06OR23177 (JSA)
and DE-FG02-04ER41302.  WD was also supported by DOE OJI grant
DE-SC0001784 and Jeffress Memorial Trust, grant J-968.  The work of TL
was performed under the auspices of the U.S.~Department of Energy by
LLNL under Contract DE-AC52-07NA27344.  The work of AWL was supported
in part by the Director, Office of Energy Research, Office of High
Energy and Nuclear Physics, Divisions of Nuclear Physics, of the
U.S. DOE under Contract No. DE-AC02-05CH11231

\appendix

%%%%%%%%%%
\section{Casimirs of SU(3) }
\label{app:su3operators}
\noindent
In order to classify the states of the nuclei into irreps of 
flavor-SU(3), the quark-level sources that generate the nuclear correlation
functions are acted on with the quadratic and cubic Casimir operators
of SU(3),
\begin{eqnarray}
  \hat C_2 & = & \sum_a\ \hat T^a \hat T^a
  \ \ ,\ \ 
  \hat C_3\ =\ \sum_{abc}\ d_{abc} \ \hat T^a \hat T^b \hat T^c
  \ \ \ .
  \label{eq:casimirs}
\end{eqnarray}
The Casimir operators acting on an irrep of SU(3) that
has a tensor representation with m upper and n lower indices,
$\hat\theta^{a_1...a_m}_{b_1...b_n}$ of dimensionality
\begin{eqnarray}
d(m,n) \ =\ {1\over 2}(m+1)(n+1)(m+n+2)
\ \ \ 
\end{eqnarray}
have eigenvalues
\begin{eqnarray}
  c_2(m,n) & = & {1\over 3}(m^2+n^2+mn) + m + n
  \nonumber\\
  c_3(m,n) & = & {1\over 18}(2m+n+3)(2n+m+3)(m-n)
  \ \ \ ,
  \label{eq:casimiraction}
\end{eqnarray}
the values of which  are given in Table~\ref{tab:casimirs}
for the relevant irreps.
\begin{table}
\begin{center}
\begin{minipage}[!ht]{16.5 cm}
  \caption{
    The values of the quadratic and cubic Casimir operators in  SU(3), $c_2(m,n)$ and $c_3(m,n)$.
  }
\label{tab:casimirs}
\end{minipage}
\setlength{\tabcolsep}{1em}
\begin{tabular}{|c|cccc|}
\hline
      irrep & m & n & $c_2$ & $c_3$\\
      \hline
      {\bf 1} & 0 & 0 & 0 & 0 \\
      {\bf 3} & 1 & 0 & ${4\over 3}$ & ${10\over 9}$ \\
      $\overline{\bf 3}$ & 0 & 1 & ${4\over 3}$ & $-{10\over 9}$ \\
      {\bf 6} & 2 & 0 & ${10\over 3}$ & ${35\over 9}$ \\
      $\overline{\bf 6}$ & 0 & 2 & ${10\over 3}$ & $-{35\over 9}$ \\
      {\bf 8} & 1 & 1 & 3 & 0 \\
      {\bf 10} & 3 & 0 & 6 & 9 \\
      $\overline{\bf 10}$ & 0 & 3 & 6 & -9 \\
      {\bf 27} & 2 & 2 & 8 & 0 \\
      {\bf 28} & 6 & 0 & 18 & 45 \\
      $\overline{\bf 28}$ & 0 & 6 & 18 & -45 \\
      {\bf 35} & 4 & 1 & 12 & 18 \\
      $\overline{\bf 35}$ & 1 & 4 & 12 & -18 \\
      {\bf 64} & 3 & 3 & 15 & 0 \\
      {\bf 81} & 5 & 2 & 20 & 30 \\
      $\overline{\bf 81}$ & 2 & 5 & 20 & -30 \\
      {\bf 125} & 4 & 4 & 24 & 0 \\
\hline
\end{tabular}
%noalign{\smallskip\hrule}\cr}
\begin{minipage}[t]{16.5 cm}
\vskip 0.0cm
\noindent
\end{minipage}
\end{center}
\end{table}     
%

%%%%%%%%%%%%%%%%%%%%%%%%%%%%%%%%%
\section{The Expected Continuum  States in  the  Finite Lattice Volumes}
\label{sec:scatt_states}
\noindent
Given the single-hadron and two-body energies
that have been extracted in Secs.~\ref{sec:B} and \ref{sec:BB}, 
the continuum  states that are expected to arise 
in the three-body sectors with  given  quantum numbers
can be estimated.
Similarly, the
information obtained for the three-body systems extracted in
Sec.~\ref{sec:BBB} 
allows for an estimate of the 
continuum states in the
four-body sector, and so forth in higher-body systems.
In the figures in the main text, this information has been
presented as the infinite-volume thresholds for the various
possible continuum  channels. 
Here, we present an example  of the expected spectrum of states in the 
\hef\ system in the different lattice volumes used in this work.

\begin{figure}[h]
  \begin{center}
  \includegraphics[width=0.99\textwidth]{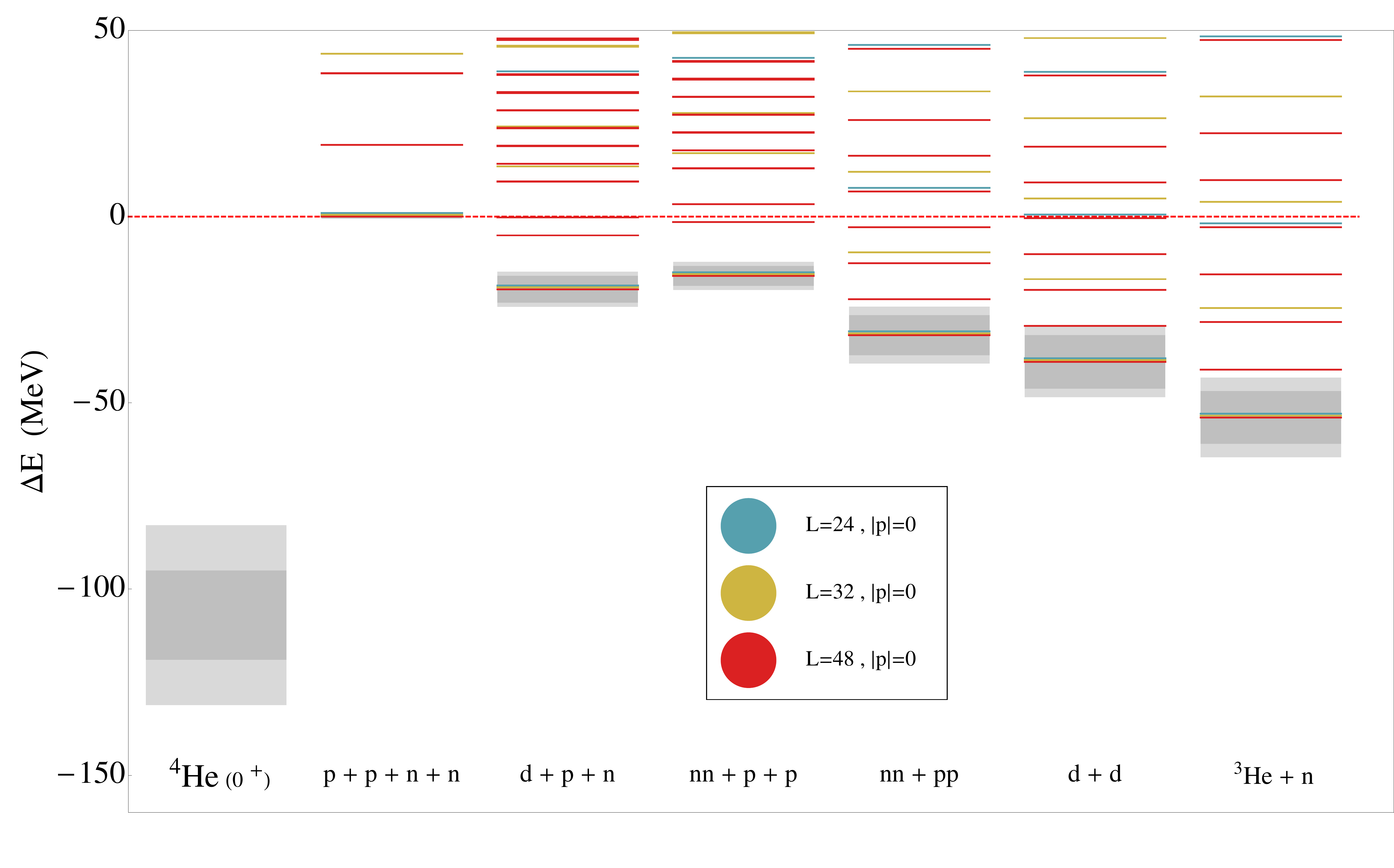}
    \caption{Expected energy levels in the $J^\pi=0^+$ \hef\ 
sector. 
The blue, green and red lines in each column denote the location of
non-interacting continuum levels in  
the \cfga, \cfgb and \cfgc ensembles, respectively.
The location of the states in the \cfga and \cfgb
ensembles have been displaced slightly for demonstrative purposes.
}
    \label{fig:ap2:1}
  \end{center}
\end{figure}

For a non-interacting 
two-component system, comprised of nuclei $A_1$ and $A_2$, 
the individual components have only back-to-back momenta,
\begin{equation}
  E^{(\rm cont.)}_{A_1,A_2}=\sqrt{M_{A_1}^2+|{\bf p}|^2}+\sqrt{M_{A_2}^2+|{\bf
      p}|^2}
\,.
\end{equation}
For three or more cluster continuum  states (for
example $d+p+n$ in the \hef\ channel), labeling the clusters $A_1$,
$A_2,\ldots, A_n$, 
the system has  energies permitted by momentum conservation
\begin{equation}
  E^{(\rm cont.)}_{A_1,A_2,\ldots,A_n}=\delta^{(3)}\left(\sum_{i=1}^n
    {\bf p}_i\right)
\sum_{i=1}^n\sqrt{M_{A_i}^2+|{\bf p_i}|^2}\,,
\end{equation}
with the obvious generalization to systems with a nonzero center-of-mass momentum. 
These considerations ignore the
interactions between the clusters, which will modify the position of
the corresponding energy levels. For two-body clusters, it is expected
that there will be ${\cal O}(1/L^3)$ shifts in the continuum energies,
but for higher-body clusters the form of the energy shifts is not
known.
In fig.~\ref{fig:ap2:1} we present
the expected (ignoring interactions)
FV energy levels 
in the \hef\  sector
for each of the volumes used in this work.  

With more accurate LQCD calculations and additional interpolating operators, we
aim to 
investigate these states in the
future.
However, this makes clear the difficulty in extracting excited states in nuclei
from this type of calculation.  The continuum states rapidly accumulate as the
lattice volume becomes large, and isolating nuclear excited states above the
lowest-lying continuum states will be challenging with current technology and algorithms.

%%%%%%%%%%%%%%%%%%%%%%%

\end{document}